\documentclass[10pt,a4paper]{article}
\usepackage[utf8]{inputenc}
\usepackage[english]{babel}
\usepackage{amsmath}
\usepackage{amsfonts}
\usepackage{amssymb}
\usepackage{graphicx}
\usepackage{xcolor,tikz,datetime,multirow,lscape,multicol,float,soul}
\usepackage[outline]{contour}
\contourlength{.2pt}
\usepackage[round]{natbib}
\usepackage[affil-it]{authblk}
\usepackage[breaklinks=true,bookmarks=true,colorlinks=true,linkcolor=blue,citecolor=blue]{hyperref}

\newsavebox{\astrutbox}
\sbox{\astrutbox}{\rule[-5pt]{0pt}{20pt}}

\usepackage{graphicx}
\usepackage{newtxtext}
\usepackage{newtxmath}
\usepackage{natbib}
\usepackage{hyperref}
\hypersetup{
    colorlinks = true,
    urlcolor   = blue,
    citecolor  = black,
}
\usepackage{tikz}
\usepackage{xcolor}
\usepackage{siunitx}
\usepackage[separate-uncertainty=true]{siunitx}
\usepackage{subcaption}
\usepackage{amsmath}
\usepackage{amssymb}
\usepackage{mathtools}

\usepackage{soul,xcolor,multirow,ifthen,float,morefloats}

\usepackage{geometry}
\newlength\halflineskip
\newlength\affilskip
\usepackage{fancyhdr}
\usepackage[font=small,labelfont=sc]{caption}
\usepackage{titlesec}
\titleformat{\section}
  {\bf\large}
  {\thesection}{1em}{\vspace*{-.2cm}}
\usepackage{etoolbox}
\AtBeginEnvironment{tabular}{\small}

\usepackage{array}

\newcommand{\RomanNumeralCaps}[1]
\linenumbers

\newcommand\emt[1]{\ensuremath{#1}}
\newcommand\bd[1]{\emt{\boldsymbol{#1}}}                          
\newcommand\surf{\emt{{\cal S}}}                          
\newcommand\Int{\emt{\displaystyle\int}}                 

\title{Incipient motion of a single particle on a regular substrate in an oscillatory flow}

\author[1,2]{{\bf Timo J.J.M. van Overveld}}
\author[3]{{\bf Marco Mazzuoli}}
\author[4]{{\bf Markus Uhlmann}}
\author[2]{\\{\bf Herman J.H. Clercx}}
\author[2]{{\bf Matias Duran-Matute}}
\affil[1]{Transport Phenomena Group, Department of Chemical Engineering, Delft University of Technology, Van der Maasweg 9, 2629 HZ Delft, the Netherlands}
\affil[2]{Fluids and Flows Group, Department of Applied Physics and Science Education, Eindhoven University of Technology, PO Box 513, 5600 MB Eindhoven, the Netherlands}
\affil[3]{Department of Civil, Chemical and Environmental Engineering, University of Genoa, Via Montallegro 1, 16145 Genova, Italy}
\affil[4]{Institute for Water and Environment, Karlsruhe Institute of Technology, 76128 Karlsruhe, Germany}

\date{May, 2025}
\begin{document}
\maketitle

\begin{abstract}
    We investigate and model the initiation of motion of a single particle on a structured substrate within an oscillatory boundary layer flow, following a mechanistic approach.
By deterministically relating forces and torques acting on the particle to the instantaneous ambient flow, the effects of flow unsteadiness are captured, revealing rich particle dynamics.
Laboratory experiments in an oscillatory flow tunnel characterise the initiation and early stages of motion, with particle imaging velocimetry measurements yielding the flow conditions at the motion threshold. The experiments validate and complement results from particle-resolved direct numerical simulations, combining an immersed boundary method with a discrete element method that incorporates a static friction contact model.
Within the parameter range just above the motion threshold, the mobile particle rolls without sliding over the substrate, indicating that motion initiation is governed by an unbalanced torque rather than a force. Both experimental and numerical results show excellent agreement with an analytical torque balance including hydrodynamic torque derived from the theoretical Stokes velocity profile, and contributions of lift, added mass, and externally imposed pressure gradient. 
In addition to static and rolling particle states, we identify a \emph{wiggling} regime where the particle moves but does not leave its original pocket.
Our deterministic approach enables prediction of the phase within the oscillation cycle at which the particle starts moving, without relying on empirical threshold estimates, and can be extended to a wide range of flow and substrate conditions, as long as turbulence is absent and interactions with other mobile particles are negligible.
\end{abstract}



\newpage
\section{Introduction}\label{sec:Introduction}
The capability to predict the onset of particle motion is fundamental to numerous natural and industrial processes, ranging from sediment transport in rivers, estuaries, and coastal waters to slurry flows in dredging and mining operations \citep{vanrijn2008principles,wierschem2008ripple}.
Identifying the threshold conditions for the transport of pollutants, such as microplastics ($1\mu$m$-5$mm), in environmental settings also requires a deep understanding of particle dynamics, which, for slightly buoyant particles, are particularly sensitive to hydrodynamic forces \citep{kane2020seafloor}. 
Therefore, a detailed examination of the early stages of particle motion is necessary \citep{buffington1997systematic,nielsen1992coastal,vowinckel2016entrainment,eyal2021does}.
The motion threshold has been extensively studied, driven by applications in civil and coastal engineering \citep{wilcock1993critical,petit2015dimensionless,agudo2012incipient}. 

\begin{figure}
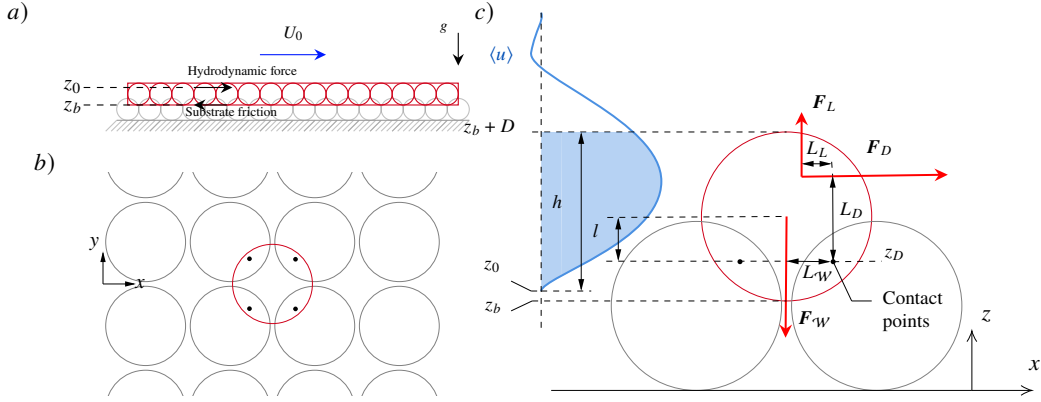

\raisebox{5cm}{\small$a)$}
\raisebox{3cm}{\small$b)$}
\begin{subfigure}[b]{0.4\textwidth}
    \centering
    \raisebox{0.5cm}{\scalebox{0.55}{\hspace*{-16.5cm}
    \input{Figures/layer}
    }
    }
    \scalebox{0.45}{\hspace*{-11.5cm}
    \input{Figures/topview_v1}
    }
\end{subfigure}
\raisebox{5cm}{\small$c)$}
\begin{subfigure}[b]{0.6\textwidth}
    \centering
    \scalebox{0.5}{\hspace*{-4.5cm}
    \input{Figures/sideview_v3}
    }
\end{subfigure}
\caption{(a) Side view of a hypothetical uniform layer of mobile spheres lying on a regular substrate, i.e. the Shields approach. (b) Top view and (c) side view of the present particle arrangement, where grey spherical particles form a fixed substrate, and the red sphere represents the mobile particle. The blue curve illustrates the Stokes velocity profile. Red arrows indicate the hydrodynamic drag, lift, and effective weight forces (added mass and imposed pressure gradient forces are not shown in this sketch). The corresponding lever arms are shown as black double-headed arrows, referenced to the downstream-side contact points (which coincide in this side view). The symbols are categorised and described in table~\ref{tab:geometry}.}
\label{fig:overview}
\end{figure}

The criterion introduced by \citet{shields1936anwendung} 
is often used to determine the conditions of incipient sediment transport. %
According to Shields' criterion, the motion of a sediment layer exposed to steady flow begins once the hydrodynamic force acting tangentially to the bottom surface exceeds the frictional resistance between the mobile layer and the underlying substrate, which is proportional to the layer's submerged weight (see figure~\ref{fig:overview}(a)). 
This balance leads to the definition of the Shields parameter, given by the ratio between tangential and vertical forces. At the threshold of incipient motion, this parameter equals the sediment friction coefficient. 
The critical value of the Shields parameter depends solely on the granular Reynolds number, which quantifies the relative contributions of inertial and viscous forces at the sediment grain scale. 
By definition, the Shields parameter is inherently a statistical measure due to the randomness of the bottom geometry, as well as flow-particle and particle-particle interactions. In practice, the threshold is typically determined empirically for an ensemble of grains on a rough bed, with critical values accompanied by subjective definitions of different stages of motion \citep{breusers1971begin}.

Recently, there has been a growing emphasis on developing more precise criteria for the motion threshold of a single particle and characterising particle dynamics immediately following the motion onset. These efforts have primarily involved combinations of detailed experiments \citep{agudo2012incipient,charru2007motion} and semi-analytical methods \citep{agudo2017shear,retzepoglu2019effect}.
Unlike the statistical nature of the Shields criterion, these approaches provide well-defined thresholds for the motion of individual spherical particles, directly linking them to specific flow conditions, material properties, local bed structure, and the influence of other surrounding particles \citep{agudo2014neighbors,agudo2017shear}.

It should be stressed that the previously mentioned studies focus exclusively on unidirectional steady flows. However, the insights gained from these studies have only limited applicability in settings with transient ambient flows.
Oscillatory flows, in particular, are critical in many engineering and environmental settings, such as in the periodic agitation of submerged particles in multiphase mixtures or the back-and-forth motion near sediment bedforms induced by coastal waves \citep{sleath1984sea,nielsen1992coastal}. 
Oscillatory flows are well-known to be more complex than unidirectional flows, even under laminar conditions. Nonlinear residual flows, such as steady streaming flows, can significantly affect particle dynamics on small and (very) large time scales compared to the period of oscillation \citep{overveld2022numerical}. The complex particle-fluid interactions can lead to self-organisation into a wide range of patterns, including chains and bands in diluted systems \citep{overveld2023pattern} or vortex ripples in dense systems \citep{sleath1984sea}.  
Despite their importance, oscillatory flows have received far less attention than unidirectional flows. To our knowledge, a similarly rigorous analytical approach for oscillatory flows has yet to be developed to determine the conditions at the onset of particle motion.

Overall, the unique description of the physics in an oscillatory flow needs one additional dimensionless parameter compared to the unidirectional case, since the system has an additional degree of freedom. This is analogous to general particle motion in oscillatory flows \citep{klotsa2007interaction,overveld2022numerical,overveld2022effect}.
Consequently, reported values for the Shields criterion (or its single-particle equivalent described by \citet{agudo2017shear}) will depend not only on the granular Reynolds number, but also on an additional dimensionless parameter related to the oscillatory flow component, such as the frequency-dependent viscous length scale relative to the exposed particle's diameter \citep{klotsa2007interaction,mazzuoli2016formation,overveld2022numerical,overveld2023pattern}.

Due to the larger parameter space, the dynamics of particle motion in oscillatory flows may differ significantly from those in unidirectional flows. In the latter, once a particle starts moving, it is slightly lifted from its pocket in the bed, exposing a larger surface area to the ambient flow and increasing the drag force induced by the ambient flow, accelerating the particle even further. This feedback mechanism produces a sharp transition between stationary and dynamic states, assuming laminar flow conditions without fluctuations.
In contrast, oscillatory flows introduce several additional complexities, including unsteady forces such as added mass and Basset history forces, which induce complex dynamics \citep{overveld2023pattern}. 
Moreover, the velocity profile in an oscillatory flow is often non-monotonic, like for a Stokes boundary layer over a flat bottom. As a result, both the drag force and associated lever arm are transient quantities, yielding a complex time-varying torque. 
The particle's dynamics can become even more intricate after the onset of motion. Depending on the flow's oscillation period relative to the particle's settling time, the particle may not settle in or even reach the next pocket in the bed before the flow reverses.

From a conceptual perspective, we classify particle behaviour into several progressive modes of motion. Initially, the particle may remain \textit{static} when the hydrodynamic forces are too weak to induce any movement. As the forcing increases, the particle may exhibit a \textit{wiggling} motion, characterised by brief periodic movements, typically near moments of maximum drag or shear stress. By definition, the amplitude of this motion is small, as the particle remains within a single pocket of the bed, falling back into its original position and coming to rest before the ambient flow reverses direction. This motion is expected to be periodic under laminar flow conditions where fluctuations are absent.
As the driving forces continue to increase, the particle may start to \textit{roll} between pockets in the bed, always maintaining contact with at least one other particle. Rolling is expected to precede any slipping or sliding, consistent with predictions for unidirectional flow \citep{agudo2017shear}. Even under laminar flow conditions, this rolling motion can be aperiodic, with the particle approaching a neighbouring pocket but not coming to rest before the flow direction reverses. 
When the driving force is sufficiently large to lift the particle temporarily from the substrate, it may \textit{hop} between pockets \citep{topic2022}. 
As the driving force increases further, the particle may eventually become \textit{suspended} for extended periods, moving freely without touching the substrate.

In this work, we comprehensively explore the threshold of particle motion under oscillatory flow conditions and the subsequent dynamics post-initiation.
We first analyse the governing equations, provide theoretical predictions for the relevant dimensionless parameters, and make quantitative predictions for the conditions at the onset of motion.
We then extend our study using laboratory experiments, exploring the parameter space to identify the threshold for different flow conditions using a combination of particle tracking and particle image velocimetry (PIV).
We complement our experimental findings with direct numerical simulations (DNS) to give insight into the transient forces acting on the particle and to explore the three-dimensional flow field near the onset of motion.
The flow field around the particles in the simulations is fully resolved, with the particles accounted for using an immersed boundary method. To faithfully reproduce the motion onset, the numerical framework is supplemented by a particle-particle contact model that accounts for the build-up and release of elastic energy around contact points when the particle is still at rest.
Throughout our broad approach, we particularly pay attention to the implications of the additional degree of freedom, distinguishing it from the more commonly studied unidirectional flow cases.

The work is structured as follows: we give an overview of the system in Sec.~\ref{sec:System}, followed by the experimental approach in Sec.~\ref{sec:ExperimentalApproach} and the numerical approach in Sec.~\ref{sec:NumericalApproach}. The results with accompanying discussion and comparison to other criteria are given in Sec.~\ref{sec:Results} and concluding remarks in Sec.~\ref{sec:Conclusion}.

\section{Formulation of the problem}\label{sec:System}
We first present a systematic overview of the physical system consisting of a single mobile, spherical particle resting on top of one fixed layer of monosized spherical particles, hereafter referred to as \emph{substrate}.
The oscillatory flow of an incompressible Newtonian fluid is driven by an externally imposed harmonic pressure gradient over a horizontal flat wall, resulting in an oscillatory boundary layer (OBL). 
Figures~\ref{fig:overview}(b)~and~(c) provide accompanying sketches of the particle and flow configuration.
For a regular substrate, the relevant dimensional parameters are the particle diameter $D$,
the amplitude of velocity oscillations far from the bottom $U_0$, %
the angular oscillation frequency $\omega$, the kinematic viscosity of the fluid $\nu$, the fluid density $\rho_f$, the particle density $\rho_s$, the gravitational acceleration $g$, the exposure height of the particle to the ambient flow $h$, and the elevation $l$ of the particle centre relative to its contact points with the bed. %
These quantities and other main parameters are listed in table~\ref{tab:geometry}.

\begin{table}
    \centering
    \begin{tabular}{lll}
        \hline
        \textbf{Category} & \textbf{Symbol} & \textbf{Description} \\
        \hline
        \textbf{General and material properties} 
            & $\rho_f$ & Fluid density \\
            & $\rho_s$ & Particle density \\
            & $\nu$ & Fluid kinematic viscosity\\
            & $g$ & Gravitational acceleration \\

        \hline
        \textbf{Oscillation parameters} 
            & $U_0$ & Amplitude of free-stream velocity oscillation \\
            & $\omega$ & Angular frequency of oscillation \\
            & $\delta_S = \sqrt{2\nu/\omega}$ & Stokes boundary layer thickness \\

        \hline
        \textbf{Geometric parameters} 
            & $D$ & Particle diameter \\
            & $z_0$ & Virtual origin of Stokes velocity profile \\
            & $z_b$ & Bottom elevation of mobile particle \\
            & $z_D$ & Contact point elevation \\
            & $h = z_b + D - z_0$ & Mobile particle exposure height\\
            & $l = z_b + D/2 - z_D$ & Particle centre elevation above contact point \\

        \hline
        \textbf{Forces and lever arms} 
            & $\boldsymbol{F}_D$ & Hydrodynamic drag force\\
            & $L_D$ & Lever arm of drag force \\
            & $\boldsymbol{F}_{\cal W}$ & Effective submerged particle weight \\
            & $L_{\cal W}$ & Lever arm of effective weight \\
            & $\boldsymbol{F}_L$ & Effective hydrodynamic lift force \\
            & $L_L$ & Lever arm of lift force \\
            & $\boldsymbol{F}_\mathrm{AM}$ & Added mass force\\
            & $\boldsymbol{F}_{\nabla p}$ & Imposed external pressure gradient\\
        \hline
    \end{tabular}
    \caption{Overview of the main parameters describing material properties, oscillation characteristics, geometric configuration (cf. Fig~\ref{fig:overview}), and relevant forces and associated lever arms.}
    \label{tab:geometry}
\end{table}

\subsection{Governing equations}    

    \subsubsection{Fluid motion}
        The flow is governed by the continuity equation
        \begin{equation}\label{eq:theory_mass}
            \boldsymbol{\nabla}\cdot\boldsymbol{u} = 0,
        \end{equation}
        and by the incompressible Navier-Stokes equation
        \begin{equation}\label{eq:theory_momentum}
            \frac{\partial \boldsymbol{u}}{\partial t} + \left(\boldsymbol{u}\cdot\boldsymbol{\nabla}\right)\boldsymbol{u} = -\frac{1}{\rho_f}\boldsymbol{\nabla} p + \nu\nabla^2\boldsymbol{u} 
            + U_0\omega\cos(\omega t)\widehat{\boldsymbol{x}}, 
        \end{equation}
        where $p$ is the pressure, $t$ is the time, and $\boldsymbol{u}=(u,v,w)$ is the fluid velocity with the components referred to the Cartesian coordinates $\bd{x}=(x,y,z)$. The unit vector $\widehat{\boldsymbol{x}}$ points in the $x$-direction, aligned with the oscillatory flow, and $z$ denotes the wall-normal coordinate pointing upwards. The last term in \eqref{eq:theory_momentum} corresponds to the imposed oscillatory pressure gradient. 
        
        In the laminar regime, the velocity in the oscillatory boundary layer over a flat, smooth wall located at $z=z_0$, is given by the Stokes boundary layer solution      \begin{equation}\label{eq:theory_solution_bounded}
            \boldsymbol{u} = U_0\left[\sin(\omega t)-\exp\left(-\frac{z-z_0}{\delta_S}\right)\sin{\left(\omega t - \frac{z-z_0}{\delta_S}\right)}\right]\widehat{\boldsymbol{x}},
        \end{equation}
        where $\delta_S=\sqrt{2\nu/\omega}$ denotes the boundary layer thickness, i.e. the characteristic viscous length scale. 
        
        The presence of bottom roughness adds a remarkable complexity to the problem, even when the roughness elements are monosized spherical particles arranged regularly on a horizontal wall \citep{mazzuoli2016formation,agudo2017shear}. Notably, roughness can amplify nonlinear effects, such as steady streaming \citep{lyne1971unsteady,riley2001steady}, and thereby influence the stability of the oscillatory boundary layer, eventually causing a transition to turbulence~\citep{mazzuoli2016transition,kaptein2020existence}.
        Presently, we consider a compact square arrangement of the bottom substrate.
        
        The influence of substrate roughness on the undisturbed flow is reflected in the velocity field overlapping with the top layer of the substrate. In other words, the flow does not vanish at the top of the substrate but instead penetrates into it to some extent. When modelling the flow using a Stokes profile, this effective penetration is accounted for by adjusting $z_0$, the virtual wall location, as illustrated in figure~\ref{fig:overview}(c).
        In practice, $z_0$ is chosen such that far above the substrate, the Stokes velocity profile matches the velocity field obtained from the DNS or experiments.
        Note that the mean velocity at $z=z_0$ in the DNS or experiments does not necessarily vanish.  
        The influence of substrate roughness on the flow experienced by the mobile particle is quantified by the particle exposure height
        \begin{equation}\label{eq:H}
            h=z_b+D-z_0,
        \end{equation}
        where $z_b$ is the bottom elevation of the mobile particle.

    \subsubsection{Particle motion}       
        The translational velocity $\boldsymbol{u}_s$ of a spherical particle is described by Newton's second law
        \begin{equation}\label{eq:theory_particle_trans}
            \rho_sV_s \frac{d\boldsymbol{u}_s}{dt} = \boldsymbol{F}_B + \boldsymbol{F}_C + \boldsymbol{F}_S,
        \end{equation}
        where
        $V_s=\pi D^3/6$ is the particle volume, %
        $\boldsymbol{F}_B$ is the resultant of body forces, %
        $\boldsymbol{F}_C$ is the resultant of inter-particle contacts due to interactions with the substrate, and %
        $\boldsymbol{F}_S$ is the resultant of surface forces
        \begin{equation}
            \boldsymbol{F}_S=\displaystyle\int_{\cal S}\left(-(p+P)\boldsymbol{n} + \boldsymbol{\tau}_\nu\right) dS,
        \end{equation}
        $P$ denoting the imposed pressure, $\boldsymbol{\tau}_\nu$ the viscous stress tangential to the sphere surface ${\cal S}$, and $\boldsymbol{n}$ the surface-normal unit vector.
        In the $x$-direction, following Maxey-Riley-Gatignol's approach, the hydrodynamic force $\boldsymbol{F}_S$ can be modelled as the sum of the drag force and contributions from added mass, Basset force, and imposed pressure gradient \citep{maxey1983equation, gatignol1983faxen}. 
        While not strictly exact, this decomposition provides a reasonable approximation for small particles, where drag is predominantly viscous and the effects of flow unsteadiness, captured by the added mass, Basset, and pressure gradient terms, can be effectively treated as additional contributions to the steady-state drag model.
        In the $z$-direction, the hydrodynamic force corresponds to the lift force. %

        The net body force $\boldsymbol{F}_B$, proportional to the volume occupied by the particle, results from the sum of contributions due to gravity and buoyancy
        \begin{equation}\label{eq:theory_particle_external}
            \boldsymbol{F}_B = - \left(\rho_s-\rho_f\right)V_sg\hat{\boldsymbol{z}},
        \end{equation}
        where $g$ is the gravitational acceleration pointing in the negative $z$\nobreakdash-direction, $-\hat{\boldsymbol{z}}$.
        
        The rotation of a spherical particle is described by Euler's second law
        \begin{equation}\label{eq:theory_particle_rot}
            \frac{\rho_s\pi D^5}{60}\frac{d\boldsymbol{\Omega}_s}{dt} = \boldsymbol{T}_C + \boldsymbol{T}_S,
        \end{equation}
        where $\boldsymbol{\Omega}_s$ denotes the particle angular velocity. The torques acting on the particle arise from interactions with the substrate, $\boldsymbol{T}_C$, and hydrodynamic viscous stresses
        \begin{equation}
            \boldsymbol{T}_S=\displaystyle\int_{\cal S} \left(\boldsymbol{x}-\boldsymbol{x}_c\right)\times\boldsymbol{\tau}_\nu\; dS,
            \label{eq:Ts}
        \end{equation}
        with $\boldsymbol{x}_c$ denoting the particle centre coordinates.  
        For a rolling particle in the compact square arrangement (cf. figure~\ref{fig:overview}(b)), rotation occurs around the contact point at elevation $z_D$. The distance between $z_D$ and the particle centre elevation is given by
        \begin{equation}\label{eq:L}
            l = z_b+\frac{D}{2}-z_D,
        \end{equation}
        as shown in figure~\ref{fig:overview}(c).

\subsection{Dimensional considerations}\label{sec:system_dimensional}   
    Let us now consider the following dimensionless variables of the hydrodynamic problem, indicated with asterisks,
    \begin{equation}\label{eq:theory_dimlesvars}
        \boldsymbol{u^*}=\dfrac{\boldsymbol{u}}{U_0}, \quad 
        t^*=\omega t, \quad
        \boldsymbol{x}^* = \dfrac{\boldsymbol{x}}{\delta_S}
        \quad
        \boldsymbol{\nabla}^* = \delta_S\boldsymbol{\nabla}, \quad
        p^*=\dfrac{p}{\rho_f \delta_S U_0\omega},
    \end{equation}
    which were also considered by \citet{mazzuoli2016transition}. 
    Moreover, the following dimensionless quantities characterising the particle dynamics are introduced %
    \begin{align}\label{eq:theory_dimlesvars2}
        \boldsymbol{x}_s^* &= \dfrac{\boldsymbol{x}_s}{D},
        \quad
        \boldsymbol{u}_s^* = \frac{\boldsymbol{u}_s}{U_0}, \quad 
        \boldsymbol{\Omega}_s^* = \frac{D\boldsymbol{\Omega}_s}{U_0}, 
        \quad
        \boldsymbol{F}_C^* = \frac{\boldsymbol{F}_C}{{\cal W}_s},
        \quad
        \boldsymbol{T}_C^* = \frac{2\boldsymbol{T}_C}{{\cal W}_sD}, \nonumber \\
        \quad
        \boldsymbol{F}_S^* &= \frac{\boldsymbol{F}_S}{\tau_{\nu0}\pi D^2}, 
        \quad
        \boldsymbol{T}_S^* = \frac{2\boldsymbol{T}_S}{\tau_{\nu0}\pi D^3}, 
        \quad
        \boldsymbol{F}_B^* = \frac{\boldsymbol{F}_B}{{\cal W}_s},
        \quad
        \boldsymbol{T}_B^* = \frac{2\boldsymbol{T}_B}{{\cal W}_sD},
    \end{align}
    where the contact forces are normalised by the particle submerged weight ${\cal W}_s=(\rho_s-\rho_f)gV_s$, and the hydrodynamic forces are normalised by the characteristic viscous shear stress $\tau_{\nu0}=\rho_fU_0\omega\delta_S/2$ multiplied by the area of the particle surface $\pi D^2$.
    
    We now use \eqref{eq:theory_dimlesvars}~and~\eqref{eq:theory_dimlesvars2} to nondimensionalise the governing equations for both the fluid phase and the particle motion. 
    Hence, the dimensionless form of \eqref{eq:theory_mass}~and~\eqref{eq:theory_momentum} reads
    \begin{equation}
        \boldsymbol{\nabla}^*\cdot \boldsymbol{u}^*= 0,
    \end{equation}
    and 
    \begin{equation}\label{eq:theory_momentum_dimless}
        \frac{\partial \boldsymbol{u}^*}{\partial t^*} + \frac{1}{2}\mathrm{Re}_\delta\left(\boldsymbol{u}^*\cdot\boldsymbol{\nabla}^*\right)\boldsymbol{u}^* = -\boldsymbol{\nabla}^*p^* + \frac{1}{2}\nabla^{*2}\boldsymbol{u}^* + \cos(t^*)\hat{\boldsymbol{x}},
    \end{equation}
    respectively, where 
    \begin{equation}\label{eq:Redelta}
        \mathrm{Re}_\delta = \frac{U_0 \delta_S}{\nu}
    \end{equation}
    is the Reynolds number of the oscillatory boundary layer.
    Substitution into \eqref{eq:theory_particle_trans}~and~\eqref{eq:theory_particle_rot} yields
    \begin{equation}\label{eq:theory_particle_trans_dimless}
        s\frac{d\boldsymbol{u}_s^*}{dt^*} = \frac{1}{\Gamma}\left(\boldsymbol{F}_C^* - \hat{\boldsymbol{z}}\right) + 3\delta\boldsymbol{F}_S^*,
    \end{equation}
    for the particle translation,
    and
    \begin{equation}\label{eq:theory_particle_rot_dimless}
        \frac{s}{5}\frac{d\boldsymbol{\Omega}^*_s}{dt^*} = \frac{1}{\Gamma}\boldsymbol{T}_C^*+3\delta\boldsymbol{T}_S^*,
    \end{equation}
    for the particle rotation, where
    \begin{equation}\label{eq:densityratio}
        s = \frac{\rho_s}{\rho_f}
    \end{equation}
    is the particle-fluid density ratio,     
    \begin{equation}\label{eq:delta}
        \delta = \frac{\delta_S}{D} 
    \end{equation}
    is the normalised viscous length scale, and
    \begin{equation}
        \Gamma = \frac{\rho_fU_0\omega}{(\rho_s-\rho_f)g}    
    \end{equation}
    is the ratio between the amplitude of the oscillatory acceleration far from the bottom and the specific gravitational acceleration. 
    From this point onward, all variables are nondimensional unless otherwise indicated, and the asterisks are omitted for notational convenience. %

    According to \eqref{eq:theory_particle_trans_dimless}, when lift forces (i.e., the vertical component of $\boldsymbol{F}_S$) are relatively small, vertical motion is primarily determined by the balance between contact forces and effective weight, with the dimensionless acceleration being inversely proportional to $s\Gamma$.
    Horizontal motion is governed by the horizontal components of the contact force $\boldsymbol{F}_C$ and the hydrodynamic force $\boldsymbol{F}_S$, the latter comprising contributions from the imposed pressure gradient and drag, which are proportional to $\delta/s$. 

    The dynamic problem involves nine distinct dimensional parameters ($U_0$, $D$, $\omega$, $\rho_f$, $\rho_s$, $\nu$, $g$, $h$, $l$), 
    and is, therefore, uniquely characterised by six independent dimensionless parameters.
    Four of these already appeared in the governing equations: $\delta$, $\mathrm{Re}_\delta$, $s$, and $\Gamma$. 
    In addition, the dynamics of the exposed particle depend on two geometric parameters, $H=h/D$ and $L = l/D$, which relate to the substrate arrangement and are defined by \eqref{eq:H} and \eqref{eq:L}, respectively.
    
    Combinations of these numbers result in other dimensionless quantities that are frequently considered in sediment transport problems, such as the particle mobility number     
    \begin{equation}\label{eq:psi}
        \psi = \frac{\rho_fU_0^2}{(\rho_s-\rho_f)gD} = \frac{1}{2}\delta\mathrm{Re}_\delta\Gamma,
    \end{equation}
    which is the ratio between the convective force and the submerged weight of the particle \citep{mazzuoli2016formation,mazzuoli2019direct,vittori2020sediment}. 

    Finally, the bulk flow velocity $U_0$ does not necessarily represent the local flow around the mobile particle, especially when $\delta$ is large. The flow conditions near the substrate are better captured by the Shields parameter, representing the dimensionless shear stress at the bed as
    \begin{equation}\label{eq:Shields}
        \theta = \frac{\tau_0}{(\rho_s-\rho_f)gD},
    \end{equation}
    where $\tau_0$ denotes the instantaneous bottom shear stress.
    According to the Shields criterion, this stress represents the horizontal hydrodynamic force per unit area acting on the top layer of particles resting on the substrate. This force is approximately equal to the shear stress evaluated at an elevation $z_0$, which typically lies within one particle diameter $D$ above the base of the top layer $z_b$ (see figure~\ref{fig:overview}~a). 
    In our configuration, where the mobile layer consists of a single particle and the particle-induced disturbance to the flow is negligible, $z_0$ is close to $z_b$ (see figure~\ref{fig:overview}~c). Consequently, the shear stress at $z_0$ closely approximates the actual shear stress acting at the base of the mobile particle.
    In the absence of a substrate (i.e., a single sphere lying on a smooth wall), $\tau_0$ can be computed from the Stokes profile~\eqref{eq:theory_solution_bounded}, yielding $\theta= (\sqrt{2}/2)\delta\Gamma\sin\left(t+\pi/4\right)$. 
    Finally, we stress that, in contrast to steady flow conditions, the Shields parameter $\theta$ in \eqref{eq:Shields} oscillates over time, implying that both the magnitude and the duration of the hydrodynamic forcing determine the onset of motion. 


\subsection{Model for the motion threshold}\label{sec:system_model}
    The detailed derivation underlying the model for the motion threshold is given in Appendix~\ref{sec:app_model}, while the main concepts are summarised in this section.
    
    Under the assumption that an exposed particle is more likely to roll rather than slip or slide out of a pocket in the bed \citep{agudo2017shear}, the motion threshold is governed by a torque balance. 
    This balance includes a stabilising contribution from the submerged weight and typically destabilising contributions from drag, lift, added mass, and the imposed pressure gradient. However, these torques may be out of phase, meaning some may have stabilising effects during parts of the oscillation cycle. The Basset history term is omitted from the present balance, since its contribution is expected to be small for the range of $\mathrm{Re}_\delta$ considered. In the absence of a substrate, its magnitude is approximately $10-20\%$ of drag, while precise evaluation in the present geometry remains challenging. Moreover, the phase of maximum history force, being related to the flow acceleration, does not coincide with the phase of maximum drag, further limiting its influence on the onset of particle motion.
    The dimensionless spanwise component of the particle torque $\bd{T}$ referred to an arbitrary point is defined by
    \begin{equation}
        T_{y} = T_{Sy}^\mathrm{drag}+T_{Sy}^\mathrm{lift}+T_{Sy}^\mathrm{AM}+T^{\nabla p}_{Sy}+ T_{By} + T_{Cy},
        \label{eq:torque}
    \end{equation}
    where the terms on the right-hand side denote contributions of drag, lift, added mass, pressure gradient, submerged weight, and inter-particle contacts, respectively. %
    In this section, let us evaluate the torque contributions referring to the contact point elevation at the incipient rolling conditions, i.e. when $T_{y}=0$ and the mobile particle has only two contact points aligned in the spanwise direction, such that also $T_{Cy}=0$. %
    Therefore, in the definition of the spanwise component of the hydrodynamic torque $T_{Sy}$ provided by equation~\eqref{eq:Ts}, the coordinates $(x_c,z_c)$ are replaced by $(x_D,z_D)$. %
    
    The primary challenge lies in relating the hydrodynamic torque contributions to the ambient flow, for which we adopt an approach similar to that of \citet{agudo2017shear} for steady shear flows. 
    A key assumption underlying the present model is that Stokes' solution~\eqref{eq:theory_solution_bounded}, which refers to the virtual wall elevation $z_0$, provides an adequate representation of the mean flow velocity. %
    The validity of this assumption is assessed in figure~\ref{fig:DNS_velocity_profile_comparison} (Appendix~\ref{sec:app_model}), where it is shown that it holds for the parameter values presently considered.     

    In this approach, the hydrodynamic torque on the particle due to drag is assumed to be related to the first angular moment of the horizontal plane-averaged velocity profile, defined as 
    \begin{equation}
        \langle u\rangle(z) \equiv \frac{1}{\cal A}\iint_{\cal A} u\, dx dy,
    \end{equation}
    where $\cal A$ denotes the substrate area. The plane-average accounts for spatial velocity variations induced by the substrate, which are present even in non-turbulent flow conditions.
    Notably, the absence of turbulent fluctuations in the ambient flow does not exclude the possibility that random velocity fluctuations may appear in the wake downstream of the mobile particle.
    The torque due to the hydrodynamic drag is defined as the moment of the drag force with respect to the contact point between the particle and the substrate. This torque is modelled as
    \begin{equation}
        T_D\propto \pi D^2\,u_1,
    \end{equation}
    according to \eqref{eq:app_torque}, where
    \begin{equation}
        u_1 
        = 
        \dfrac{1}{D}\int_{z_b}^{z_b+D} (z-z_D) \langle u\rangle dz, 
    \end{equation}
    is the first moment of the streamwise ambient velocity (see \eqref{eq:app_u_moments})
    with $z_b$ and $z_D$ denoting the $z$-coordinates of the lowest point of the exposed particle and of the rotation axis, respectively (see figure~\ref{fig:overview}(c)).
    Similarly, the contribution of lift to the hydrodynamic torque is based on \citeauthor{saffman1965lift}'s~\citeyearpar{saffman1965lift} expression, see \eqref{eq:app_lift_saffman}. 
    Notably, both the expressions of drag and lift in dimensionless form contain correction coefficients that depend on the particle Reynolds number 
    \begin{equation}\label{eq:Res}
        \mathrm{Re}_s 
        = 
        \dfrac{u_cD}{\nu}
        = 
        \frac{\mathrm{Re}_\delta}{\delta}\left[\sin(t)-e^{-\zeta}\sin\left(t-\zeta\right)\right],
    \end{equation}
    where $u_c$ denotes the ambient velocity at the same elevation as the particle centre, and $\zeta=(H-1/2)/\delta$ is the normalised elevation of the particle centre above the zero level of the velocity profile $z_0$.

    The ratio $\Upsilon$ between the aforementioned destabilising contributions and the stabilising contribution due to submerged weight is given by 
        \begin{equation}\label{eq:criticalcondition}
        \begin{split}
        \Upsilon 
        &\equiv 
        \dfrac{T_{Sy}^\mathrm{drag}+T^{\nabla p}_{Sy}+T_{Sy}^\mathrm{AM}+T_{Sy}^\mathrm{lift}}{T_{By}}, \\
        \end{split}
        \end{equation}
    which allows us to define the following time-dependent criterion for the initiation of rolling motion, which depends on six dimensionless numbers:
    \begin{equation}\label{eq:upsilon1}
        \Upsilon\left(t;\,\mathrm{Re}_\delta,\,\delta,\,\Gamma,\,s,\,H,\,L\right) \geq 1.
    \end{equation}
    The full expression for $\Upsilon$ is provided in \eqref{eq:upsilon}. 

    Both $H$ and $L$ are primarily determined by the geometry of the bottom substrate.
    For a substrate composed of spheres in a compact square arrangement, $L = \sqrt{2}/4 \approx 0.35$. 
    The value of $H$ is related to the effective position of the virtual origin $z_0$ of the velocity profile within the substrate crevices.
    In a similar substrate configuration but under steady flow conditions, \citet{agudo2017shear}~and~\citet{topic2022} positioned $z_0$ at a distance $0.077D$ below the top of the substrate, corresponding to $H=2L+0.077\approx0.78$. %
    In our DNS results, $H\approx 0.84$ (cf. table~\ref{tab:simparams}). %

\begin{figure}
        \centering
        \includegraphics[width=\linewidth]{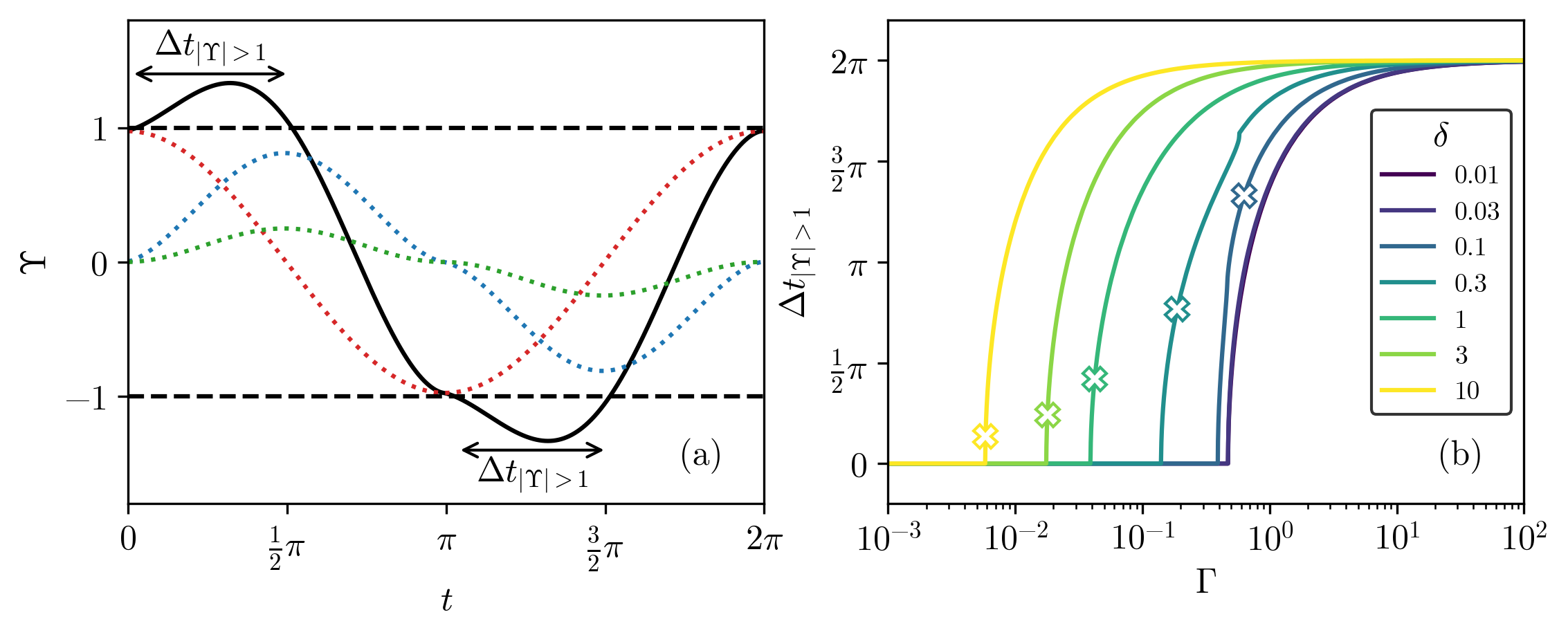}
        \caption{(a) Temporal evolution of the motion threshold parameter $\Upsilon$, as defined by \eqref{eq:criticalcondition}, for typical parameter values in the present experiments and simulations (specifically matching DNS run 3, see table~\ref{tab:simparams}): $\delta=0.12, \mathrm{Re}_\delta=41, \Gamma=0.45, s=1.81, H=0.8$ and $L=0.36$ (black line). Dotted lines represent relative contributions of hydrodynamic drag (blue), lift (green), and added mass plus imposed pressure gradient (red). (b) Normalised time interval $\Delta t_{\Upsilon>1}$ during which the motion threshold condition~\eqref{eq:upsilon1} is exceeded, as a function of $\Gamma$ (proportional to the forcing strength) for various values of $\delta$, with other parameters as in (a). Crosses mark the characteristic time scale $\tau$ \eqref{eq:time_scaling} for the particle to reach the crest of the substrate.}
        \label{fig:LHS_time}
    \end{figure}

    To get familiar with the $\Upsilon$\nobreakdash-criterion \eqref{eq:criticalcondition}-\eqref{eq:upsilon1} and distinguish the different contributions to the particle torque, figure~\ref{fig:LHS_time}(a) shows the time variation of $\Upsilon$ for constant parameters $H=0.8$, $L=0.35$, and $\Gamma=0.45$, which are representative of the present experimental and numerical conditions.
        
\subsection{Particle motion just above the motion threshold}
    Assuming that the particle rolls during the early stages of its motion, as verified in \S~\ref{sec:results_qualitative}, we define the spanwise rotation angle $\phi_s$ as the angle swept by the particle while rolling from its initial position to the crest of the substrate particle, before settling into the adjacent pocket. For the present substrate configuration, a compact square arrangement, we find $\phi_s=\pi/2-\arctan(\sqrt{2})\approx 0.62$.
    The angle $\phi_s$ may alternatively be obtained by integrating Euler's second law \eqref{eq:theory_particle_rot_dimless} twice with respect to time, which yields \begin{equation}\label{eq:theory_particle_angle_dimless_TvO}
        \phi_s = \iint_0^\tau \frac{d\Omega_{sy}}{dt} {dt}' dt
        \approx \frac{5\delta\mathrm{Re}_\delta}{4s}\tau^2\left(3 
        \delta T_{Sy}-\frac{1}{\Gamma}T_{Cy}
        \right),
    \end{equation}
    where $\tau$ denotes the characteristic (dimensionless) time for the particle to reach the substrate crest, hereafter referred to as the \textit{time to crest}.
    Close to the motion threshold, the difference between stabilising and destabilising torques is small. Based on the scaling analysis in \S\ref{sec:system_dimensional}, the resultant dimensionless torque (i.e. the left-hand side of \eqref{eq:theory_particle_rot_dimless}) is expected to be of order unity during the early stages of motion. Furthermore, the same analysis suggests that $3\delta T_{Sy}$ and $T_{Cy}/\Gamma$, i.e. the terms between brackets in \eqref{eq:theory_particle_angle_dimless_TvO}, are typically of the same order of magnitude. 
    Then, the characteristic time to crest scales as
    \begin{equation}\label{eq:time_scaling}
        \tau \sim \sqrt{\frac{s}{\delta\mathrm{Re}_\delta}} = \sqrt{\frac{s}{2K_C}},
    \end{equation}
    where $K_C=U_0/(\omega D)$ is the Keulegan-Carpenter number, describing the relative importance of hydrodynamic drag to inertial forces in oscillatory flows. %
    
    Figure~\ref{fig:LHS_time}{b} shows the normalised time interval $\Delta t_{|\Upsilon|>1}$, representing the fraction of the oscillation period during which the motion threshold \eqref{eq:upsilon1} is met.
    The prediction of $\Delta t_{|\Upsilon|>1}$ is crucial for determining whether a particle can accelerate sufficiently to overcome substrate barriers and settle into neighbouring pockets before flow reversal. %
    At low forcing (small $\Gamma$), the threshold is typically not exceeded ($\Delta t_{|\Upsilon|>1}=0$). As $\Gamma$ increases, $\Delta t_{|\Upsilon|>1}$ grows rapidly and asymptotically approaches $2\pi$, corresponding to continuous particle motion throughout the oscillation cycle.
    The dependence on $\delta$ reflects the effect of flow uniformity at the particle scale, where larger values of $\delta$ (more shear-like flow) promote particle motion at lower $\Gamma$, for otherwise identical parameters.
    The figure also shows the predicted time to crest \eqref{eq:time_scaling}, marked by crosses. Below each cross, the duration during which $|\Upsilon|\geq1$ is too short for the particle to roll over the crest of the substrate particle, resulting in wiggling motion. Above the cross, the particle can roll out of its pocket and reach a neighbouring one. 
    Notably, for small $\delta$, the time to crest diverges. In this regime, the oscillation frequency is sufficiently high that flow reversal occurs before the particle can accelerate sufficiently, preventing it from rolling over the substrate.

\section{Experimental approach}\label{sec:ExperimentalApproach}
The experiments are carried out in an oscillatory flow tunnel (OFT) and serve a dual purpose. First, they enable the characterisation of flow conditions and particle behaviour as a function of the problem's parameters close to the motion threshold. Second, they validate and inform the numerical simulations.

\subsection{Experimental setup}
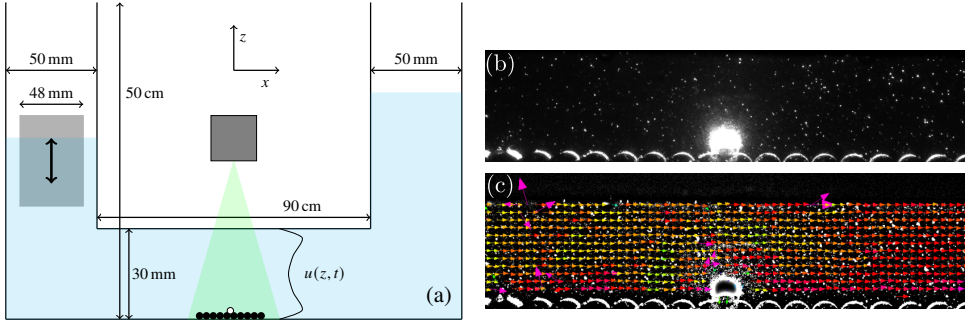
\begin{figure}
        \scalebox{0.6}{\begin{tikzpicture}
    \draw[black, thick] (0,0) -- (10,0);
    \draw[black, thick] (2,2) -- (8,2);
    \draw[black, thick] (0,0) -- (0,7);
    \draw[black, thick] (2,2) -- (2,7);
    \draw[black, thick] (8,2) -- (8,7);
    \draw[black, thick] (10,0) -- (10,7);
    
    \filldraw[black,opacity=0.3] (0.3,4.5) -- (1.7,4.5) -- (1.7,2.5) -- (0.3,2.5);

    \filldraw[cyan,opacity=0.15] (0,0) -- (10,0) -- (10,5) -- (8,5) -- (8,2) -- (2,2) -- (2,4) -- (0,4) -- (0,0);

    \draw[black, thick, ->] (5,5.5) -- ++(1,0);
    \draw[black, thick, ->] (5,5.5) -- ++(0,1);
    

    \draw[black, ultra thick, <->] (1,3) -- (1,4);
    
    \draw[black, thick, <->] (0,5.5) -- (2,5.5);
    \draw[black, thick, <->] (0.3,4.75) -- (1.7,4.75);
    \draw[black, thick, <->] (8,5.5) -- (10,5.5);
    \draw[black, thick, <->] (2.7,0) -- (2.7,2);
    \draw[black, thick, <->] (2.5,0) -- (2.5,7);
    \draw[black, thick, <->] (2,2.25) -- (8,2.25);
    
    

    \node at (1,5) {\SI{48}{mm}};
    \node at (1,5.75) {\SI{50}{mm}};
    \node at (9,5.75) {\SI{50}{mm}};
    \node at (3.25, 1) {\SI{30}{mm}};
    \node at (3.05, 5) {\SI{50}{cm}};
    \node at (6.5,2.5) {\SI{90}{cm}};

    \draw plot [thick,smooth,tension=0.5] coordinates {(6,0) (6.5,0.2) (6.3,0.6) (6.2,1) (6.3,1.4) (6.5,1.8) (6,2)};

    \node at (7,1) {$u(z,t)$};
    \node at (5.7,3.7+1.5) {$x$};
    \node at (5.2,4.7+1.5) {$z$};

    \node at (9.5, 0.5) {\Large (a)};

    \filldraw[black,fill=gray] (4.5,3.5) rectangle ++(1,1);
    \filldraw[green,fill=green,opacity=0.15] (5,3.5) -- (4,0) -- (6,0) -- (5,3.5);

    \foreach \x in {-5,...,4}
        {\filldraw[black] (5+0.15*\x,0.075) circle (2pt);}
    
    \filldraw[black,fill=white] (5-0.075,0.075*2.5) circle (2pt);

\end{tikzpicture}}
        \begin{subfigure}[b]{0.49\textwidth}
            \centering
            \includegraphics[width=\textwidth]{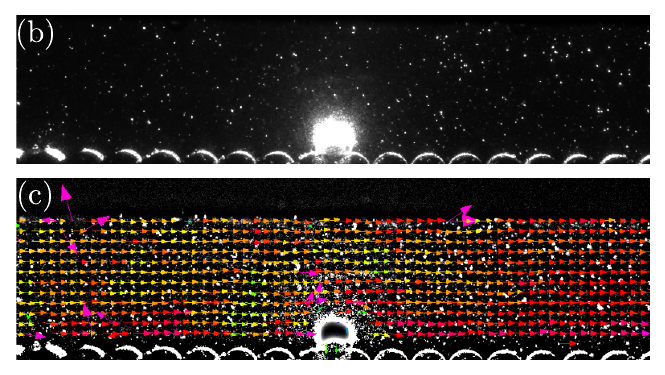}
        \end{subfigure}
        \caption{(a) Schematic of the oscillatory flow tunnel (OFT), where a harmonically oscillating piston (grey) drives flow between two parallel flat plates. On the bottom, a single mobile particle (white) lies on top of a fixed monolayer (black), with a vertical laser sheet (green) illuminating the cross-section parallel to the oscillation direction. 
        (b) Experimental snapshot ($\delta\approx0.12$ and $s=1.81$) showing the substrate, exposed particle, and laser-illuminated tracer particles.
        (c) A contrast-enhanced snapshot overlaid with velocity vectors obtained using PIV, scaled and colour-coded by magnitude (green to red for increasing velocity magnitude).}
        \label{fig:IPM_setup}
    \end{figure}

    Figure~\ref{fig:IPM_setup} shows a schematic overview of the setup. We use a custom-made OFT consisting of an acrylic tank with inner dimensions $(100\times 50\times 30)\,\SI{}{cm}$ in the streamwise, spanwise, and vertical directions, respectively. A U-shaped lid is suspended within this tank, creating a central section with two parallel flat plates spaced $\SI{30}{mm}$ apart. Each end of the tank contains a vertical section with a free surface, measuring $\SI{50}{mm}$ in width. The tank is filled with tap water with mass density $\rho_f=\SI{0.999\pm0.001 e3}{kg/m^3}$ and kinematic viscosity $\nu=\SI{1.05\pm0.01 e-6}{m^2/s}$. Special care has been taken to ensure a watertight seal between the lid and the side walls of the outer tank at the front and back (out of plane). 
    A partially submerged piston with a width of $\SI{48}{mm}$ is placed in the left column, covering approximately $77\%$ of the free surface area (as seen from above). The piston is driven by a PID-controlled linear motor (LinMot P01-37x120F/100x180-HP) in a vertical oscillatory motion. This motion modulates the water level in the left column, generating a hydrostatic pressure difference between the left and right columns, which in turn drives the flow through the central horizontal section.
    The piston does not fully cover the free surface in the lateral direction, but it is sufficiently large to generate the necessary flow through vertical oscillations of the water level. A tight seal or full interfacial coverage is not required, as this would involve extreme forces and special care to prevent friction.

    The substrate consists of a monolayer of spheres with a diameter of $\SI{5.950\pm0.005}{mm}$, arranged in a square lattice with $\SI{6.0}{mm}$ centre-to-centre spacing, forming a $25\times26$ particle grid aligned with the oscillation direction. The spheres are securely positioned in circular holes ($\SI{3.0}{mm}$ in diameter) on a bottom plate, with a \SI{5}{mm}-high surrounding frame preventing the layer from shifting. 
    A single spherical particle with diameter $D=\SI{5.950\pm0.005}{mm}$ is placed on top of the substrate approximately in the centre of the grid. Its density is either $\rho_s=\SI{1.09\pm0.05 e3}{kg/m^3}$ or $\rho_s=\SI{1.81\pm0.05 e3}{kg/m^3}$ for light and heavy particles, respectively.

    We use PIV to characterise the oscillating flow field, specifically to measure the mean flow velocity $U_0$ far from the bed (e.g., as in figure~\ref{fig:IPM_ParticlePosition}) and to obtain the velocity profile near the exposed particle (cf. figure~\ref{fig:DNS_velocity_profile_comparison}).
    A vertical two-dimensional slice of the velocity field is captured above the bed and around the exposed particle. A vertical laser sheet, approximately $\SI{5}{mm}$ wide (similar to the particle's diameter), is centred on the exposed particle and aligned with both the grid and the oscillation direction.
    The laser operates in a double-pulsed mode, emitting two pulses separated by $\SI{5.0}{ms}$, with pulse pairs generated at a frequency of \SI{15}{Hz}. A RedLake MegaPlus II camera equipped with a Nikon 28 mm f/2.8 lens records the experiment from the side.
    Figure~\ref{fig:IPM_setup}(b) shows a typical snapshot, where the vertical laser sheet illuminates the substrate at the bottom of the image, the mobile particle on top, and the tracer particles dispersed in the fluid for PIV analysis.

    The PIV recordings are analysed with the image processing software PIVview2C (v3.9.3). For each image pair, instantaneous velocity vectors are computed within interrogation windows of $32 \times 32$ pixels, with 50\% overlap between adjacent interrogation windows \citep{prasad2000particle}. Each window typically contains 10-20 tracer particles. 
    Figure~\ref{fig:IPM_setup}(c) shows a typical frame after PIV analysis. A high-pass filter and pixel value threshold are applied to the original image to enhance the contrast of the tracer particles. The coloured arrows represent velocity vectors within each interrogation window.

    Finally, several factors contribute to the eventual lack of perfect symmetry of the flow in the OFT, including the levelling of the tank and substrate, as well as inertial effects due to the piston's motion, such as vortex shedding. As a result, the velocity field above the substrate is unlikely to be perfectly symmetric, leading to minor variations in the mean velocity. These asymmetries can further affect particle motion, particularly near the motion threshold, where small differences in torques may cause the threshold to be exceeded in only one direction, resulting in a preferred direction of particle movement. 

\subsection{Measurement approach}
    For a given particle-fluid combination and substrate geometry, the density ratio $s$ and the geometrical parameter $L$ remain constant across all experiments. 
    The parameter $H$, primarily determined by the substrate geometry (as discussed in \S~\ref{sec:system_model}), is also assumed to remain approximately constant.
    Three independent degrees of freedom remain ($\delta$, $\mathrm{Re}_\delta$, $\Gamma$), all depending on the flow conditions. 
    For comparison with DNS results, we also consider the particle mobility number $\psi=\delta \mathrm{Re}_\delta \Gamma/2$ (see \eqref{eq:psi}).
    In each experiment, the oscillation frequency $\omega$ is fixed while gradually increasing the piston's stroke, thereby increasing the fluid velocity amplitude $U_0$. The constant frequency implies that the value of $\delta$ remains constant within each experiment but varies across different trials as the frequency is adjusted. 
    At each stroke value, the piston oscillates for ten periods before its peak-to-peak amplitude is slightly increased: by $\SI{5}{mm}$ at lower amplitudes and by $\SI{1}{mm}$ near the motion threshold.
    The oscillation period ranges between $1.0$ and $\SI{1.8}{s}$, corresponding to the Stokes boundary layer thickness $\delta_S\approx 0.58-\SI{0.78}{mm}$, which is significantly smaller than the $\SI{30}{mm}$ spacing between the parallel plates. This ensures that the velocity profile above the bottom remains unaffected by the OFT's upper boundary \citep{overveld2022effect}.
    The ranges of dimensionless parameters explored near the motion threshold are listed in table~\ref{tab:simparams}.

\section{Numerical approach}\label{sec:NumericalApproach}
Direct numerical simulations (DNS) are conducted to extend the experimental results and to determine the dynamics of the mobile particle during its early stages of motion.
The incompressible Navier-Stokes equations \eqref{eq:theory_momentum} are solved numerically using a second-order central-difference scheme to discretise the spatial derivatives and a Runge-Kutta-based fractional-step method to advance in time. The no-slip/no-penetration boundary conditions at the fluid-solid interfaces of both the fixed and mobile particles are enforced using an immersed boundary method (IBM) developed for particulate flows by \citet{uhlmann2005immersed}. This method employs a direct forcing approach, adding a localised volume force term to the momentum equations.
The additional forcing term is explicitly computed at each time step as a function of the particle position and velocity, providing the stress distribution at each particle's surface.
The dynamics of the mobile particle are determined by Newton's second law for its translation and by Euler's second law for its rigid-body rotation, according to \eqref{eq:theory_particle_trans}~and~\eqref{eq:theory_particle_rot}, respectively.
The hydrodynamic forces and torques result from the zeroth and first angular moments of the stresses, respectively, integrated over the particle surface. The contribution of inter-particle contacts is given by the sum of the contact forces between a considered particle and its neighbours, computed using a discrete element method (DEM) similar to that proposed by \citet{kidanemariam2014interface}. %
The present approach, which is described in Appendix~\ref{sec:contact}, accounts for inter-particle static friction, simultaneously incorporating elastic, inelastic, and frictional contributions to the tangential force at each contact point.
Correct evaluation of the inter-particle contact force is necessary for describing the early stages of particle motion because the mobile particle may either roll or slide. 
Table~\ref{tab:simparams} and table~\ref{tab:simdomain} show the values of the parameters characterising the flow conditions and the computational domain, respectively. 
The density ratio is fixed at $s=1.813$ in all simulations, matching experiments with heavier particles. The lighter particles, with density ratios closer to unity, could not be accurately simulated due to stability limitations of the numerical approach.

\begin{table}
    \centering
    \begin{tabular}{lccccccccc}
        Run no. & $\delta$ & $\mathrm{Re}_\delta$ & $\mathrm{Re}_D=\mathrm{Re}_\delta/\delta$ & $\Gamma$ & $H$ & $L$ & $\psi$ & $\psi_{\mathrm{tr},w}$ \\
        \hline
        1 & $0.12$ & $164$ & $1400$ & $0.10$ & $0.84$ & $0.36$ & $0.74~\textendash~1.11$ & $1.10$ \\
        2 & $0.12$ & $188$ & $1600$ & $0.13$ & $0.84$ & $0.36$ & $1.46$ & $-$ \\
        3 & $0.12$ & $41$ & $340$ & $0.45$ & $0.82$ & $0.36$ & $1.11$ & $1.11$ \\
        4 & $0.12$ & $164$ & $1400$ & $0.10~\textendash~0.11$ & $0.84$ & $0.36$ & $0.99~\textendash~1.11$ & $0.99$ \\
        5 & $0.12$ & $188$ & $1600$ & $0.13$ & $0.84$ & $0.36$ & $1.46$ & $-$ \\
        6 & $0.96$ & $164$ & $170$ & $0.014~\textendash~0.19$ & $0.86$ & $0.39$ & $1.11~\textendash~2.23$ & $1.55$ \\
        \hline
        Exp. & $s$ & $\delta$ & $\mathrm{Re}_\delta$ & $\Gamma$  & $\psi$ \\
        \hline
        Heavy part. & $1.81$ & $0.106~\textendash~0.125$ & $115~\textendash~162$ & $0.102~\textendash~0.133$ & $0.73~\textendash~1.11$\\
        Light part. & $1.09$ & $0.097~\textendash~0.130$ & $12~\textendash~33$ & $ 0.12~\textendash~0.28$ & $0.08~\textendash~0.46$\\
    \end{tabular}
    \caption{Values of the dimensionless parameters that characterise the DNS cases, alongside experimental ranges near the motion threshold (corresponding to the symbols in figure~\ref{fig:IPM_MotionPhase}). The Reynolds number based on the particle diameter is denoted by $\mathrm{Re}_D$. The final column reports the threshold value of the mobility number $\psi_{\mathrm{tr},w}$ for the onset of wiggling motion.
    In the DNS, the density ratio is fixed at $s=1.813$ and the value of $\psi$ is explicitly prescribed. In the experiments, $\psi$ is not an independent control parameter but follows from the other dimensionless parameters.}
    \label{tab:simparams}
\end{table}

\begin{table}
    \centering
    \begin{tabular}{lccccccccc}
        Run no. & $\delta$ & $\Lambda_{x}/\delta$ & $\Lambda_{y}/\delta$ & $\Lambda_{z}/\delta$ & $n_{x}$ & $n_{y}$ & $n_{z}$ & $D/\Delta x$ & $\omega\Delta t$ \\
        \hline
        1-3 & $0.12$& $201.9$ & $67.3$ & $50.5$ & $768$ & $256$ & $192$ & 31 & $4\cdot10^{-4}$ \\
        4-5 & $0.12$& $201.9$ & $67.3$ & $50.5$ & $1536$ & $512$ & $384$ & 62 & $4\cdot10^{-4}$ \\
        6 & $0.96$& $25.2$  & $8.4$  & $25.2$ & $384$  & $128$ & $384$ & 16 & $2\cdot10^{-4}$ \\
    \end{tabular}
    \caption{Simulation domain parameters, where $\Lambda$ represents the size of the computational domain, while $n$ indicates the number of grid points used to discretise the domain along the corresponding axis. $\Delta x$ and $\Delta t$ denote the grid spacing and time step used in the simulations, respectively.}
    \label{tab:simdomain}
\end{table}

%
%
%

\section{Results and discussion}\label{sec:Results}
\subsection{Description of the early stages of particle motion}\label{sec:results_qualitative}

\begin{figure}
    \centering
    \raisebox{5cm}{\small$a)$}
    \begin{subfigure}[b]{0.68\textwidth}
        \centering
        \includegraphics[width=\textwidth]{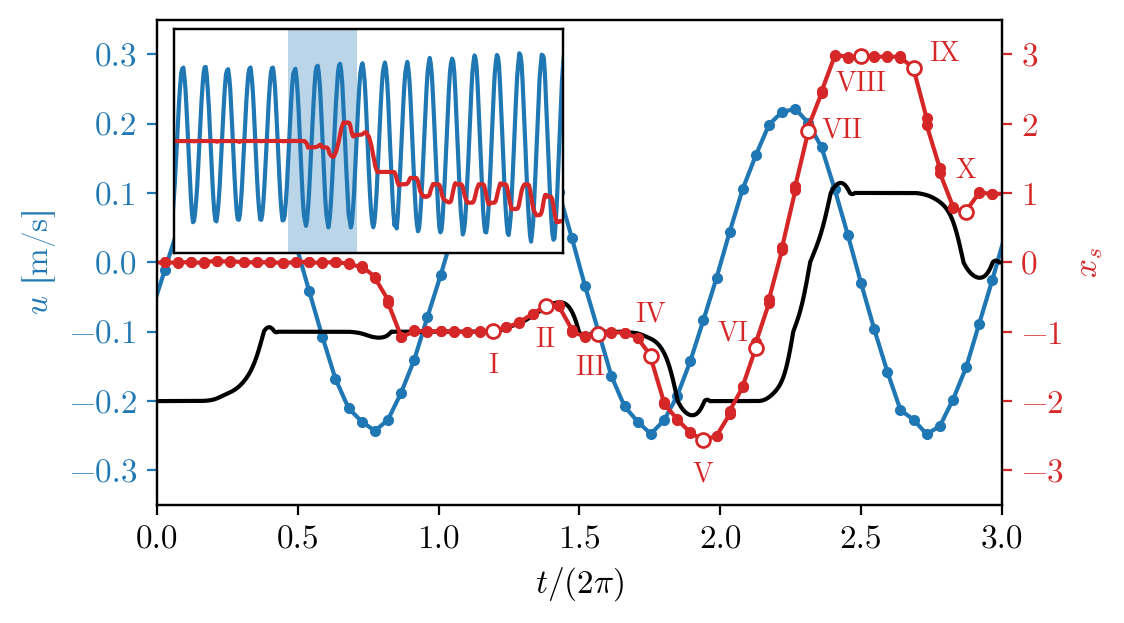}
    \end{subfigure}
    \raisebox{5cm}{\small$b)$}
    \begin{subfigure}[b]{0.25\textwidth}
        \centering
        \includegraphics[width=\textwidth]{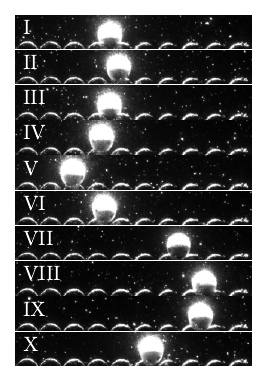}
    \end{subfigure}
    \caption{(a) Temporal evolution of the space-averaged flow velocity above the substrate, $u$ (blue), and the (normalised) particle position, $x_s$ (red), over approximately three oscillation periods, obtained from the OFT experiments ($s=1.81$, $\delta\approx0.11$, $\mathrm{Re}_\delta\approx 150$, $\Gamma\approx0.13$, $\psi\approx 1.1$). 
    The velocity amplitude is gradually ramped up to identify the conditions for incipient motion. The complementary DNS data (black) shows the trajectory of a rolling particle just above the motion threshold for run 4 ($\psi=1.1$).
    Roman numerals mark every eight frames, corresponding to the snapshots in (b). The inset shows the particle motion and velocity over a longer duration, with the highlighted region corresponding to the main plot.} 
    \label{fig:IPM_ParticlePosition}
\end{figure}

The space-averaged flow velocity within the OFT is calculated using the PIV velocity fields (e.g., as shown in figure~\ref{fig:IPM_setup}), excluding regions near the top plate, the substrate, and the exposed particle to minimise boundary effects.
Figure~\ref{fig:IPM_ParticlePosition}a shows the space-averaged flow velocity (blue curve), oscillating harmonically with an amplitude $U_0\approx\SI{0.23\pm0.01}{m/s}$, where the error follows from PIV correlations and subsequent averaging procedure, for $\delta=0.11$ and $s=1.81$.
Figure~\ref{fig:IPM_ParticlePosition}b shows some frames of the particle motion. %

The particle centre position is determined by filtering out tracers and identifying the median position of pixels with intensities above half the maximum value. The red curve in figure~\ref{fig:IPM_ParticlePosition}a shows the particle position over time under conditions near the motion threshold.
The DNS results (black curve) closely match the experimental results, exhibiting similar motion characteristics, including the phase of motion initiation, typical excursion lengths, and movement directions. This agreement suggests that, in the absence of significant velocity fluctuations in the ambient flow, deterministic predictions of the incipient particle motion are possible. 
Quantitative differences arise due to the system's sensitivity to small perturbations near the motion threshold, where minor positional variations alter the particle's exposure to the flow, leading to significant long-term deviations in its trajectory, such as settling into a different pocket in the substrate.

For the parameter values considered in the experiments, the particle motion was observed to begin around the maximum velocity phases (frames I, IV, IX). In some cases, the particle rolled without escaping its initial pocket (II), a behaviour we refer to as \emph{wiggling motion}. 
In others, sustained motion continued until the flow reversal (V), after which the particle rolled across multiple pockets in the opposite direction (VI, VII), and occasionally overshot (V, VII, X) before settling. %
The range of flow velocities (captured in the forcing parameters $\mathrm{Re}_\delta$, $\Gamma$, and $\psi$) for which wiggling motion is observed is relatively narrow, as even a slight increase of the hydrodynamic torque intensity or duration above the critical threshold can cause the particle to roll over the substrate.
As the particle leaves its initial position and falls into one of the neighbouring pockets, the effects of the non-linearities associated with the inter-particle contacts reflect on the hydrodynamic force, which becomes impossible to predict deterministically. %
Therefore, from now on, only the early stages of particle motion are considered. %

\begin{figure}
    \centering
    \includegraphics[width=0.6\textwidth]{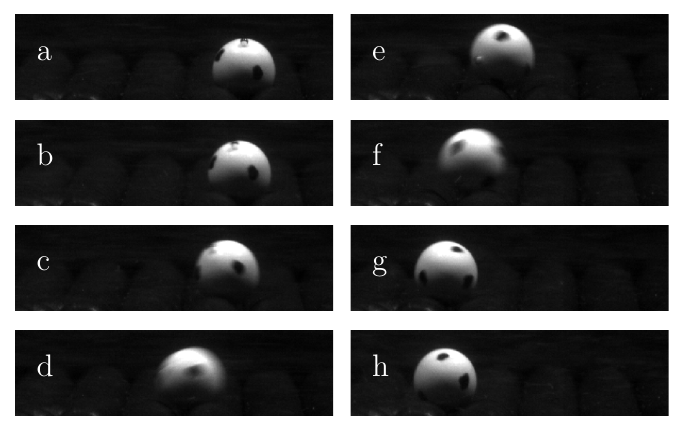}
    \caption{Consecutive snapshots from an experiment with $s=1.81$, $\delta\approx 0.12$, $\mathrm{Re}_\delta\approx190$, and $\Gamma\approx0.13$, showing that the particle (with black dots drawn on its surface) rolls when it moves over the substrate.}
    \label{fig:IPM_particlerotation}
\end{figure}

While figure~\ref{fig:IPM_ParticlePosition}b provides a detailed view of the particle translation, it does not resolve its rotation due to the laser sheet illumination. Therefore, supplementary experiments are conducted with the laser turned off, as shown in figure~\ref{fig:IPM_particlerotation}.
These snapshots support that the particle rotates throughout its motion, also when it moves between the pockets of the substrate, which is consistent with \citeauthor{agudo2017shear}'s~\citeyearpar{agudo2017shear} experiments in steady flow conditions, where rolling always characterised the initiation of motion.

\begin{figure}
    \centering
    \includegraphics[width=0.8\textwidth]{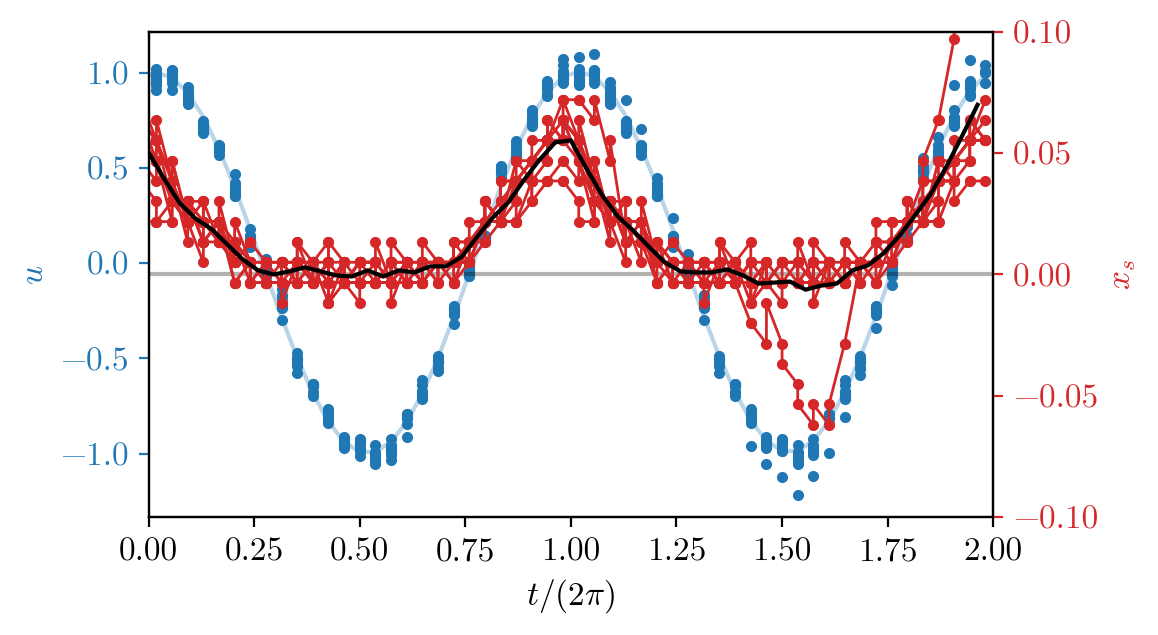}
    \caption{Particle position during ten consecutive oscillation cycles with constant flow amplitude, projected onto two oscillation cycles, from an experiment with $s=1.09$ and $\delta=0.127$. The spatially averaged flow velocity (blue dots) and its sinusoidal fit (blue curve) are shown alongside the exposed particle position (red dots) and its mean trajectory over the ten cycles (black curve).}
    \label{fig:IPM_MotionPhase}
\end{figure}

To confirm the deterministic nature of the motion threshold and wiggling, figure~\ref{fig:IPM_MotionPhase} shows the streamwise centre position of a wiggling particle over ten oscillation cycles, projected onto two cycles for clarity. When the particle comes to rest before the flow reverses, its motion remains highly repeatable.
This is reflected in the tight clustering of red dots and minimal fluctuations in the mean position, which also serve as the basis for estimating the phases of motion onset and cessation in \S~\ref{sec:results_phase}. %
In contrast, if the particle has not settled in time, it remains exposed after flow reversal and is likely to remain displaced from its equilibrium position in the subsequent half-cycle, resulting in a trajectory that is not exactly predictable. %
Additionally, a slight asymmetry in the setup, as discussed in \S~\ref{sec:ExperimentalApproach}, leads to asymmetric wiggling, with motion predominantly directed towards one side. While this asymmetry does not notably affect the overall repeatability of the motion, it introduces a small directional preference in the particle's trajectory.

To further quantitatively investigate the nature of the particle motion, DNS run~$4$ replicates the experimental conditions, with three main objectives: $(i)$ examining contact-point dynamics to verify that the mobile particle does not slide, $(ii)$ characterising the local flow field around the mobile particle, and $(iii)$ quantifying the hydrodynamic force acting on the particle. %

\citet{topic2022} showed that a spherical particle on a rough substrate may either roll, maintaining continuous contact with the substrate, or slide, if the tangential force at the contact point exceeds the Coulomb static friction threshold.
In the former case, the minimum torque required to initiate rotation can be predicted based on the hydrodynamic torque, as described in \S~\ref{sec:system_model}.
In the latter case, including sliding, the particle rotation angle no longer matches the angular displacement of its centre, invalidating the model described in \S~\ref{sec:system_model}.

\begin{figure}
    \centering
    \raisebox{3cm}{\small$a)$}
    \includegraphics[width=0.3\textwidth]{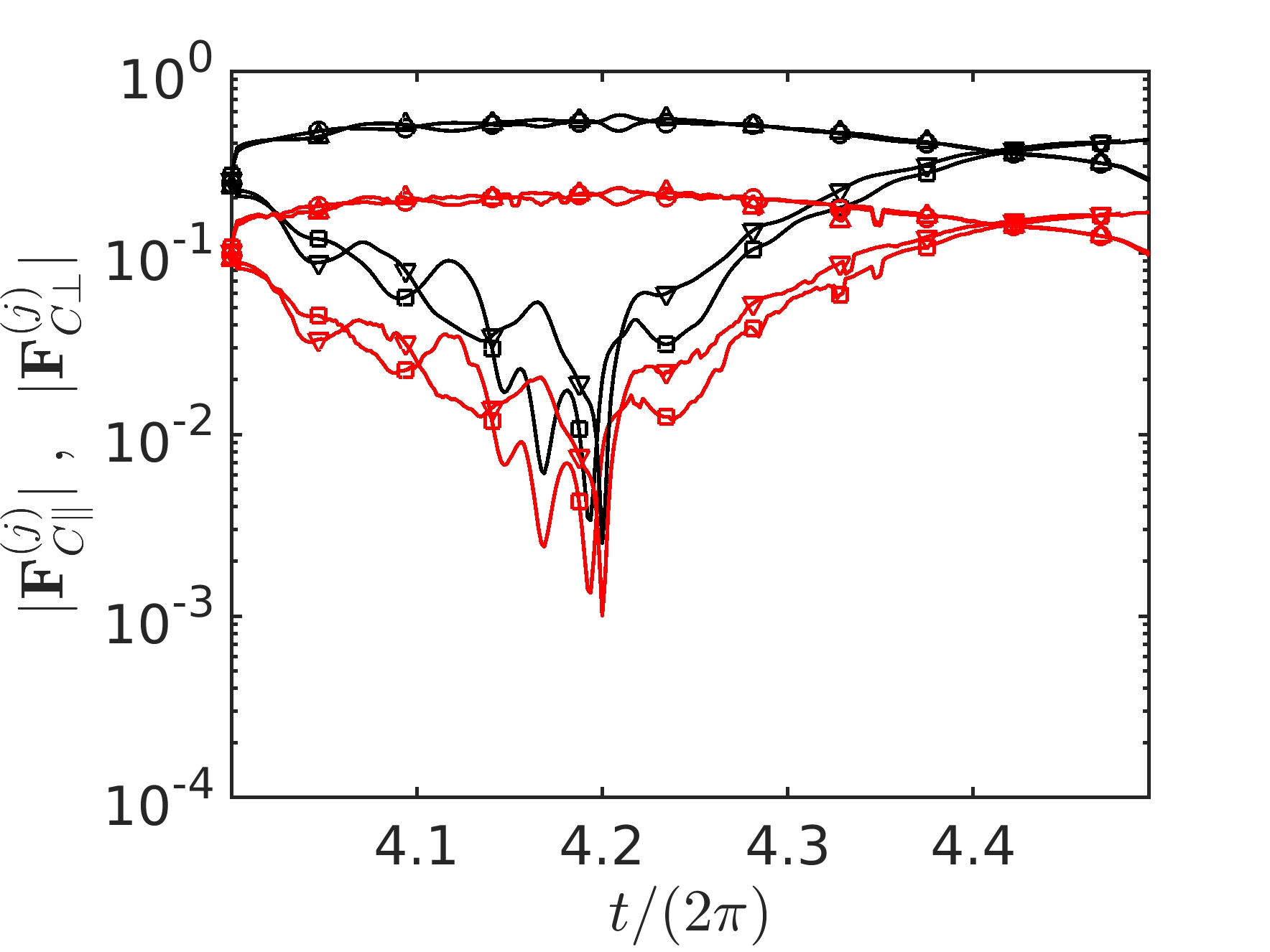}
    \raisebox{3cm}{\small$b)$}
    \includegraphics[width=0.3\textwidth]{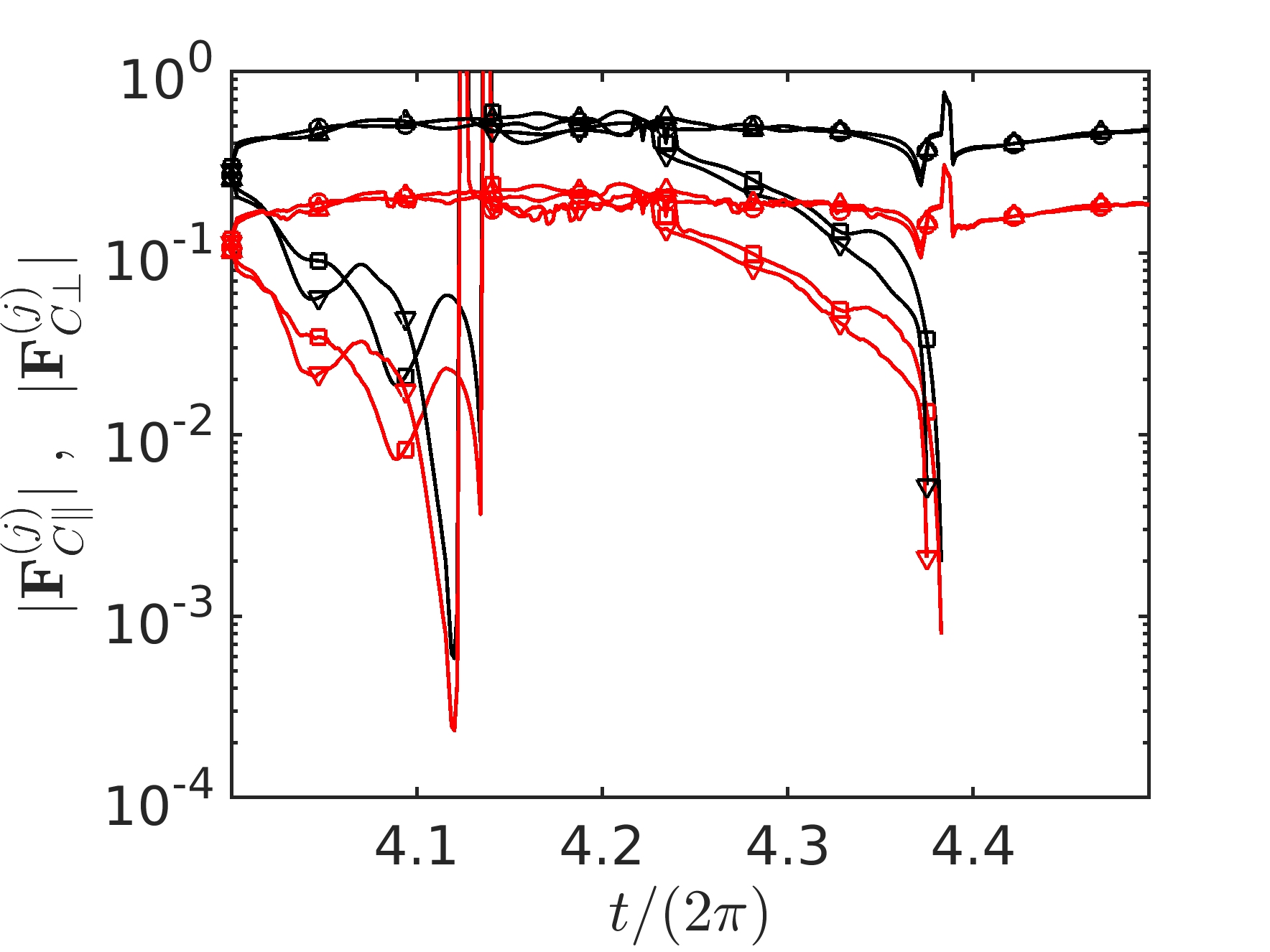}
    \raisebox{3cm}{\small$c)$}
    \includegraphics[width=0.3\textwidth]{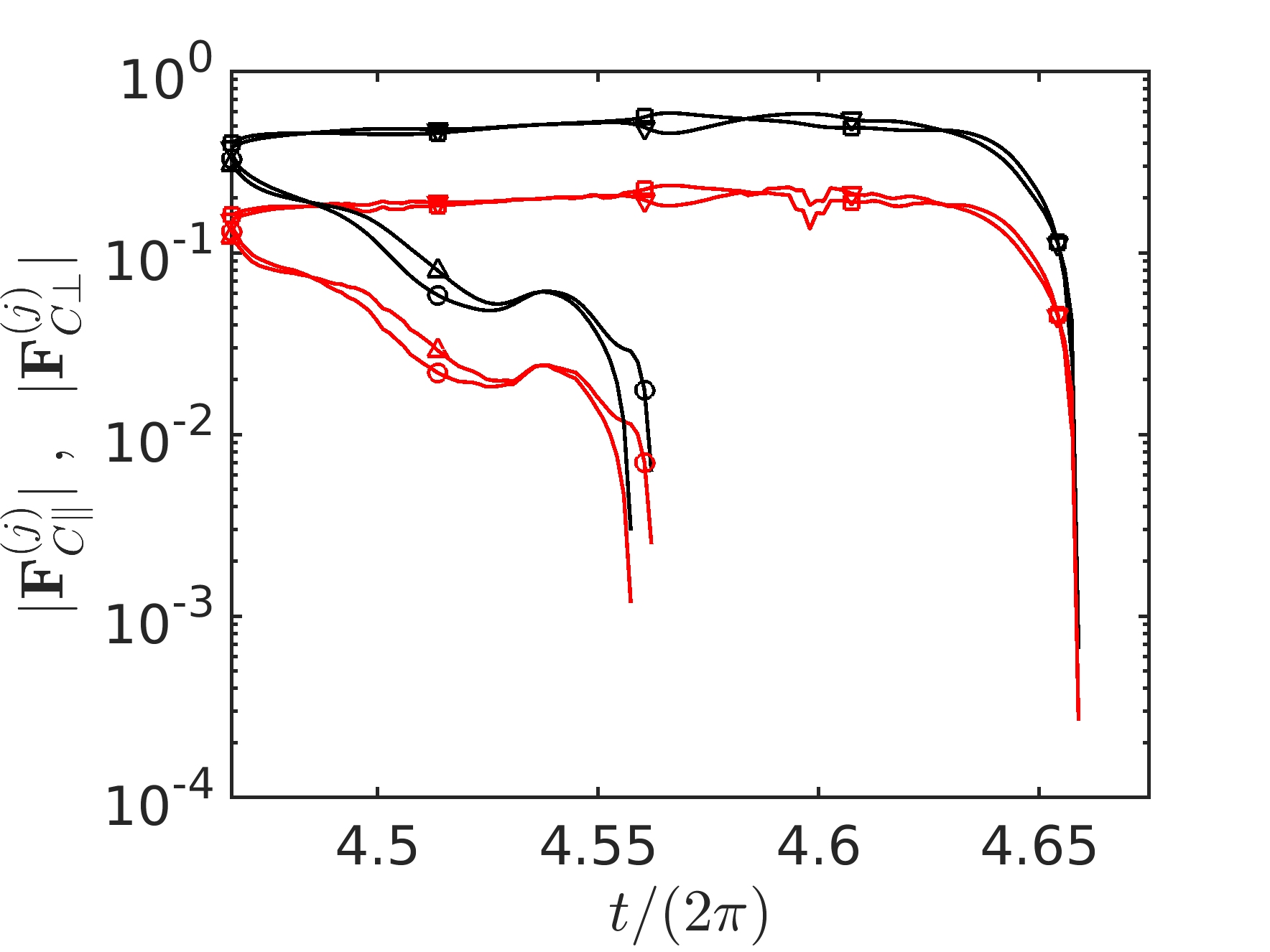}
    \raisebox{3cm}{\small$d)$}
    \includegraphics[width=0.3\textwidth]{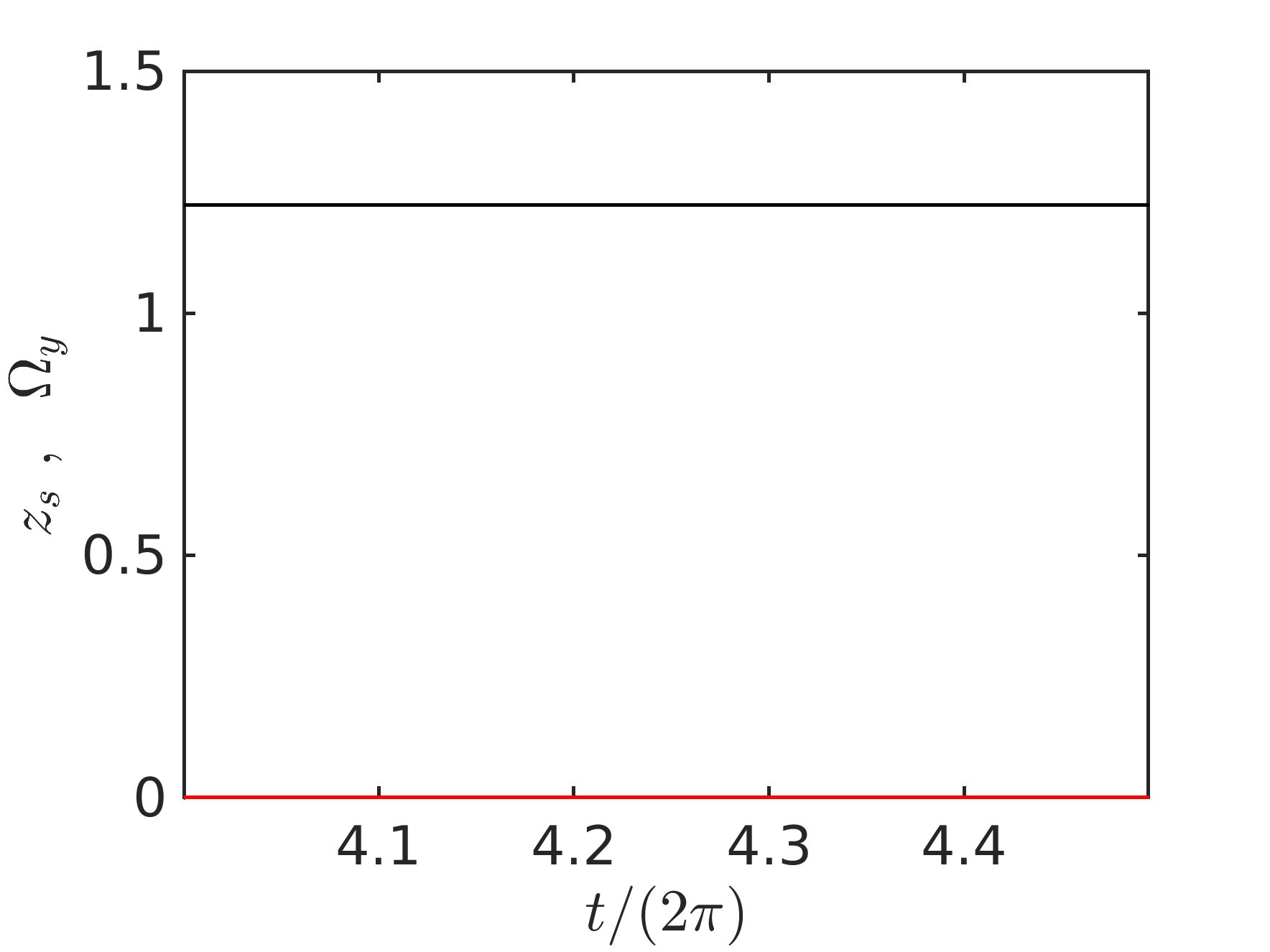}
    \raisebox{3cm}{\small$e)$}
    \includegraphics[width=0.3\textwidth]{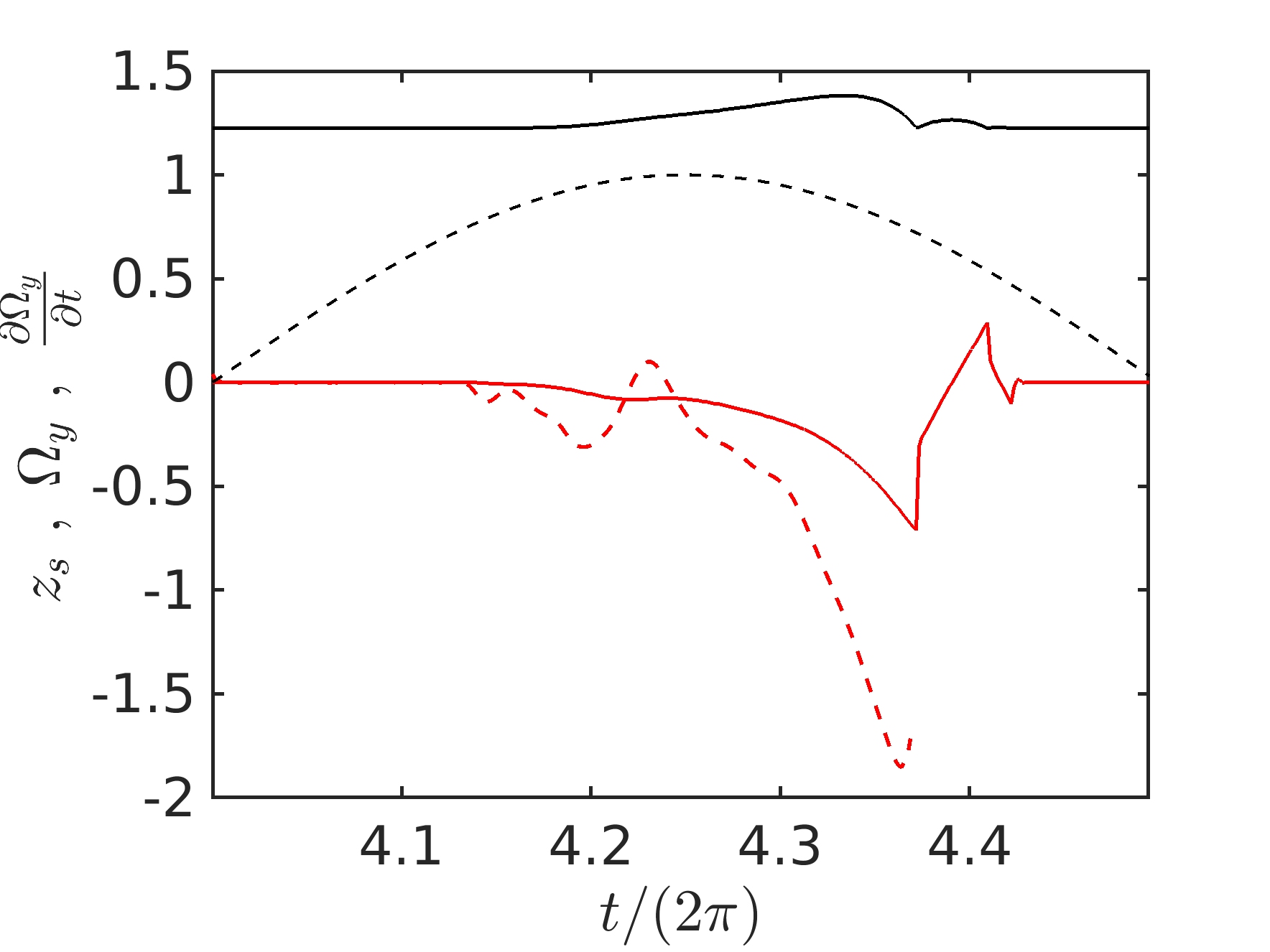}
    \raisebox{3cm}{\small$f)$}
    \includegraphics[width=0.3\textwidth]{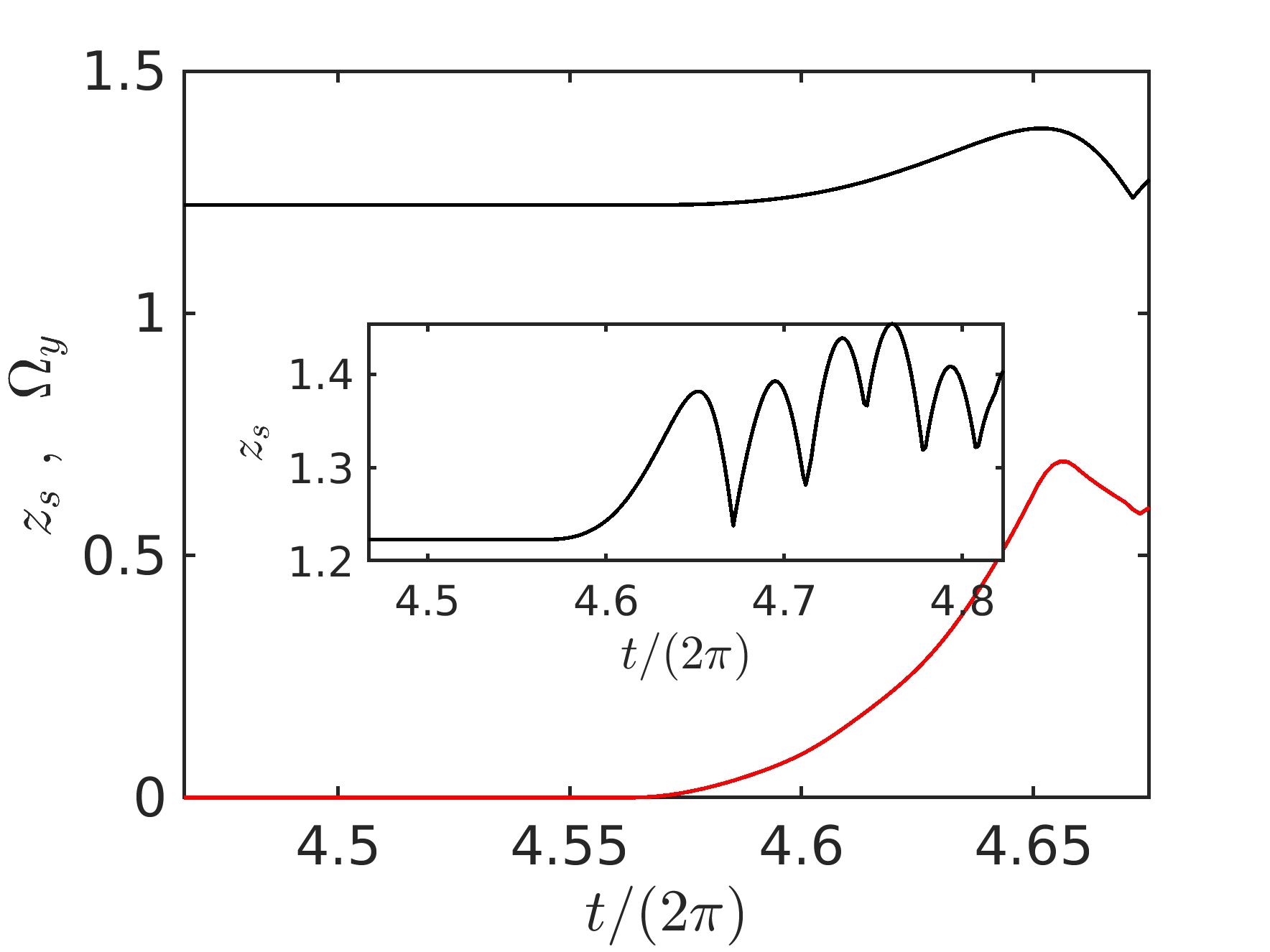}
    \begin{picture}(0,0)(0,0)
      \put(-170,34){
      \scalebox{.16}{
      \input{Figures/inset1}
      }}
    \end{picture}
    \caption{Force and motion characteristics for three representative cases from run~4 (with $\mathrm{Re}_\delta=164$ and $\delta=0.12$): a stationary particle at $\psi=0.99$ ($\Gamma=0.10$, panels $a$ and $d$), a wiggling particle at $\psi=1.1$ ($\Gamma=0.11$, panels $b$ and $e$), and a rolling particle at $\psi=1.46$ ($\Gamma=0.15$, panels $c$ and $f$).
    Panels $(a,b,c)$ show the normal ($F_{C\perp}$, black) and tangential ($F_{C\parallel}$, red) components of the contact force between the mobile (red) particle and particles $j=$ \#1 ($\square$), \#2 ($\triangledown$), \#3 ($\bigcirc$), \#4 ($\vartriangle$). Forces are normalised by the particle's submerged weight (cf.~\eqref{eq:theory_dimlesvars2}). The inset in panel~$(d)$ defines the contact point indices.
    Panels $(d,e,f)$ show the time evolution of the mobile particle's wall-normal centre position $z_s$ (black) and spanwise angular velocity $\Omega_y$ (red).     
    In panel~$(e)$, the dashed black indicates the velocity far above the substrate, while the dashed red curve represents the particle's angular acceleration until it comes to rest in a pocket.
    Panel~$(f)$ includes an inset showing a longer time interval.}
    \label{fig:ctc_force}
\end{figure}

Figure~\ref{fig:ctc_force}(a-c) presents the contact forces (split into normal and tangential components) at the four contact points of the mobile particle during the early motion stages.
The three simulations have constant parameters $\mathrm{Re}_\delta=164$ and $\delta=0.12$, while the mobility number $\psi$ increases, corresponding to a stationary case ($\psi=0.99$, panels $a$ and $d$), a case with wiggling motion ($\psi=1.1$, panels $b$ and $e$), and a case where the particle rolls over the substrate ($\psi=1.46$, panels $c$ and $f$). 

Initially, the contact forces at the four contact points can be distinguished. 
As the flow accelerates from left to right, the magnitude of the contact force at points $j=3,\,4$ increase, while those at $j=1,\,2$ rapidly diminish. For sufficiently large hydrodynamic torque, these forces even vanish and the particle starts to roll (cf. figure~\ref{fig:ctc_force}b,c). %
Otherwise, when the hydrodynamic force is not strong enough to initiate rolling, the particle remains within its original pocket and the contact forces at $j=1,\,2$ recover their initial values (cf. figure~\ref{fig:ctc_force}a,d). %
Notably, the difference between the normal and tangential components of the contact force is set by the static friction coefficient. %

Importantly, no sliding is observed at the contact points, consistent with the experimental observations. %
This is most clearly reflected in the smooth temporal evolution of the angular velocity in figure~\ref{fig:ctc_force}(f), which shows the particle rolling continuously from one pocket to the next. %
These results support the conclusion that the motion threshold is governed by a torque balance, as outlined in \S~\ref{sec:system_model}.

\begin{figure}
    \centering
    \raisebox{1.8cm}{\small $a)$}
    \includegraphics[width=0.46\textwidth]{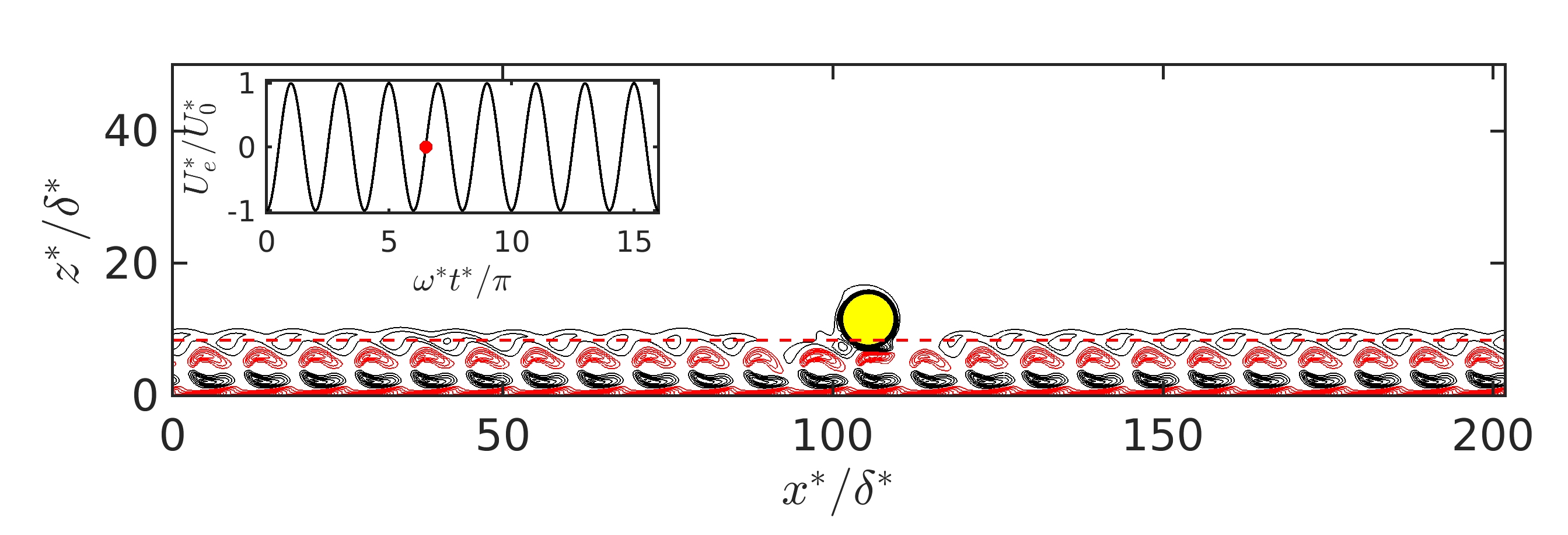}
    \raisebox{1.8cm}{\small $b)$}
    \includegraphics[width=0.46\textwidth]{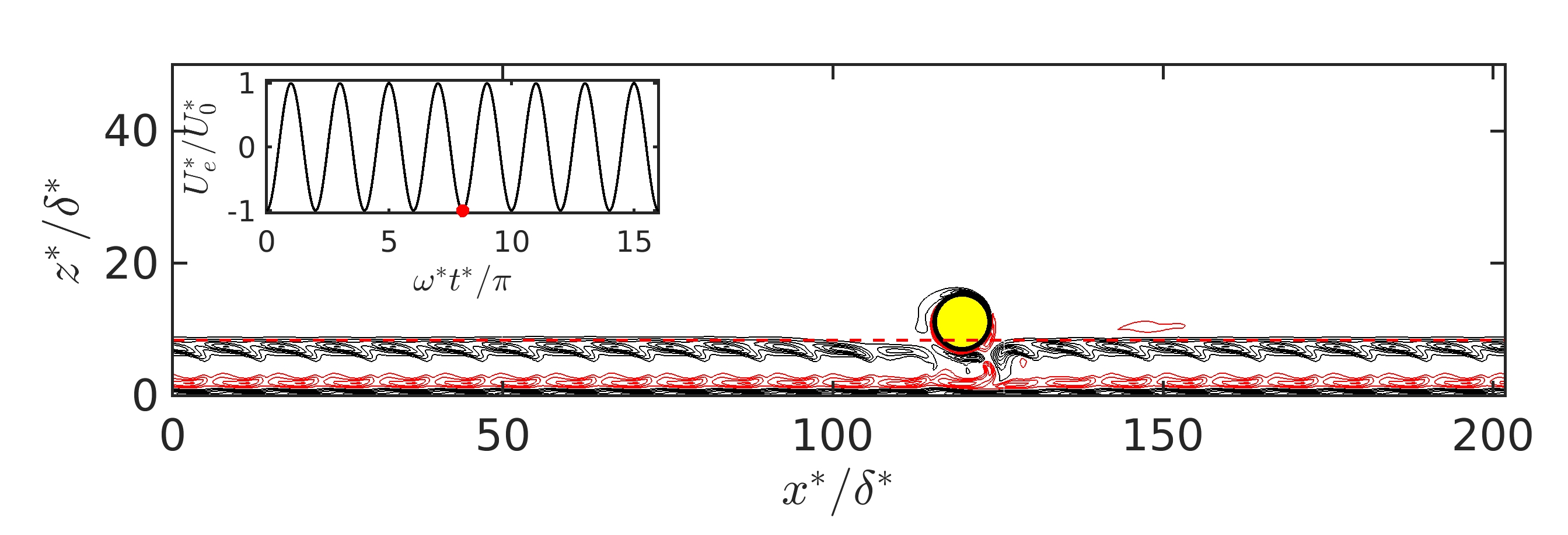}
    \raisebox{1.8cm}{\small $c)$}
    \includegraphics[width=0.45\textwidth]{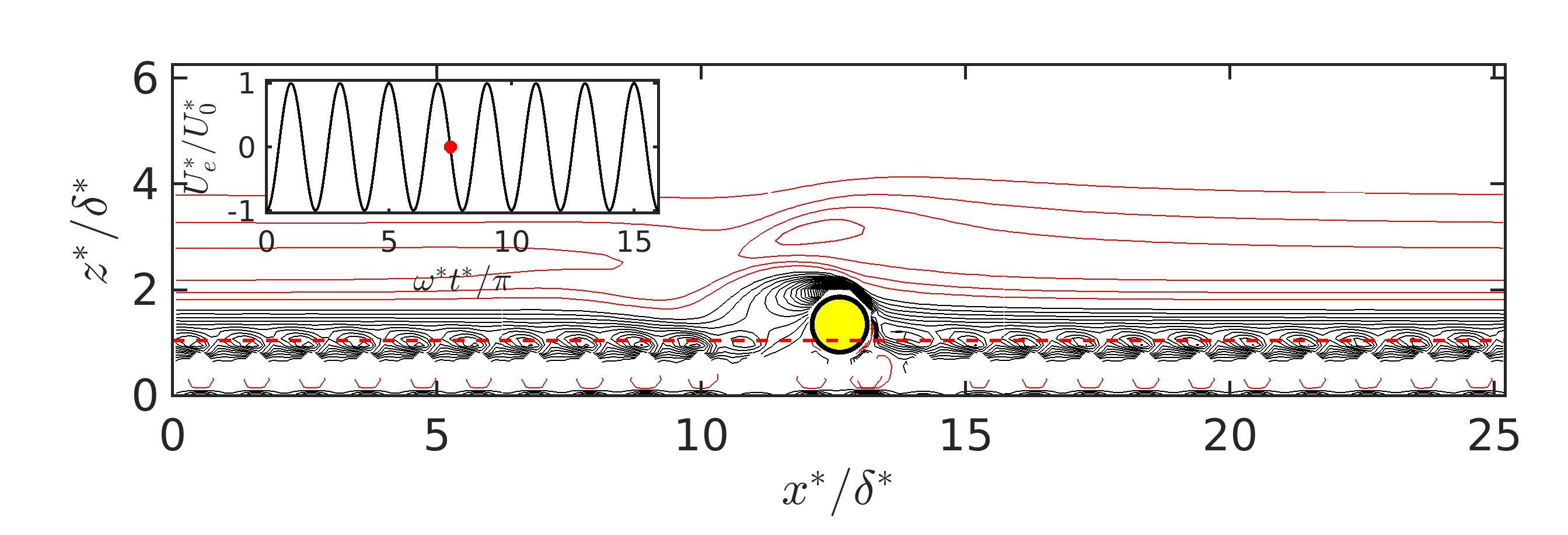}
    \:\:\raisebox{1.8cm}{\small $d)$}
    \includegraphics[width=0.45\textwidth]{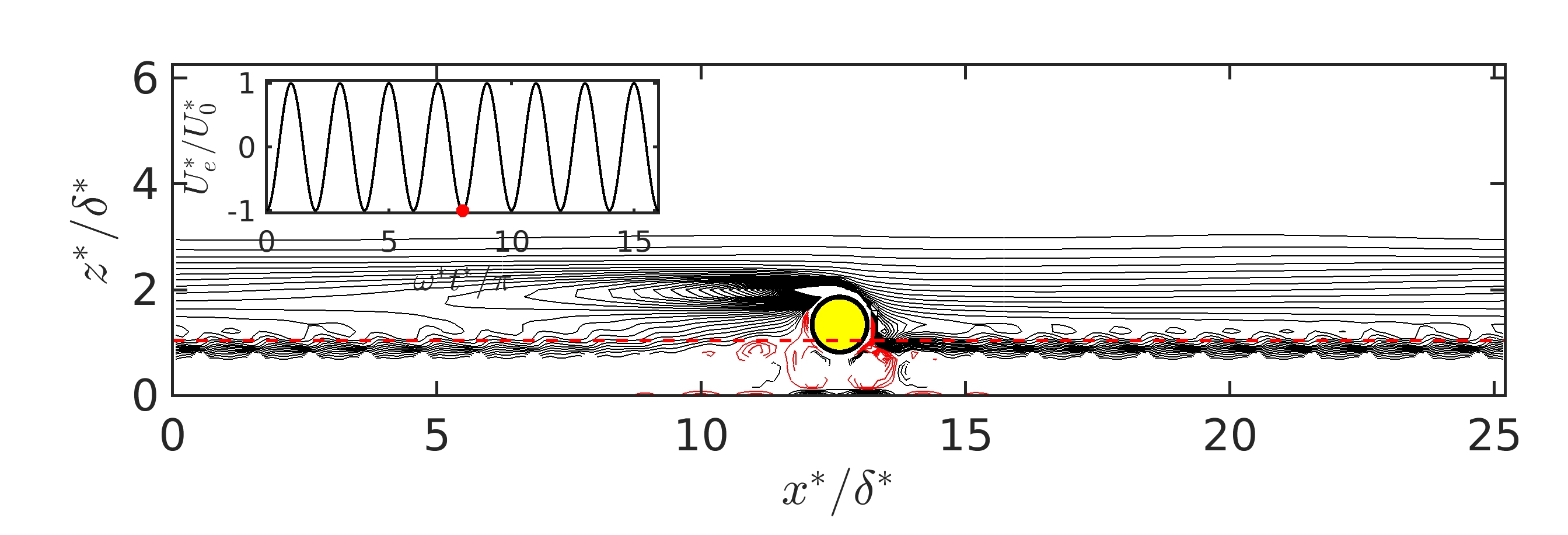}\\
    \raisebox{1.8cm}{\small $e)$}
    \includegraphics[width=0.46\textwidth]{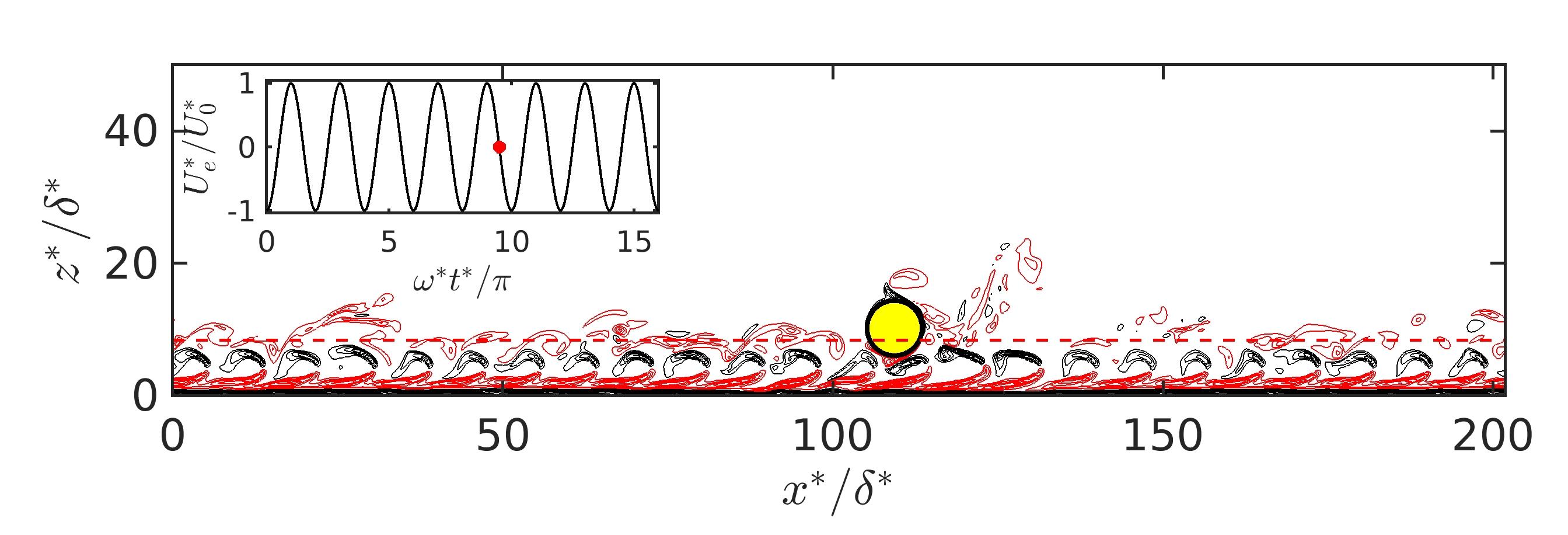}
    \raisebox{1.8cm}{\small $f)$}
    \includegraphics[width=0.46\textwidth]{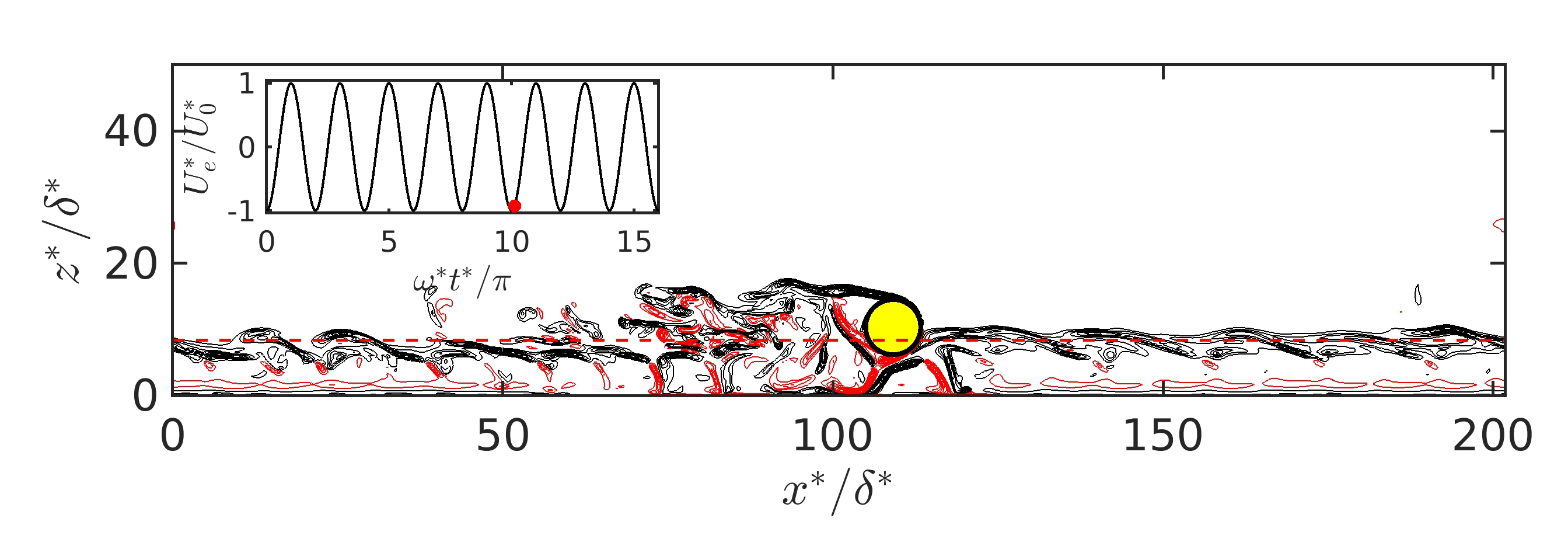}
    \caption{Contour lines of the spanwise vorticity component in the vertical symmetry plane crossing the centre of the exposed particle (yellow) for run 3 $(a,b)$ ($\delta=0.12$, $\mathrm{Re}_\delta=41$, $\mathrm{Re}_D=340$, $\Gamma=0.45$, $\psi=1.1$), run 6 $(c,d)$ ($\delta=0.96$, $\mathrm{Re}_\delta=164$, $\mathrm{Re}_D=170$, $\Gamma=0.019$, $\psi=1.5$), and run 4 $(e,f)$ ($\delta=0.12$, $\mathrm{Re}_\delta=164$, $\mathrm{Re}_D=1400$, $\Gamma=0.10$, $\psi=0.98$). The snapshots correspond to the phases of flow reversal $(a,c,e)$ and maximum velocity far above the substrate $(b,d,f)$. Contours represent positive (black) and negative (red) values of vorticity normalised by the oscillation frequency, with contour levels spaced by $0.05$ in $(a,b)$ and $0.2$ in $(c,d,e,f)$. The zero contours are omitted for clarity.}
    \label{fig:IPM_ParticlemotionDNS}
\end{figure}

\begin{figure}
    \centering
    \raisebox{2.5cm}{\small $a)$}
    \includegraphics[trim=0 6cm 0 12cm,clip,width=.41\textwidth]{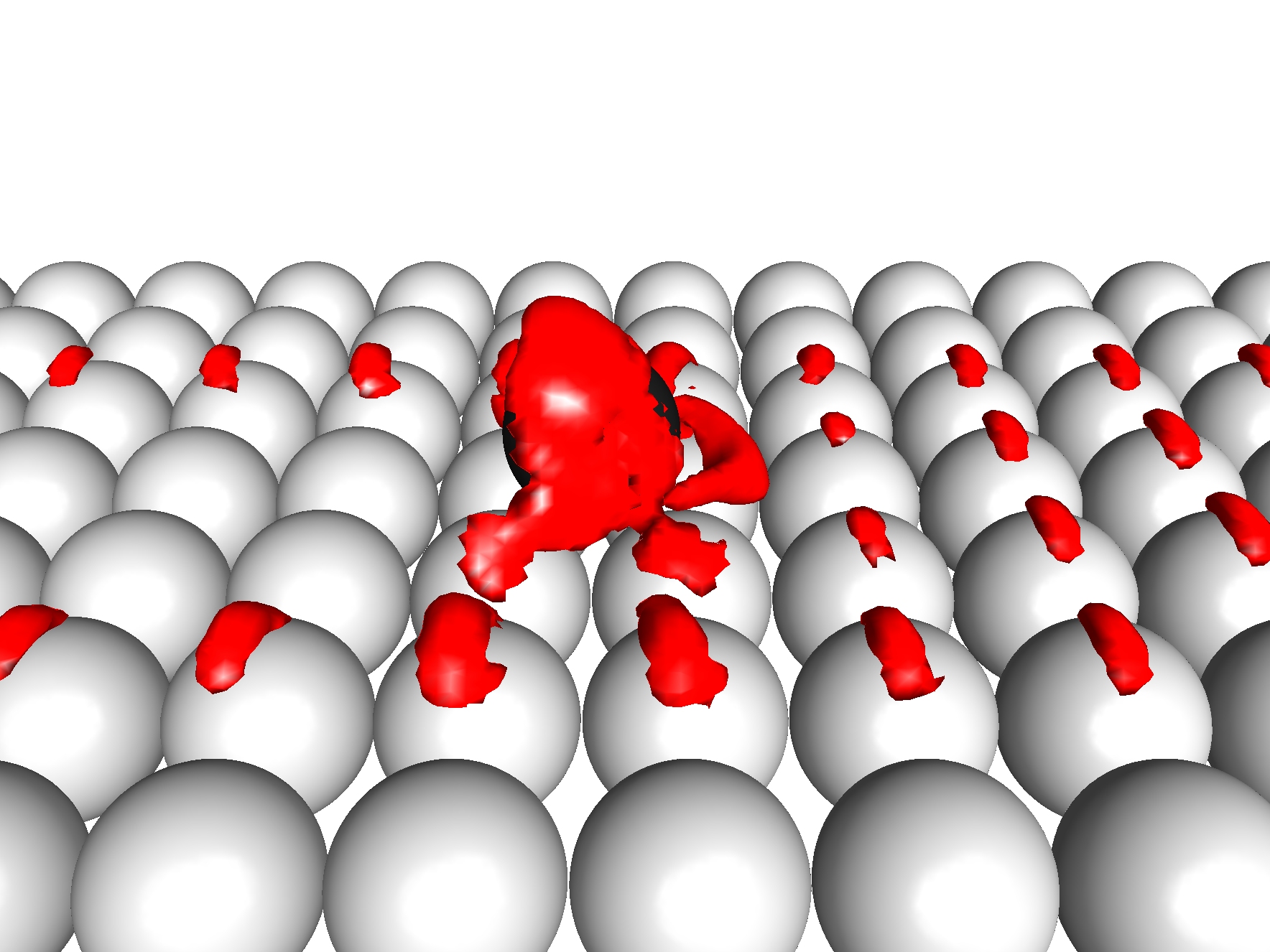}
    \raisebox{2.5cm}{\small $b)$}
    \includegraphics[trim=0 9cm 0 12cm,clip,width=.46\textwidth]{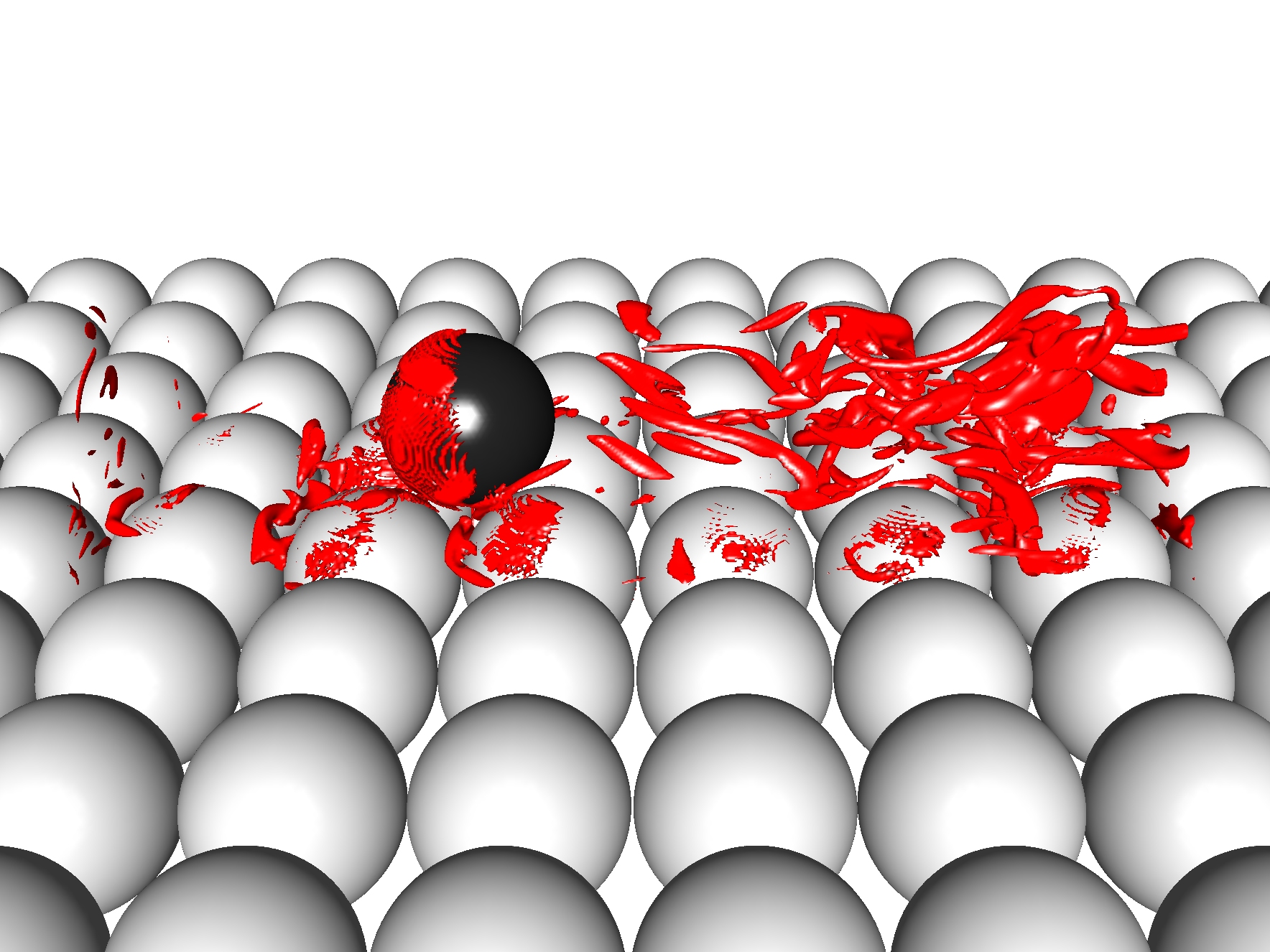}
    \caption{
    Vortex structures visualised by isosurfaces of $\lambda_2=-0.3\,U_0^{2}/\delta^{2}$, where $\lambda_2$ is the second-largest eigenvalue of a tensor derived from the velocity gradient, used to distinguish vortical regions from purely straining ones \citep{Jeong1995}. 
    The visualisations show the flow shortly before the black particle starts to roll, for (a) run~$6$ ($\delta = 0.96$, $\mathrm{Re}_\delta=164$, $\psi=1.53$, $\Gamma=0.019$) and (b) run~$4$ ($\delta = 0.12$, $\mathrm{Re}_\delta=164$, $\psi=1.1$, $\Gamma=0.11$). These runs illustrate two distinct wake regimes despite having the same value of $\mathrm{Re}_\delta$: a coherent wake without vortex shedding in (a) and a transitional regime with unsteady wake dynamics and vortex shedding in (b).}
    \label{fig:IPMNum_l2}
\end{figure}

The hydrodynamic force acting on the mobile particle is governed by a combination of viscous and advective effects, depending on the surrounding flow regime and the development of a wake downstream.
As the particle Reynolds number $\mathrm{Re}_D=\mathrm{Re}_\delta/\delta$ (see table~\ref{tab:simparams}) increases, different flow behaviours emerge that may significantly influence the particle dynamics. 
In the viscous-dominated regime, the drag force scales linearly with the ambient flow velocity, as is the case in run~$3$ ($\mathrm{Re}_\delta=41$, $\delta=0.12$, $\Gamma=0.45$), 
where the flow remains laminar and the boundary layers developing on both the substrate and the mobile particle stay attached throughout the oscillation cycle (cf. figures~\ref{fig:IPM_ParticlemotionDNS}a-b). 
As inertial effects become more pronounced, such as in run~$6$ ($\mathrm{Re}_\delta=164$, $\delta=0.96$, $\Gamma=0.019$), 
the flow separates downstream of the mobile particle, resulting in the formation of a wake that, while not strictly steady due to the transient nature of the oscillating flow, remains coherent without shedding vortices (cf. figures~\ref{fig:IPM_ParticlemotionDNS}c-d and \ref{fig:IPMNum_l2}a). In this case, the particle size is similar to the Stokes boundary layer thickness ($\delta\approx 1$), and the vicinity of the bottom significantly affects the resulting hydrodynamic force acting on the particle.
At higher particle Reynolds numbers, advective effects dominate. In run~$4$ ($\mathrm{Re}_\delta=164$, $\delta=0.12$, $\Gamma=0.10$), 
the wake becomes unsteady, leading to a transitional regime where vortices are randomly shed downstream of the mobile particle (cf. figures~\ref{fig:IPM_ParticlemotionDNS}e-f and \ref{fig:IPMNum_l2}b).
To account for these advective effects in the torque balance model described in \S~\ref{sec:System} and Appendix~\ref{sec:app_model}, the Schiller-Naumann correction \eqref{eq:cDa_appendix} is applied to the hydrodynamic drag.

\subsection{Characterisation of the parameter space}
%
The onset and modality of the early particle motion (whether the particle remains stationary, wiggles within its pocket, or rolls over the substrate) are predicted for each experimental and numerical case using the torque-balance criterion \eqref{eq:upsilon1} presented in \S~\ref{sec:System} and fully derived in Appendix~\ref{sec:app_model}.
First, figure~\ref{fig:Experiment_Simulation_PhaseSpace} maps these cases in the parameter space $(\psi,\delta,\mathrm{Re}_\delta)$.
In the experiments, the oscillation frequency is held constant while the velocity amplitude is gradually increased, such that both $\mathrm{Re}_\delta$ and $\psi$ increase simultaneously (figure~\ref{fig:Experiment_Simulation_PhaseSpace}b).
Circles indicate the cases where the particle is never set into motion throughout the oscillation cycle, while crosses and squares mark wiggling and rolling motion, respectively.
In the experiments, the values of $\psi$ separating the threshold conditions between stationary, wiggling, and rolling are approximately equispaced by $\Delta\psi=0.1$. %

For the numerical simulations, a series of independent runs is performed at increasing values of $\psi$, while keeping the same values of the other parameters indicated in table~\ref{tab:simparams}. %
We find good agreement between the numerical and experimental results at $s=1.81$, as shown in figure~\ref{fig:Experiment_Simulation_PhaseSpace}. %
For these conditions, the motion threshold is estimated at $\psi\approx 1.0$ for wiggling and $\psi\approx1.1$ for rolling.

Wiggling thus occurs within a narrow range of values of $\psi\approx1.0-1.1$, indicating that the transition from stationary to rolling is relatively sensitive to small changes in flow conditions. This behaviour was already observed in figure~\ref{fig:IPM_ParticlePosition}. 
%
It is worth stressing that the wiggling motion is peculiar to oscillatory flow and has no equivalent in steady flow conditions. %

For lighter particles ($s=1.09$), the motion threshold is reached at smaller values of $\mathrm{Re}_\delta$ at similar values of $\delta$, which corresponds to weaker hydrodynamic torque. The corresponding $\psi$-values for wiggling and rolling motion are approximately four times lower than those of the denser particles.
This variation in $\psi$-values marking the motion threshold across different experimental conditions indicates that no universal critical range of $\psi$ can predict incipient motion for all cases considered here.
The data further exhibits a significant scatter, attributed to the relatively large uncertainty in determining the critical velocity at these operating conditions.
Indeed, the dynamics of light particles are sensitive to small changes in parameter values; even changes on the order of the measurement precision can lead to significant discrepancies in $\Upsilon$. In particular, the uncertainty in $s$ ($\pm0.05$) due to measurement precision results in substantial variations in the effective weight, which scales with $s-1$.

\begin{figure}
    \centering
    \includegraphics[width=\linewidth]{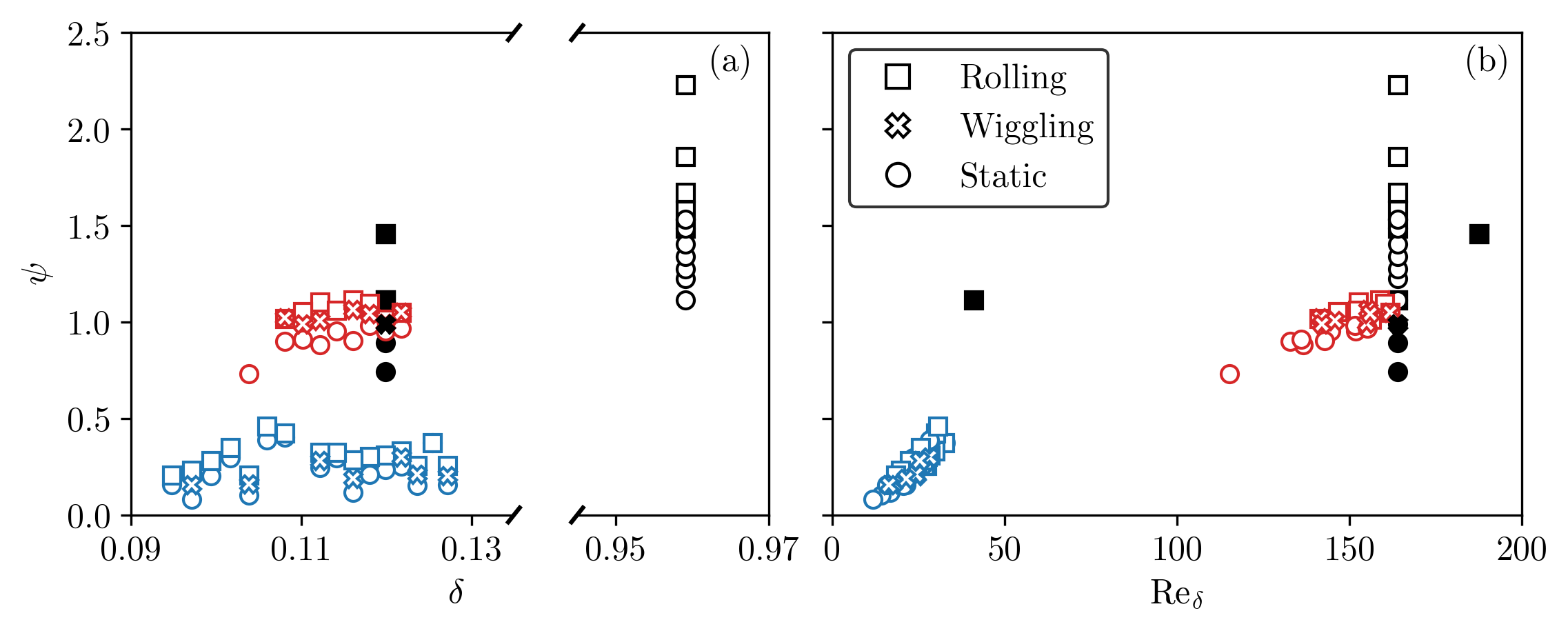}
    \caption{Phase space overview of particle behaviour as a function of $\delta$, $\mathrm{Re}_\delta$, and $\psi$. Crosses and squares mark wiggling and rolling cases, respectively, while circles indicate cases just below the motion threshold. Experimental results are shown in blue ($s=1.09$) and red ($s=1.81$), whereas black symbols represent simulation results for small ($\delta\approx0.96$, open symbols) and large ($\delta\approx0.12$, filled symbols) particles.}
    \label{fig:Experiment_Simulation_PhaseSpace}
\end{figure}

\begin{figure}
    \centering
    \includegraphics[width=\linewidth]{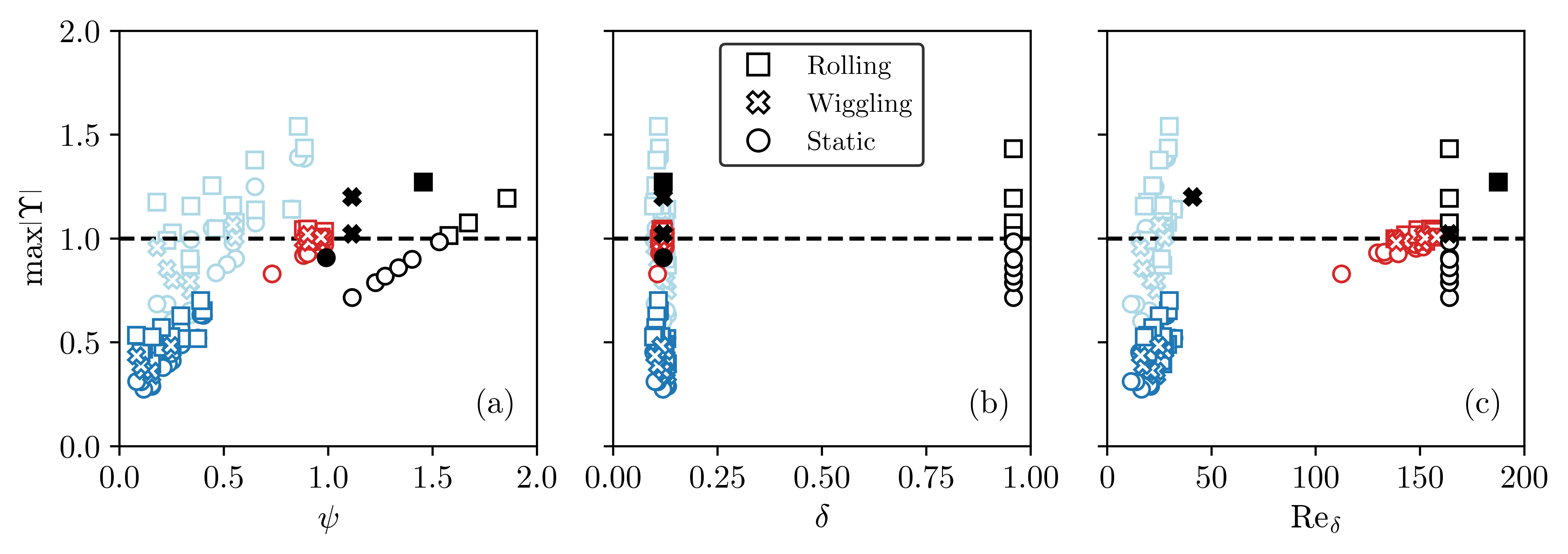}
    \caption{The threshold parameter $\Upsilon$ as a function of (a) $\psi$, (b) $\delta$, and (c) $\mathrm{Re}_\delta$, with circles, crosses, and squares as in figure~\ref{fig:Experiment_Simulation_PhaseSpace}.
    Experimental results are shown in red for $s=1.81$, and in blue for $s=1.09$ (dark symbols) and $s=1.04$ (light symbols). The latter corresponds to a slight modification of $s$ to the lower bound of measurement uncertainty, highlighting the model's sensitivity to lighter particles, due to the scaling with $s-1$. Simulation results are shown in black, with open symbols for small particles ($\delta\approx0.96$) and filled symbols for large particles ($\delta\approx0.12$).}
    \label{fig:Experiment_model_comparison}
\end{figure}

Figure~\ref{fig:Experiment_model_comparison} presents the experimental and numerical maximum values of $\vert\Upsilon\vert$ attained during the oscillation cycle and calculated using \eqref{eq:criticalcondition}, and compares them to the theoretical motion threshold of $|\Upsilon|=1$. The figure shows good agreement between experiments, simulations, and the theoretical threshold. 
Figure~\ref{fig:Experiment_model_comparison} additionally shows that a slight correction of the experimental density ratio from $s=1.09$ to $s=1.04$, within the bounds of measurement uncertainty, yields a significant shift in $\Upsilon$ in the diagram and substantially improves the prediction of incipient motion, for these lighter particles. %

\begin{figure}
    \centering
    \raisebox{5cm}{\small $a)$}
    \includegraphics[width=0.5\textwidth]{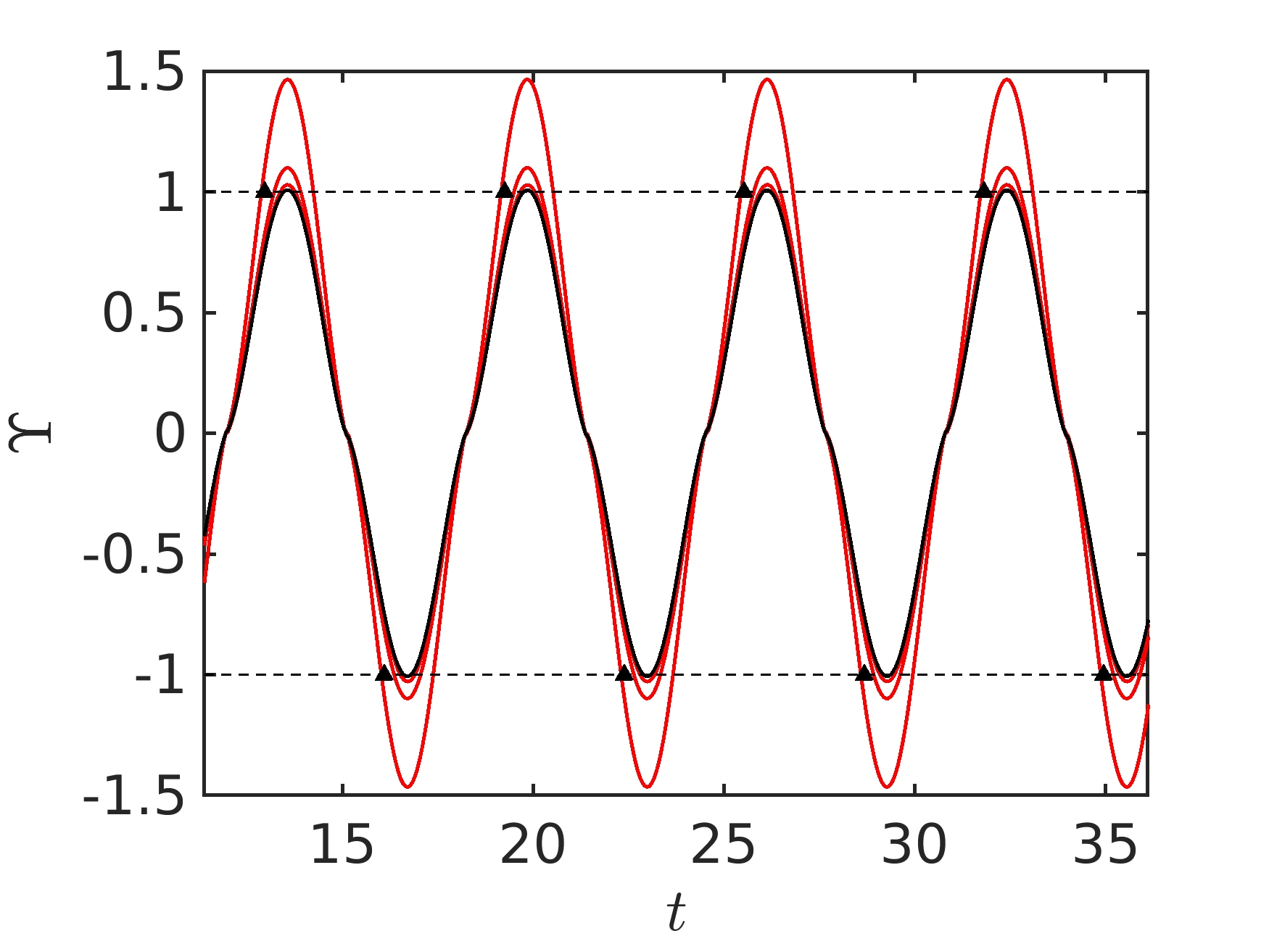}\\
    \raisebox{4.2cm}{\small $b)$}
    \includegraphics[width=0.45\textwidth]{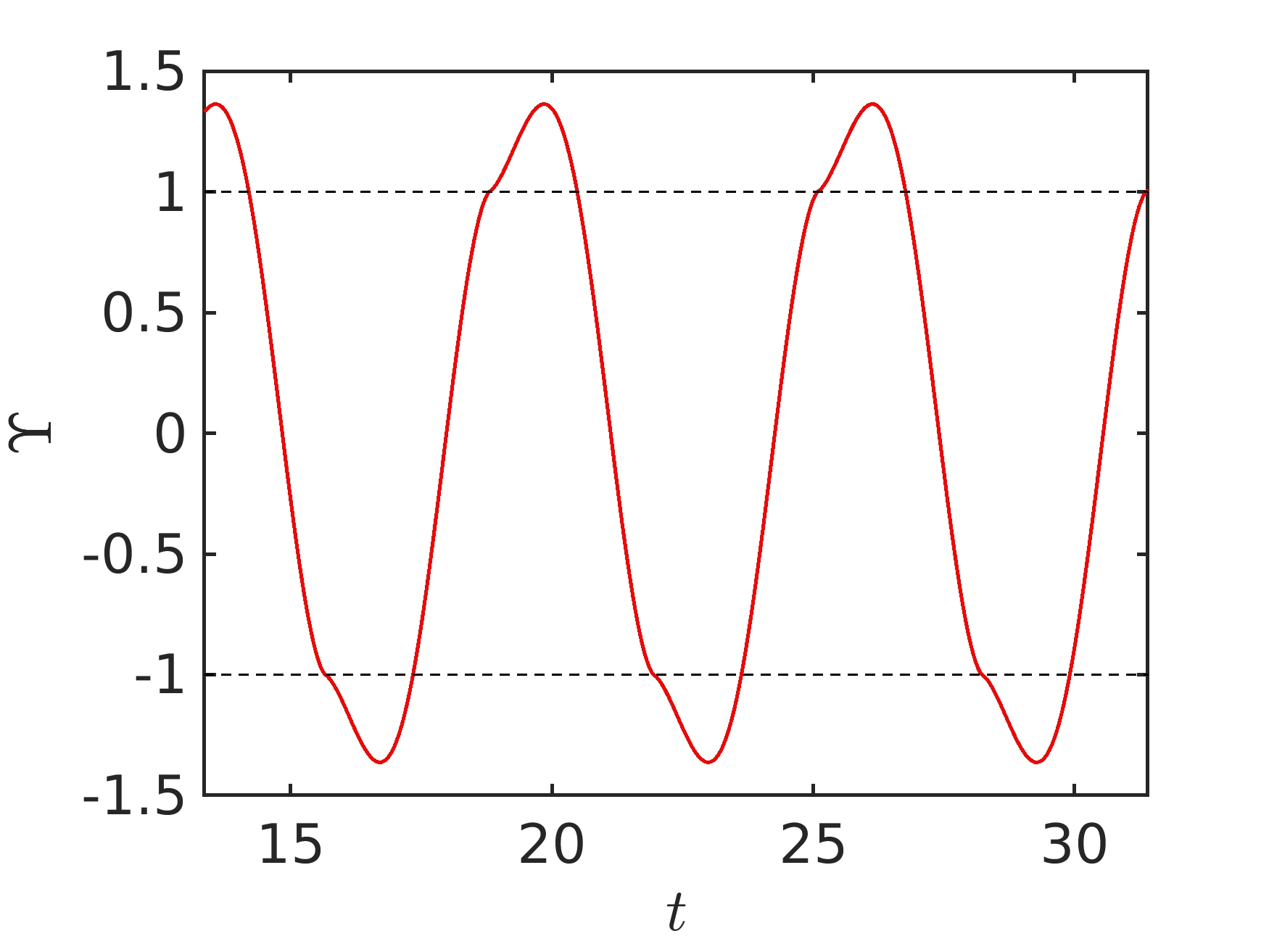}
    \raisebox{4.2cm}{\small $c)$}
    \includegraphics[width=0.45\textwidth]{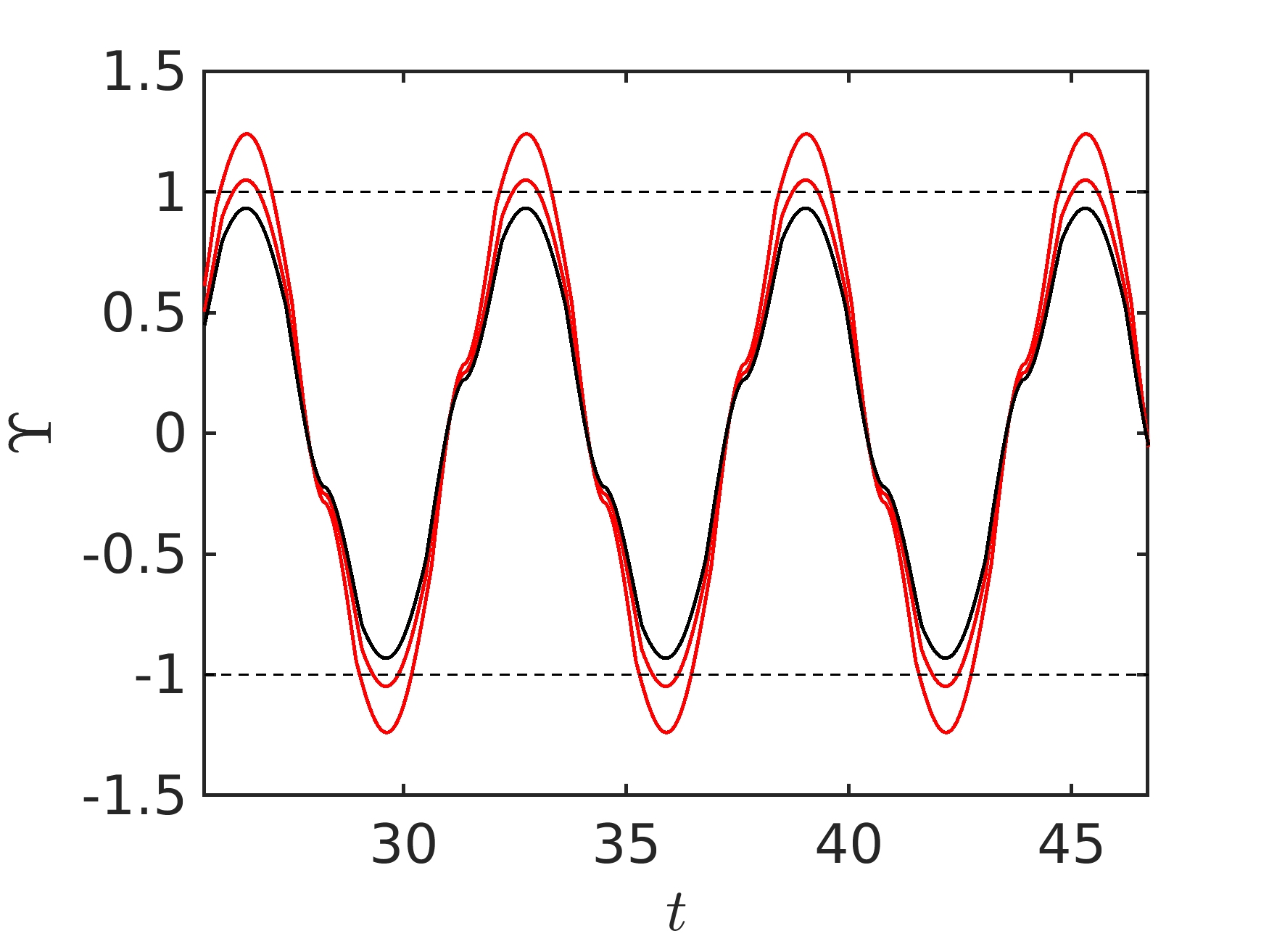}
    \caption{Computed values of $\Upsilon$ for $(a)$ run~$6$ at $\psi=1.53,\,1.58,\,1.67,\,2.23$, $(b)$ run~$3$ at $\psi=1.11$, and $(c)$ runs~$4-5$ at $\psi=0.99,\,1.11$. Black and red curves correspond to simulations in which the mobile particle is static or starts to roll, respectively. The black triangles in (a) mark the phases of incipient particle motion observed in the DNS, which intercept the corresponding model prediction (red curve). Note that the red curve for $\psi=1.58$ nearly overlaps with the black curve for $\psi=1.53$.}
    \label{fig:Upsilon_transient_DNS}
\end{figure}

The accuracy of our model is further confirmed by comparing the transient values of $\Upsilon$ over multiple oscillation cycles and identifying the instants when the particle motion starts, as shown in figure~\ref{fig:Upsilon_transient_DNS}. Notably, in simulations where the maximum value of $\Upsilon$ exceeds unity, the mobile particle is consistently observed to roll. %
A key advantage of the $\Upsilon$-criterion is that it not only predicts the possibility that the particle is set into wiggling or rolling motion without introducing empirical threshold values, but that it also determines the phase within each half-cycle at which the motion starts. %

\subsection{Phase information}\label{sec:results_phase}

\begin{figure}
    \centering
    \includegraphics[width=0.7\textwidth]{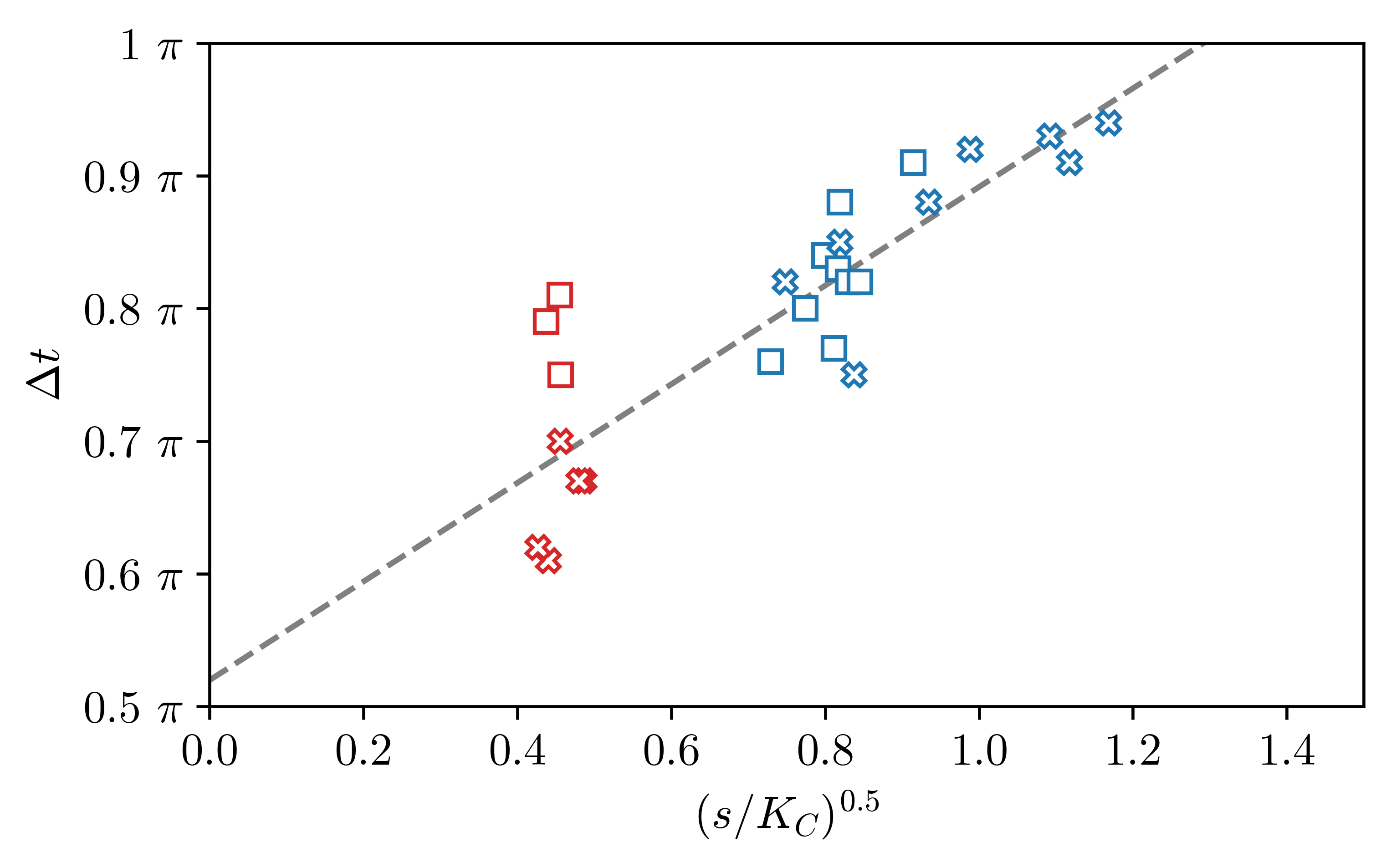}
    \caption{The duration of particle motion within each half-cycle for conditions just above the motion threshold. Red and blue symbols correspond to experiments with relatively heavy ($s=1.81$) and light ($s=1.09$) particles, respectively. Crosses and squares indicate wiggling and rolling motion, respectively. The dashed line shows a linear fit through the data.}
    \label{fig:IPM_MotionPhaseDelta}
\end{figure}

The phases of motion initiation and cessation are determined from mean particle trajectories near the motion threshold, e.g. as shown in figure~\ref{fig:IPM_MotionPhase}. The resulting phase difference $\Delta t$ in figure~\ref{fig:IPM_MotionPhaseDelta} exhibits a linear correlation with $\sqrt{s/K_C}$, similar to that of the time to crest $\tau$ predicted by \eqref{eq:time_scaling}, which is particularly evident for light particles.
Unlike in the calculation of $\Upsilon$, the results in figure~\ref{fig:IPM_MotionPhaseDelta} are much less sensitive to small variations in $s$, as the scaling here involves $s$ rather than $s-1$. %
%
%
For all the present simulations, the dimensionless particle angular acceleration 
remained of order one (e.g., see figure~\ref{fig:ctc_force}e for run~$4$), leading to an equality for the time to crest $\tau$ as given by \eqref{eq:time_scaling}. %

\begin{figure}
    \centering
    \raisebox{4.5cm}{\small $a)$}
    \includegraphics[width=0.465\textwidth]{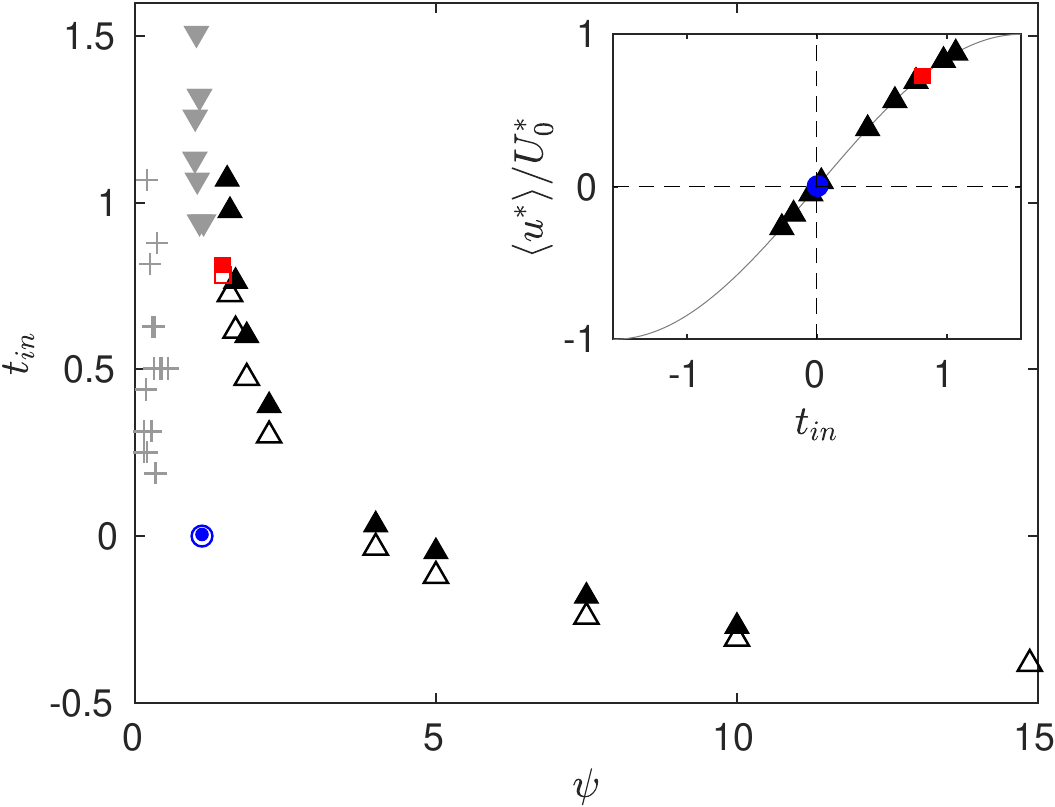}
    \raisebox{4.5cm}{\small $b)$}
    \includegraphics[width=0.465\textwidth]{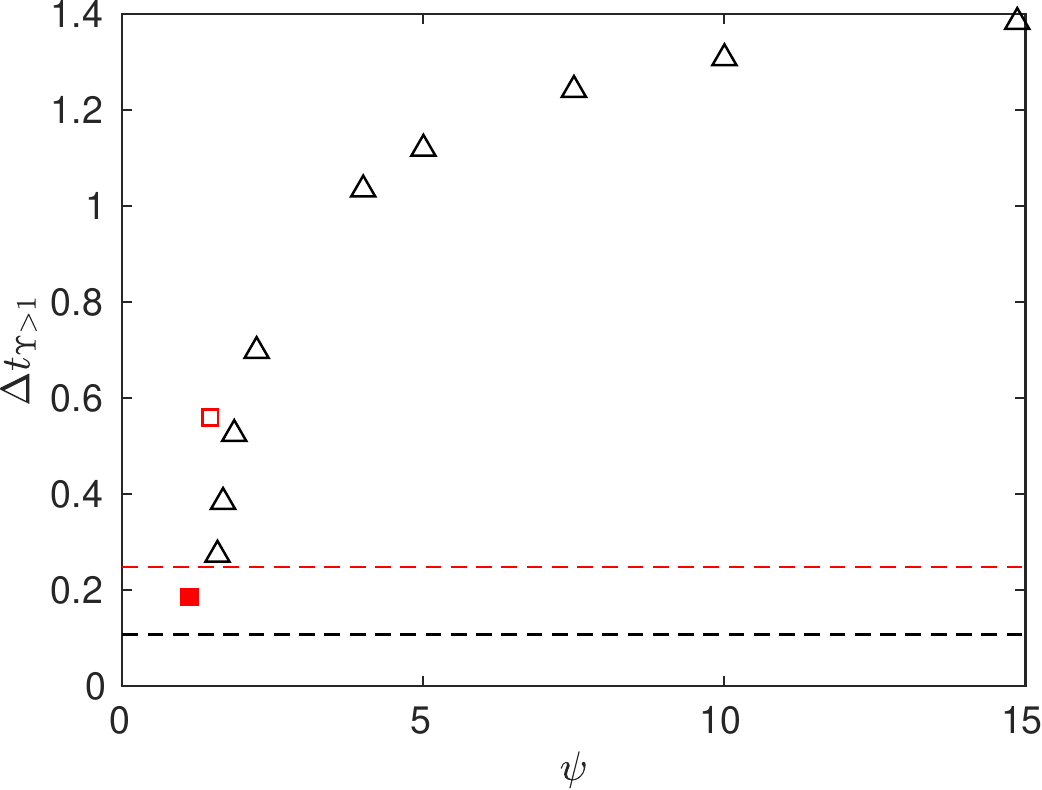}
    \caption{(a) The initiation time $t_\mathrm{in}$ at which $\vert \Upsilon\vert$ exceeds unity and the particle begins to move. DNS results (filled symbols), experimental results (filled grey symbols), and model predictions (open symbols) are compared for parameter values corresponding to run~$4$ (red squares), run~$3$ (blue circles) and run~$6$ (black triangles). The inset shows the initiation time projected onto part of the oscillation cycle centred around the flow reversal, highlighting that for some cases, initiation occurs at flow reversal (blue circles) or even before (leftmost black triangles). 
    (b) The modelled time interval during which $\vert \Upsilon\vert>1$. 
    The threshold values separating wiggling (filled symbols) from rolling particles (open symbols) are indicated by the horizontal red and black dashed lines, as predicted by the time to crest scaling \eqref{eq:time_scaling} for runs~$4$ and $6$, respectively.}
    \label{fig:t_in}
\end{figure}

To further investigate the phase of motion initiation, we analyse cases well above the motion threshold by increasing the value of $\psi$ in the DNS, as shown in figure~\ref{fig:t_in}. The experimental data for dense particles and the DNS results from run 6 (triangles) complement each other, with our model accurately predicting the initiation time observed in the DNS. 
As $\psi$ increases and the condition $|\Upsilon|>1$ is satisfied for a longer duration within the oscillation cycle, the onset of motion shifts to earlier phases. Remarkably, for sufficiently strong forcing, motion begins even before the reversal of the flow far above the substrate. This is attributed to viscous contributions within the oscillatory boundary layer, which induce a phase lead of up to $\pi/2$ in the near-bed velocity relative to the bulk flow. Since the boundary layer is relatively thick in this case ($\delta=0.96$), this phase lead has a pronounced effect on the phase of the hydrodynamic force.
For larger particles, but still in laminar flow conditions (run 3, blue circles), motion initiation in the DNS occurs precisely at the flow reversal, aligning with our model predictions.

Comparing the type of particle motion with the duration $\Delta t_{|\Upsilon|>1}$ for which $\vert \Upsilon\vert>1$ (figure~\ref{fig:t_in}b), we find that this interval generally exceeds the characteristic time scale $\tau$ from \eqref{eq:time_scaling} which separates the wiggling and rolling regimes. The only case where wiggling is observed corresponds to $\Delta t_{\Upsilon>1}<\tau$, supporting the relevance of $\tau$ as the critical time scale for a particle to roll out of a pocket in the bed. 
Additionally, the sharp increase in $\Delta t_{\Upsilon>1}$ with particle mobility (i.e., increasing $\psi$ in figure~\ref{fig:t_in}b) explains the relatively abrupt transition from static to rolling particles, with only a narrow regime of wiggling in between, as previously observed in figure~\ref{fig:Experiment_Simulation_PhaseSpace}.
Finally, the scaling $\tau\sim\delta^{-0.5}$ predicts that $\tau$ asymptotically vanishes for small particles ($\delta\gg1$), which is also the reason why it is challenging to observe the wiggling motion in the case of small particles (run 6, black triangles in figure~\ref{fig:t_in}b). The wiggling motion state is thus typically characteristic of large particles (e.g., run 3, red squares in figure~\ref{fig:t_in}b).

\subsection{Comparison with other criteria}
\begin{figure}
    \centering
    \raisebox{4.5cm}{\small $a)$}
    \includegraphics[width=0.465\linewidth]{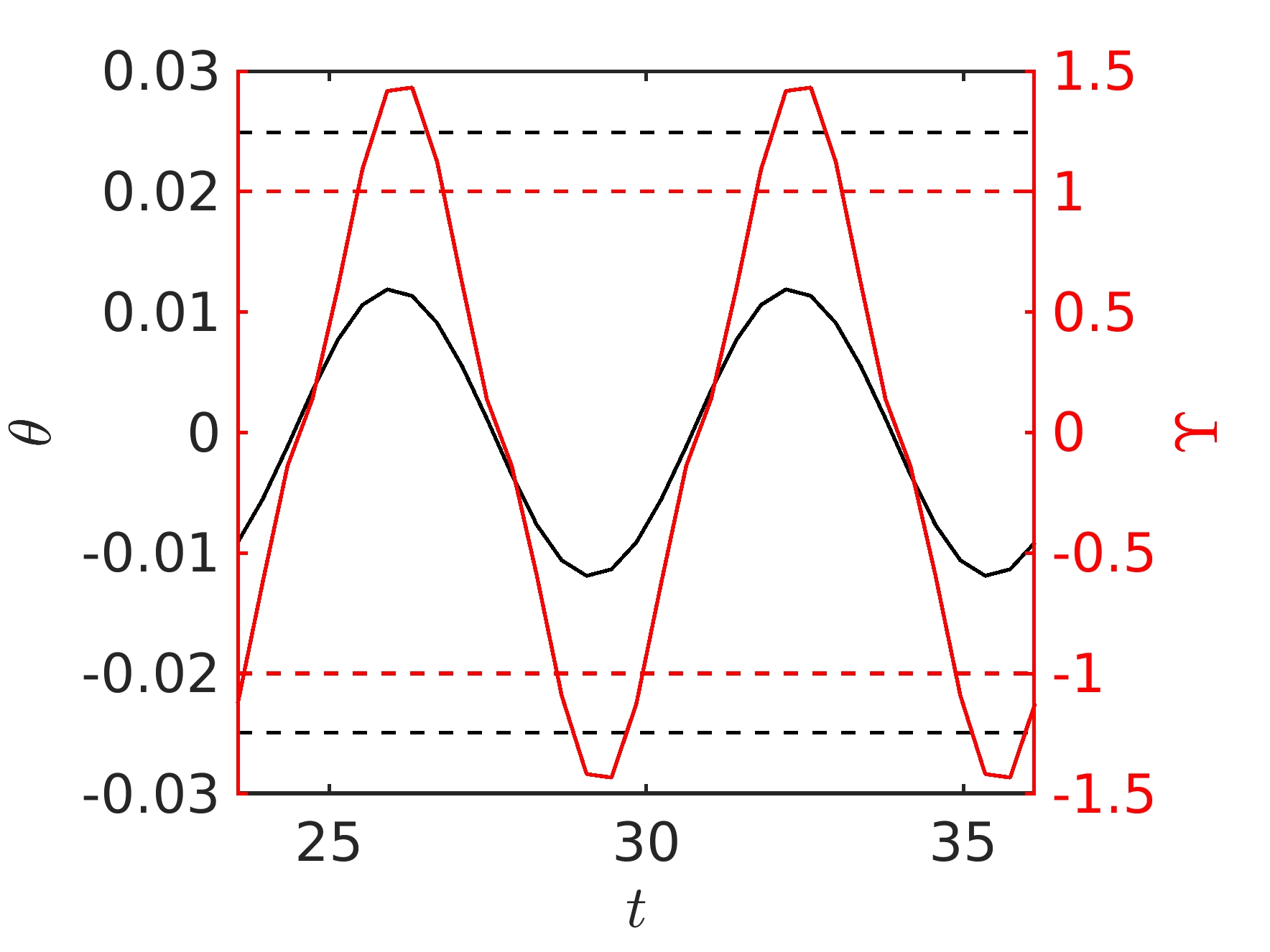}
    \raisebox{4.5cm}{\small $b)$}
    \includegraphics[width=0.465\linewidth]{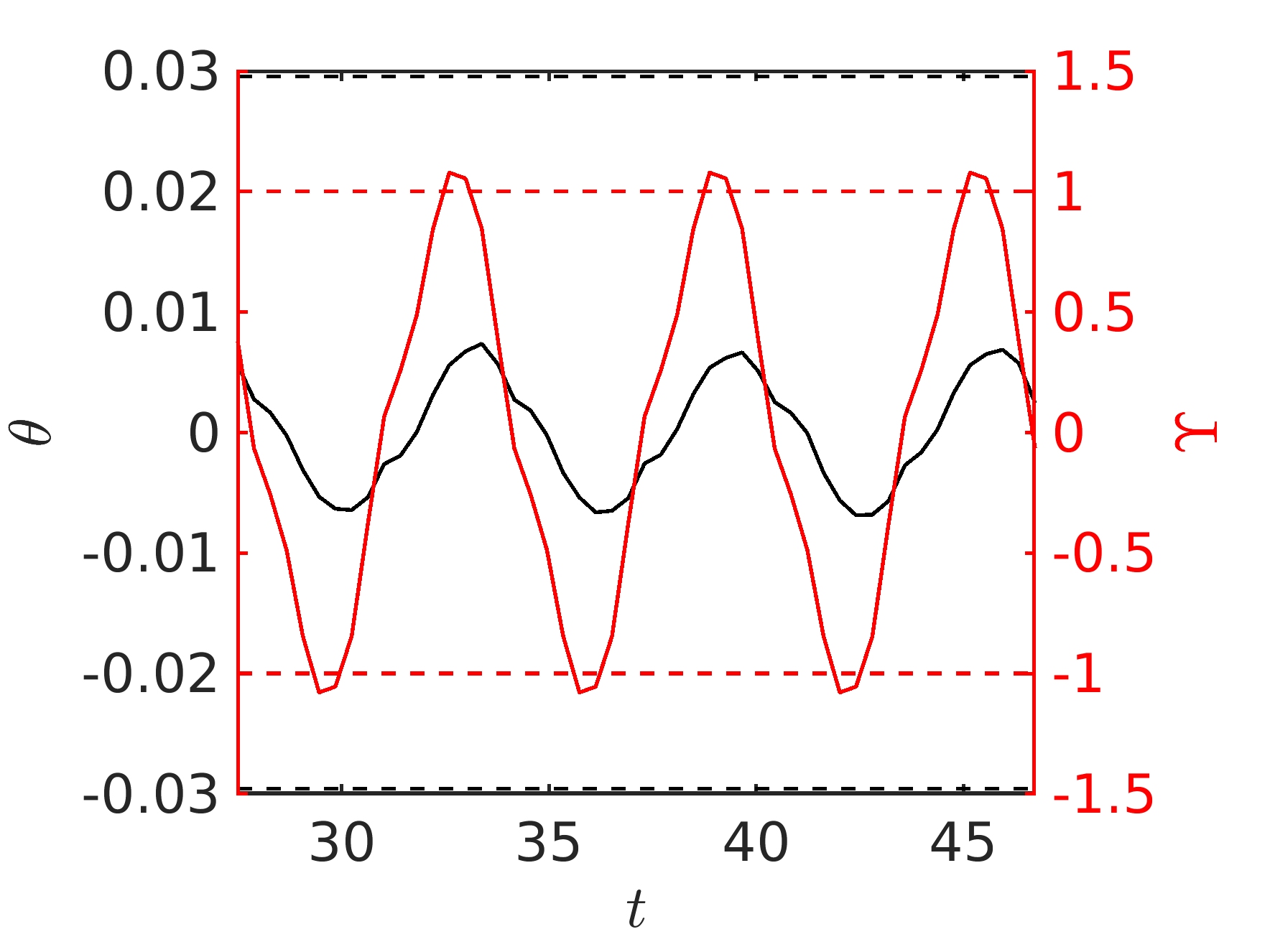}
    \caption{Time evolution of $\Upsilon$ (red curve) as computed from \eqref{eq:criticalcondition} and the Shields parameter $\theta$ (black curve) for cases just above the motion threshold in (a) run~$6$ and (b) run~$4$.
    The horizontal dashed lines indicate the respective critical values, where $\theta_{cr}$ is estimated using the `lower-limit' formula proposed by \citet{paphitis2001}. %
    }
    \label{fig:shileds}
\end{figure}

The method presented in this study enables the deterministic prediction of the initiation of rolling motion for a single sphere exposed to an oscillatory flow. This is conceptually different from predicting the incipient motion of the surface layer of a horizontal sediment layer. %
In the latter case, the arrangement of the particles in the substrate, as well as the disturbance in the flow due to neighbouring mobile particles and their contacts, introduce randomness and additional resistance. %
Hence, criteria based on the balance of horizontal forces acting on the mobile layer, such as the Shields criterion (balancing drag and bottom friction), are expected to fail to predict the incipient motion of individual sediment particles. %
In an OBL, the imposed oscillatory pressure gradient by itself can lead to the initiation of particle motion, making the Shields parameter $\theta$, as defined in \eqref{eq:Shields}, incomplete since it only accounts for the contribution of drag. %
Figure~\ref{fig:shileds} compares incipient motion predictions based on $\theta$ and $\Upsilon$ (as computed from \eqref{eq:criticalcondition}) for runs~$6$ and $4$. %
The critical Shields parameter $\theta_{cr}$ was estimated using the `lower-limit' formula obtained by \citet{paphitis2001}, based on experimental OBL data. %
While both quantities relate to the onset of motion, their time evolution differs significantly, especially for run 4, making it difficult to infer the exact phase of incipient rolling motion directly from $\theta$. More importantly, the threshold $\theta_{cr}$ substantially overestimates the actual onset of motion by up to a factor of five, as indicated by the difference between the black curve and the black dashed threshold line in figure~\ref{fig:shileds}. Thus, it mispredicts the actual conditions for particle transport in unsteady flows. %
Beyond the Shields parameter, \citet{frank2015} proposed to consider also the Sleath parameter, which normalises the force due to the imposed pressure gradient by the particle weight, to predict incipient sediment transport within an OBL with relatively small $\delta$. %
However, none of the models proposed so far accounted for the variation in the velocity distribution during the oscillation cycle and the consequent effects on the torque. %
The present mechanistic approach additionally eliminates the need to determine empirical threshold values for different criteria, which typically depend on flow conditions. %

\section{Concluding remarks}\label{sec:Conclusion}
The threshold and characteristics of particle motion on top of a structured substrate driven by hydrodynamics have been studied, focusing on a uniform monolayer of spheres arranged in a square grid with a single mobile particle on top. While prior results have primarily focused on steady unidirectional flows, this work concentrates on the added complexity introduced by oscillatory flows. The problem is subsequently governed by an additional degree of freedom: the oscillation frequency introduces a time scale that enters the problem through the frequency-dependent viscous length scale $\delta$, corresponding to the normalised thickness of the oscillatory boundary layer. 

Our model for the motion threshold, based on a balance of torques, provides excellent quantitative agreement with both our experimental and numerical data. The model effectively captures the threshold differences due to variations in particle size relative to the viscous length scale, which sets the relative importance of the pressure gradient and hydrodynamic drag. Notably, the model performs well despite relying solely on theoretical velocity profiles and derived forces in the streamwise direction, with only two tunable constants.
Our mechanistic approach provides an accurate estimate of the motion threshold without requiring consideration of the shear stress at the substrate.
In other words, it eliminates the need to determine empirical threshold values for different criteria, which typically depend on flow conditions.

Above the threshold, the particle motion in oscillatory flows is richer than in steady flows, as explored by \citet{agudo2012incipient}~and~\citet{topic2022}. In the steady case, only the motion threshold itself is relevant: once exceeded, a positive feedback loop between hydrodynamic drag and exposure to the flow starts \citep{agudo2017shear}.
In contrast, the additional degree of freedom in the oscillatory case adds a dynamic complexity to the particle behaviour, emerging as a competition between the particle inertia as it rolls towards a neighbouring pocket and the flow reversal time. When the reversal occurs too quickly, the particle may move but fail to reach the next pocket, resulting in a wiggling state that is not observed in the steady case.

Our findings suggest that the present approach can be extended to various flow and substrate conditions, e.g. higher Reynolds numbers, and even to flows that are not strictly laminar but in the transitional regime, provided the relevant length scale characterising the boundary layer can be accurately predicted. In such cases, the pressure gradient has a dominant contribution to the destabilising torque, for which the model accurately accounts. However, the predictive power reaches its limits when hydrodynamic fluctuations become significant, such as those induced by chaotic vortex shedding in turbulent conditions. These fluctuations introduce variability in the force and torque acting on the particle, making its motion inherently unpredictable.

Finally, our work provides extensive opportunities to further explore incipient particle motion in unsteady flows and extend our modelling approach. The effect of bottom geometry warrants further study: while its role has been examined in unidirectional flows \citep{agudo2012incipient,agudo2017shear,retzepoglu2019effect}, its effect on particle dynamics in oscillatory flows remains largely unexplored. An especially intriguing aspect is the potential for cross-stream particle transport relevant to the mixing and dispersion processes of both sediment grains and microplastics. The combination of symmetrical forcing, which on average eliminates net streamwise transport, and anisotropy in the substrate geometry could lead to lateral transport and particle-particle interactions. This mechanism may contribute to the formation of sediment bedforms, such as ripples, alongside other self-organisation processes like steady streaming flows \citep{mazzuoli2016formation,klotsa2009chain,overveld2023pattern}.



\subsection*{Funding}
Marco Mazzuoli was supported under the subaward no.~SUB00004180 of the University of Florida, Gainesville, FL, USA (PTE Federal award no.~N00173-21-2-C900). The simulations were partly performed using the CINECA HPC facilities under the ISCRA-b project SEAWAVES no.~HP10BPMJZR.

\subsection*{Declaration of interests}
The authors report no conflict of interest.

\subsection*{Data availability statement}
The data that support the findings of this study are (soon) openly available in 4TU.ResearchData. 

\subsection*{Author ORCIDs}
T.J.J.M. van Overveld https://orcid.org/0000-0002-3510-7751 \\ 
M. Mazzuoli https://orcid.org/0000-0002-4689-1078\\
M. Uhlmannn https://orcid.org/0000-0001-7960-092X\\
H.J.H. Clercx https://orcid.org/0000-0001-8769-0435\\
M. Duran-Matute https://orcid.org/0000-0002-1340-339X


\appendix
\section{Analytical model for the torque balance}\label{sec:app_model}
This appendix presents the full derivation of the analytical model describing the torque balance on the mobile spherical particle, evaluated about its contact points. To avoid ambiguity, all quantities are expressed in dimensional form. 
The model distinguishes between stabilising and destabilising contributions. %
The onset of particle motion is assessed by evaluating the ratio of these contributions, yielding a criterion for incipient rolling.
The destabilising torques are associated hydrodynamic forces such as drag, lift, added mass, and the imposed pressure gradient, the latter two being related to the flow unsteadiness. The stabilising torque is primarily due to the submerged weight of the particle, which helps preserve its stable position.%

\subsection{Drag and lift force models}
First, we analyse the hydrodynamic drag and lift forces acting on the mobile particle, which are expressed as
\begin{align}
    F_D &= \Int_\surf (\mathbb{T}\cdot\bd{n})\cdot\bd{e}_x dS
          = \pi D\Int_{z_b}^{z_b+D} f_x dz\:\:,
    \\
    F_L &= \Int_\surf (\mathbb{T}\cdot\bd{n})\cdot\bd{e}_z dS 
          = \pi D\Int_{x_c-D/2}^{x_c+D/2} f_z dx\:\:,\label{eq:fz}
\end{align}
where $\mathbb{T}$ is the fluid stress tensor, $x_c$ the streamwise coordinate of the particle centre, and $f_x$ ($f_z$) the streamwise (wall-normal) component of the force density per unit length, namely the hydrodynamic force acting on vertical (horizontal) strips of infinitesimal thickness $dz$ ($dx$), respectively. %
We further define the $n$-th moment $u_n$ of the ambient streamwise velocity $\langle u\rangle(z)$ relative to the elevation $z=z_D$ of the contact point, as %
\begin{equation}\label{eq:app_u_moments}
    u_n = \dfrac{1}{D}\Int_{z_b}^{z_b+D} (z-z_D)^n \langle u\rangle dz.
\end{equation}
Assuming $f_x$ is proportional to the local undisturbed velocity, the drag force and its torque about $z=z_D$ follow as
\begin{align}
    F_D &\propto \pi D\Int_{z_b}^{z_b+D} \langle u\rangle dz 
    = \pi D^2u_0\:\:,
    \label{eq:hypu0}\\
    T_D &\propto \pi D\Int_{z_b}^{z_b+D} (z-z_D)\langle u\rangle dz 
    = \pi D^2 u_1\:\:,
    \label{eq:app_torque}
\end{align}
which define the drag lever arm $L_D$ as
\begin{align}
    L_D=\dfrac{T_D}{F_D} = \dfrac{u_1}{u_0}.
    \label{eq:lDdef}
\end{align}

In the case of a uniform ambient flow, the non-uniformity of $f_x$ over the particle height is solely due to the varying orientation (altitude angle) of the local surface element, since the strip surface remains constant at $\pi D dz$. %
However, in more complex flows, particularly if convective transport dominates (e.g., at moderate-to-high Reynolds numbers or turbulent wake conditions), the actual value of $L_D$ may deviate from \eqref{eq:lDdef}.

Similarly, the lift lever arm $L_L$ is defined by
\begin{align}\label{eq:Ll}
    L_L=\dfrac{T_L}{F_L}\:\:,
\end{align}
where the lift torque about the particle centre is given by
\begin{align}
    T_L &= \pi D\Int_{x_c-D/2}^{x_c+D/2} f_z(x-x_D) dx.
\end{align}

For a spherical particle resting on a flat wall ($z_D=z_b=0$) exposed to a linear shear flow, %
the zeroth velocity moment reads
\begin{align}
    u_0 &= \dfrac{1}{D}\int_0^{D} 2\dfrac{u_c}{D}z dz = u_c \:,
\end{align}
where $u_c$ denotes the ambient flow velocity at the particle centre elevation. %
This quantity plays a central role in the model, as it allows generalisation to unsteady flow conditions beyond the canonical oscillatory boundary layer (OBL).
In the low-Reynolds-number limit, the drag force reduces to Stokes' drag, 
\begin{equation}\label{eq:hydrodynamicdrag_shear}
    F_D = 
    3\pi\rho_f\nu u_0D = 3\pi\rho_f\nu u_cD,
\end{equation}
which is consistent with the assumption \eqref{eq:hypu0}. The corresponding lever arm equals $L_D=(2/3)D$. %

For more complex flow conditions, where $u_0$ deviates from $u_c$, empirical corrections are introduced.
Following \citet{agudo2017shear} for a particle resting on a rough, structured substrate, corrections account for (i) the effective elevation of the particle ($z_D-z_b$) and (ii) the shielding effect of the substrate (e.g., related to $z_0-z_b$).
The normalised geometrical parameters $H=h/D$ and $L=l/D$, from \eqref{eq:H}~and~\eqref{eq:L}, capture these effects.
Additionally, a correction coefficient $C^\prime$ incorporates effects due to finite-size and non-linear shear flows, such that the drag force is modified to
\begin{equation}\label{eq:drag}
    F_D = 3\pi C^\prime\rho_f\nu u_0D,
\end{equation}
where $C^\prime$ depends on the particle Reynolds number 
\begin{equation}\label{eq:ResA}
    \mathrm{Re}_s = \dfrac{u_cD}{\nu}
\end{equation}
and the relative boundary layer thickness $\delta$.
For $\mathrm{Re}_s\lesssim1000$, the Schiller-Naumann correction \citep{schiller1933drag} is used:
\begin{equation}\label{eq:cDa_appendix}
    C'=    1 + 0.15\mathrm{Re}_s^{0.687}.
\end{equation}
For $\delta>1$, the correction proposed by \citet{zeng2009forces} applies for $\mathrm{Re}_s<200$:
\begin{equation}\label{eq:cDb_appendix}
    C'= 1.7005\left(1 + 0.104\mathrm{Re}_s^{0.753}\right).
\end{equation}
Presently, \eqref{eq:cDa_appendix} is applied for cases with $\delta=0.12$, and \eqref{eq:cDb_appendix} for run~$6$. %


We now turn to the lift force. For small particles, we adopt the Saffman lift force expression \citep{saffman1965lift}
\begin{equation}\label{eq:app_lift_saffman}
    F_L = 3.22 C^{\prime\prime} D \rho_f\sqrt{\frac{\nu D}{2\vert u_c\vert}} u_c^2\:\:,
\end{equation}
where $C^{\prime\prime}=1$ for $\delta\gtrsim 1$. For $\mathrm{Re}_s>40$, we use the correction proposed by \citet{mclaughlin1991inertial}~and~\citet{mei1992approximate}
\begin{equation}
    C^{\prime\prime} = 0.0741\sqrt{\mathrm{Re}_s}.
\end{equation}
To prevent overprediction at high values of $\mathrm{Re}_s$, we cap $C^{\prime\prime}$ at the Saffman value ($C^{\prime\prime}=1$) when $\mathrm{Re}_s\gtrsim 1100$, namely when advective effects on the particle dynamics became dominant (see \S~\ref{sec:torque}). %
A similar cut-off was also predicted by \citet{schiller1933drag} for drag at high $\mathrm{Re}_s$ in steady flow conditions, though it is not observed here for the OBL for the parameter values presently considered. %

Since the lift force distribution $f_z(x)$ along the streamwise coordinate is not directly known, the lift lever arm $L_L$ is estimated from our DNS results reported in the following section. Its validity is thus specific to the OBL under conditions close to ours.

\subsection{Torque balance in an oscillatory boundary layer}\label{sec:torque}

We now examine the torque acting on a particle exposed to oscillatory boundary layer (OBL) flow, in the configuration shown in figure~\ref{fig:overview}. %

\begin{figure}
    \centering
    \raisebox{4.5cm}{\small$a)$}
    \includegraphics[width=0.465\textwidth]{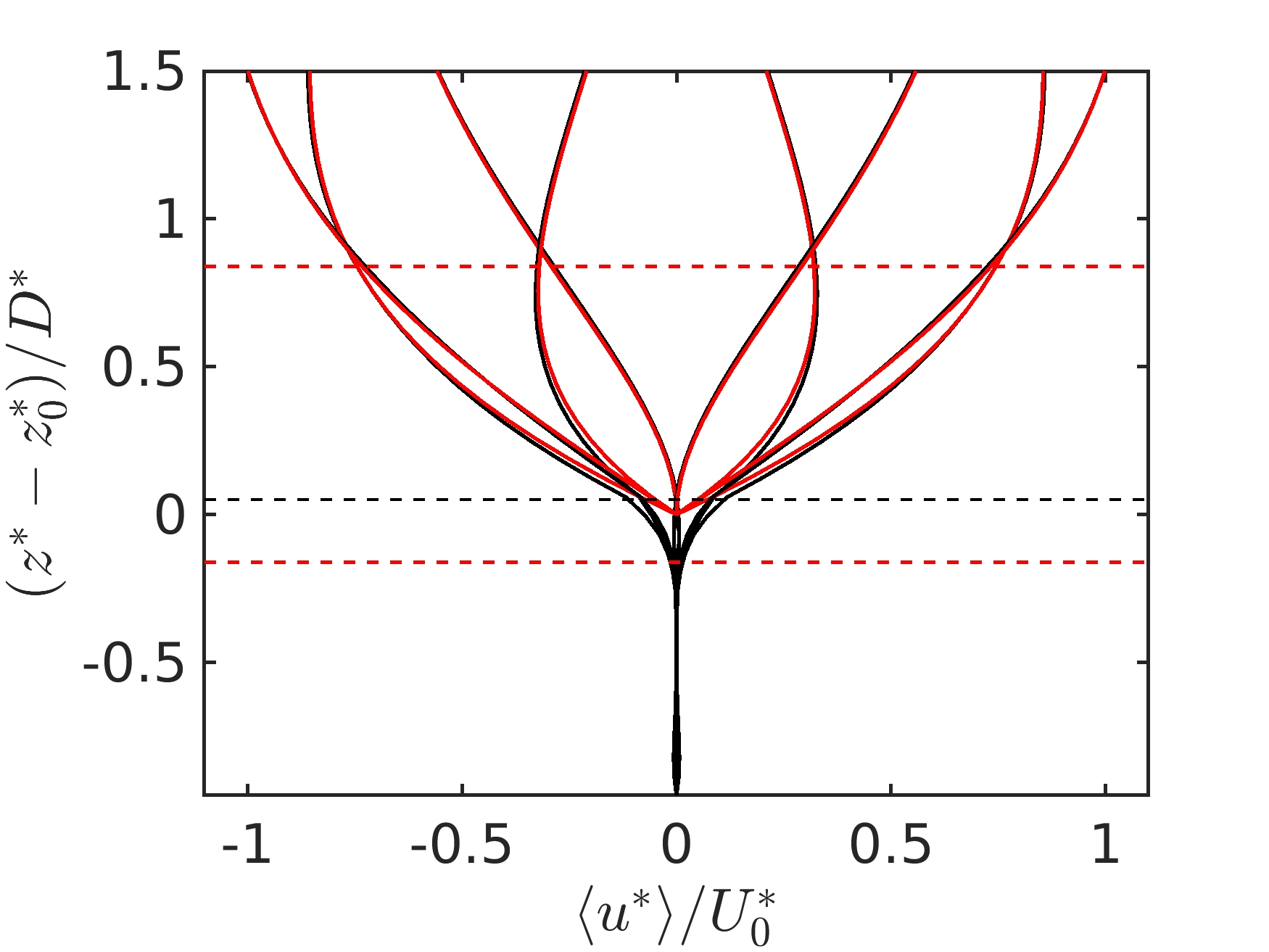}
    \raisebox{4.5cm}{\small$b)$}
    \includegraphics[width=0.465\textwidth]{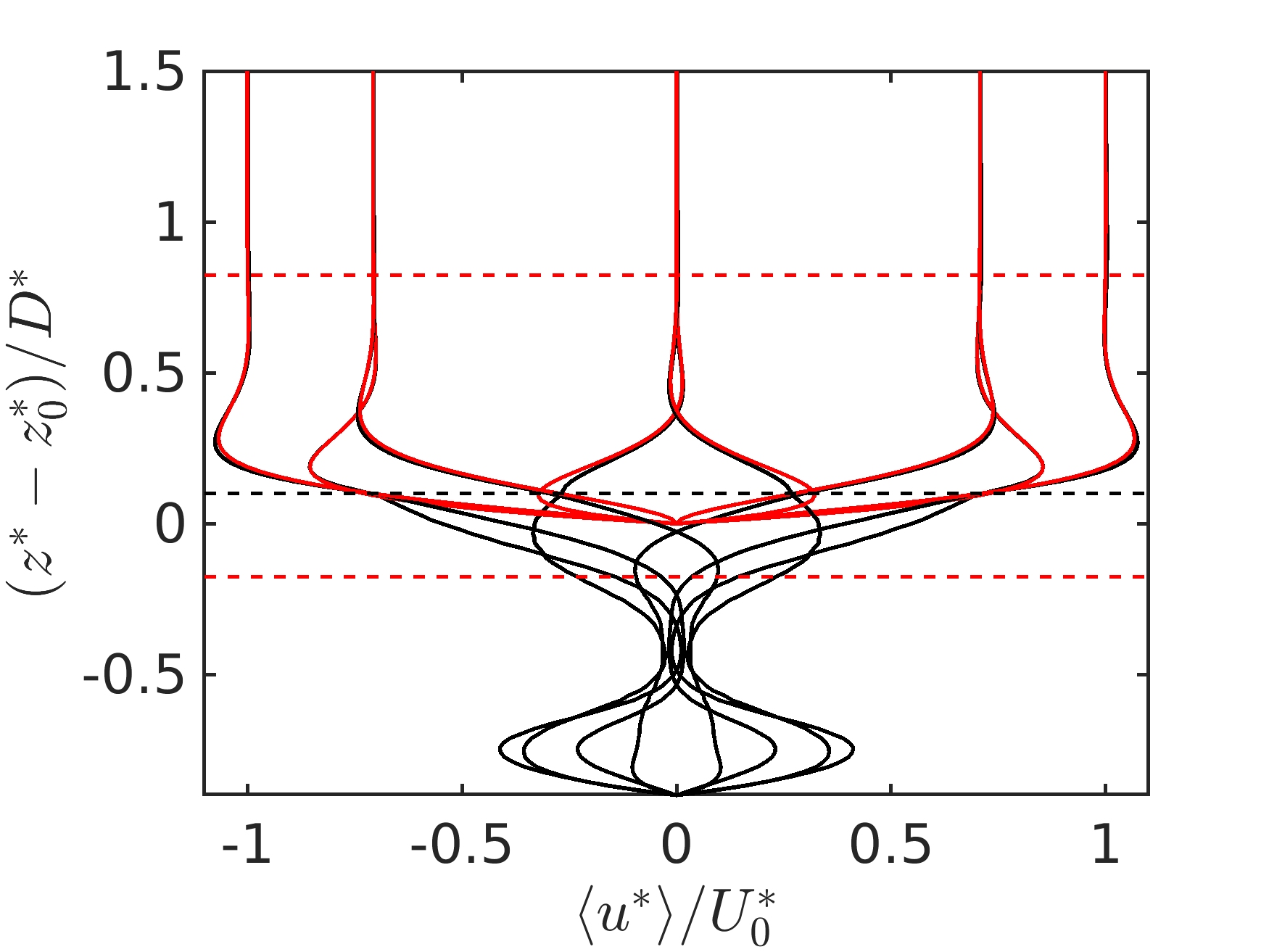}
    \raisebox{4.5cm}{\small$c)$}
    \includegraphics[width=0.5\textwidth]{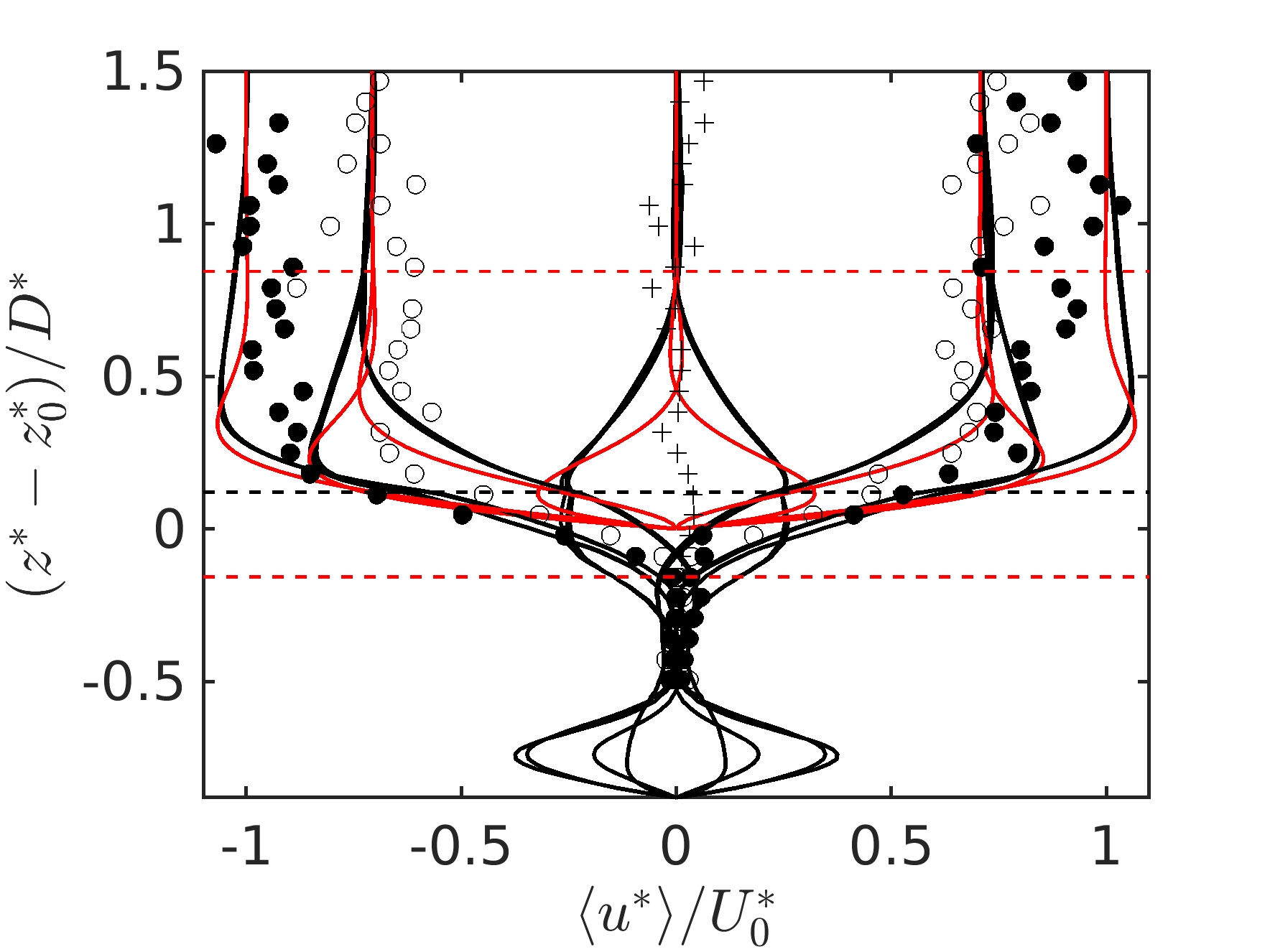}
    \caption{Comparison of streamwise velocity profiles at different phases of the oscillation cycle from DNS results for (a) run 6 ($\delta=0.96$, $\mathrm{Re}_\delta=164$), (b) run~$3$ ($\delta=0.12$, $\mathrm{Re}_\delta=41$) and (c) run~$4$ ($\delta=0.12$, $\mathrm{Re}_\delta=164$). All velocity profiles are plotted relative to the virtual wall elevation $z_0$ (black curves) and compared with the analytical Stokes solution shifted vertically by $z_0$ (red curves). 
    Black dashed lines mark the crest elevation of the fixed substrate particles, located at $z=z_0+0.06\,D$. The lower red dashed lines mark the base elevation of the mobile particle, expressed as $(z_b-z_0)/D$, which equals $-0.14$ (run~$6$), $-0.18$ (run~$3$), and $-0.16$ (run~$4$). The upper red dashed lines indicate the exposure height $h$ of the mobile particle above the substrate. In panel (c), symbols represent experimental PIV data at five oscillation phases, shown for comparison with the numerical results.}
    \label{fig:DNS_velocity_profile_comparison}
\end{figure}

Assuming that the mean velocity $\langle u\rangle$ is reasonably approximated by the Stokes solution \eqref{eq:theory_solution_bounded}, referenced from the effective zero level $z_0$, each torque contribution can be expressed in closed form using the previously introduced force models. 
Figure~\ref{fig:DNS_velocity_profile_comparison} supports this assumption, showing good agreement between DNS velocity profiles and the vertically shifted Stokes solution, particularly for low to moderate $\mathrm{Re}_\delta$. At higher values $\mathrm{Re}_D=\mathrm{Re}_\delta/D$ (e.g. run 4), deviations remain moderate and can be compensated by adjusting the effective viscous length scale. Corrections for such advective effects associated with bottom roughness at moderate Reynolds numbers are further discussed in \S~\ref{sec:modRe}.

The particle Reynolds number is now given by 
\begin{equation}\label{eq:Reparticle}
    \mathrm{Re}_s 
    = 
    \frac{\mathrm{Re}_\delta}{\delta}\frac{u_c}{U_0}
    = 
    \frac{\mathrm{Re}_\delta}{\delta}\left[\sin(t)-e^{-\zeta}\sin\left(t-\zeta\right)\right],
\end{equation}
where $\zeta=(H-1/2)/\delta$ denotes the dimensionless elevation of the particle centre above the zero level of the velocity profile $z_0$. %

\begin{figure}
        \centering
        \includegraphics[width=0.9\textwidth]{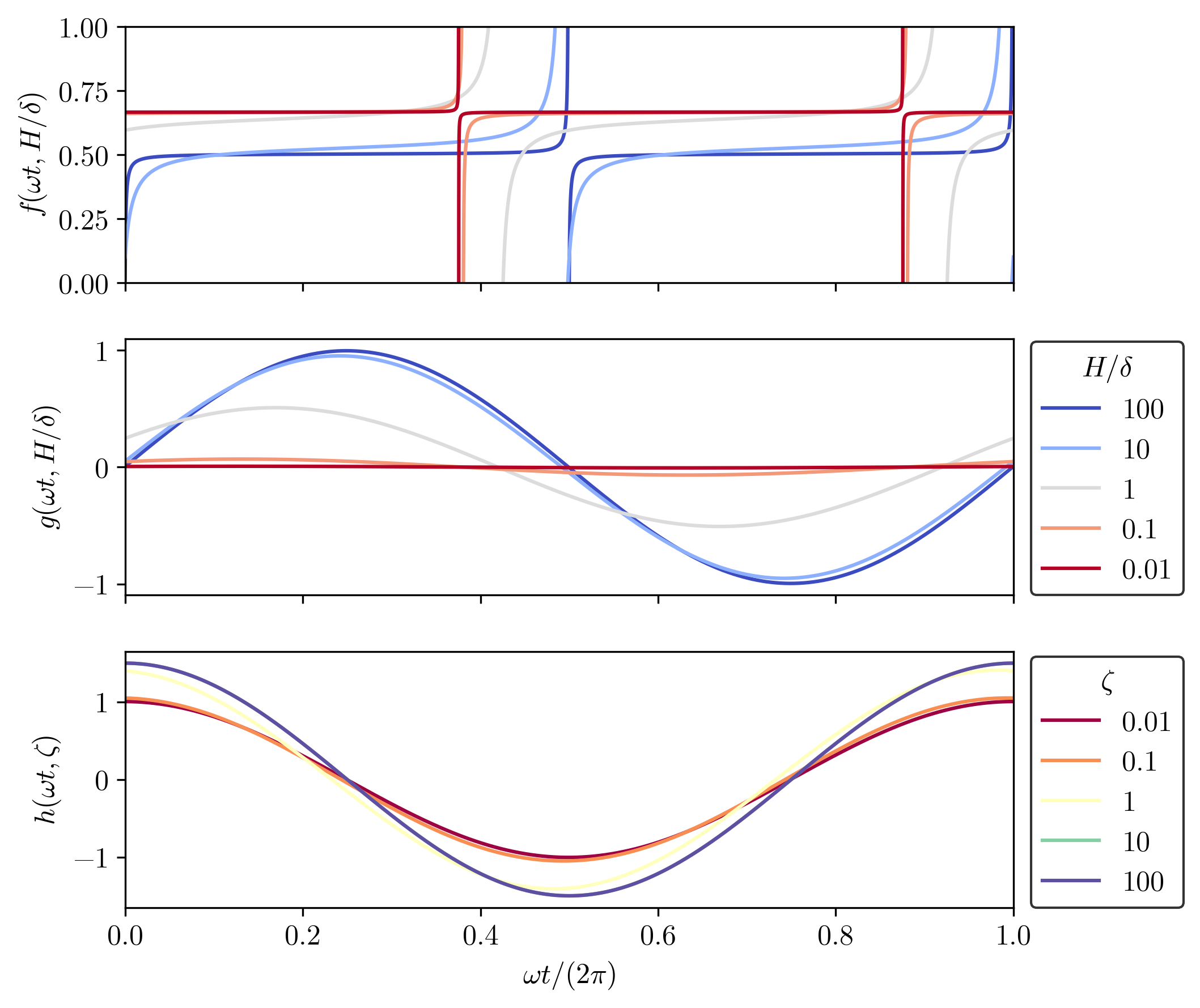}
        \caption{Temporal evolution of the dimensionless functions governing the hydrodynamic torque: the lever arm $f(\omega t,H/\delta)$, the zeroth order moment of the velocity profile $g(\omega t,H/\delta)$, and the contributions from added mass and imposed pressure gradient $h(\omega t,\zeta)$. Note that $f(\omega t,H/\delta)$ diverges at the zero crossings of $g(\omega t,H/\delta)$, corresponding to the flow reversal, where the instantaneous drag force vanishes.}
        \label{fig:fgh_func}
    \end{figure}

The zeroth moment of the velocity across the particle height is
\begin{equation}
    \dfrac{u_0(t,H,\delta)}{U_0} 
    = \dfrac{1}{U_0D}\int_{z_b}^{z_b+D} u dz 
    = \dfrac{1}{U_0D}\int_{z_0}^{z_b+D} u dz  
    = H\,g\left(\omega t,\frac{H}{\delta}\right),
\end{equation}
with $H=(z_b+D-z_0)/D$ the dimensionless exposure height of the particle. %
The function $g(\omega t,H/\delta)$, plotted in figure~\ref{fig:fgh_func}b, quantifies the effect of shielding by the substrate and is defined as
\begin{equation}\label{eq:g_fun}
    g\left(\omega t,\frac{H}{\delta}\right) = \sin(\omega t)-\frac{\sqrt{2}}{2}\frac{\delta}{H}\left[\sin\left(\omega t-\frac{\pi}{4}\right)- e^{-H/\delta}\sin\left(\omega t - \frac{H}{\delta} - \frac{\pi}{4}\right)\right].
\end{equation}
In the limit $H/\delta \gg 1$ (large particle), $u_0/U_0\approx H\sin(\omega t)$, reflecting a quasi-uniform profile. In contrast, for $H/\delta \ll 1$ (small particle), the expression reduces to $u_0/U_0\approx\left(\sqrt{2}/2\right)\left(H^2/\delta\right)\sin\left(\omega t+\pi/4\right)$, 
which is maximum during the phases of maximum bottom shear stress. The latter case further shows the torque being dominated by bottom shear stress rather than imposed pressure gradient. %

With the drag force \eqref{eq:drag} now reading
\begin{equation}
    F_D = 3\pi C'\rho_f\nu U_0DH\,g\left(\omega t,\frac{H}{\delta}\right),
\end{equation}
%
the hydrodynamic torque associated with drag is given by the product of the drag force $F_D$ and its effective lever arm $L_D$, illustrated in figure~\ref{fig:overview} and approximated by \eqref{eq:lDdef}. %
Using the Stokes boundary layer profile \eqref{eq:theory_solution_bounded}, the normalised lever arm can be written as
\begin{align}
    \frac{L_D}{D}=\frac{u_1}{D\,u_0} = H\left[f\left(\omega t,\frac{H}{\delta}\right)-1\right] + L + \frac{1}{2},
\label{eq:ld}
\end{align}
where $L=l/D = 1/2 + \left(z_b-z_D\right)/D$ accounts for the vertical offset between the particle center and the contact points, effectively incorporating the influence of the substrate geometry (cf. figure~\ref{fig:overview}b). %

The time-dependent function $f(\omega t,H/\delta)$, shown in figure~\ref{fig:fgh_func}a, describes the first moment of the velocity distribution and is given by 
\begin{equation}\label{eq:f_fun}
    f\left(\omega t,\frac{H}{\delta}\right) = \dfrac{
    \sin\left(\omega t\right) 
    +
    \left(\dfrac{\delta}{H}\right)^2\,\cos\left(\omega t\right) 
    +
    \dfrac{\delta}{H}{\mathrm{e}}^{-H/\delta}\,\left[\sqrt{2}\sin\left(\omega t-\dfrac{H}{\delta}-\dfrac{\pi}{4}\right)-\dfrac{\delta}{H}\cos\left(\omega t-\dfrac{H}{\delta}\right)\right]
    }{
    2 g\left(\omega t,\dfrac{H}{\delta}\right)
    }
    .
\end{equation}
In the limiting case of a particle resting directly on a flat surface ($L=1/2$ and $H=1$), the lever arm simplifies to $L_D/D=f(\omega t, H/\delta)$. %
For a substrate of spherical particles in a square arrangement, we obtain $L=\sqrt{2}/4$. %
In the large particle regime ($H/\delta\gg 1$), the lever arm reduces to $L_D=1/2$, while in the small particle regime ($H/\delta\ll 1$), it tends to $L_D\approx 2/3$, consistent with predictions for steady linear shear flow \citep{agudo2017shear}. %
Thus, the torque associated with the drag force, normalised according to \eqref{eq:theory_dimlesvars2}, reads
\begin{equation}
\frac{T_{Sy}^{\mathrm{drag}}}{\tau_{\nu0}\pi D^3/2} = 6 C'H\delta\left[\left(f\left(\omega t,\frac{H}{\delta}\right)-1\right)H+L+\frac{1}{2}\right]g\left(\omega t,\frac{H}{\delta}\right),
\label{eq:t_adim}
\end{equation}
where the terms inside the brackets correspond to the time-dependent lever arm $L_D$ defined in \eqref{eq:ld}.

\begin{figure}
    \centering
    \raisebox{3cm}{\small$a)$}
    \includegraphics[width=0.3\textwidth]{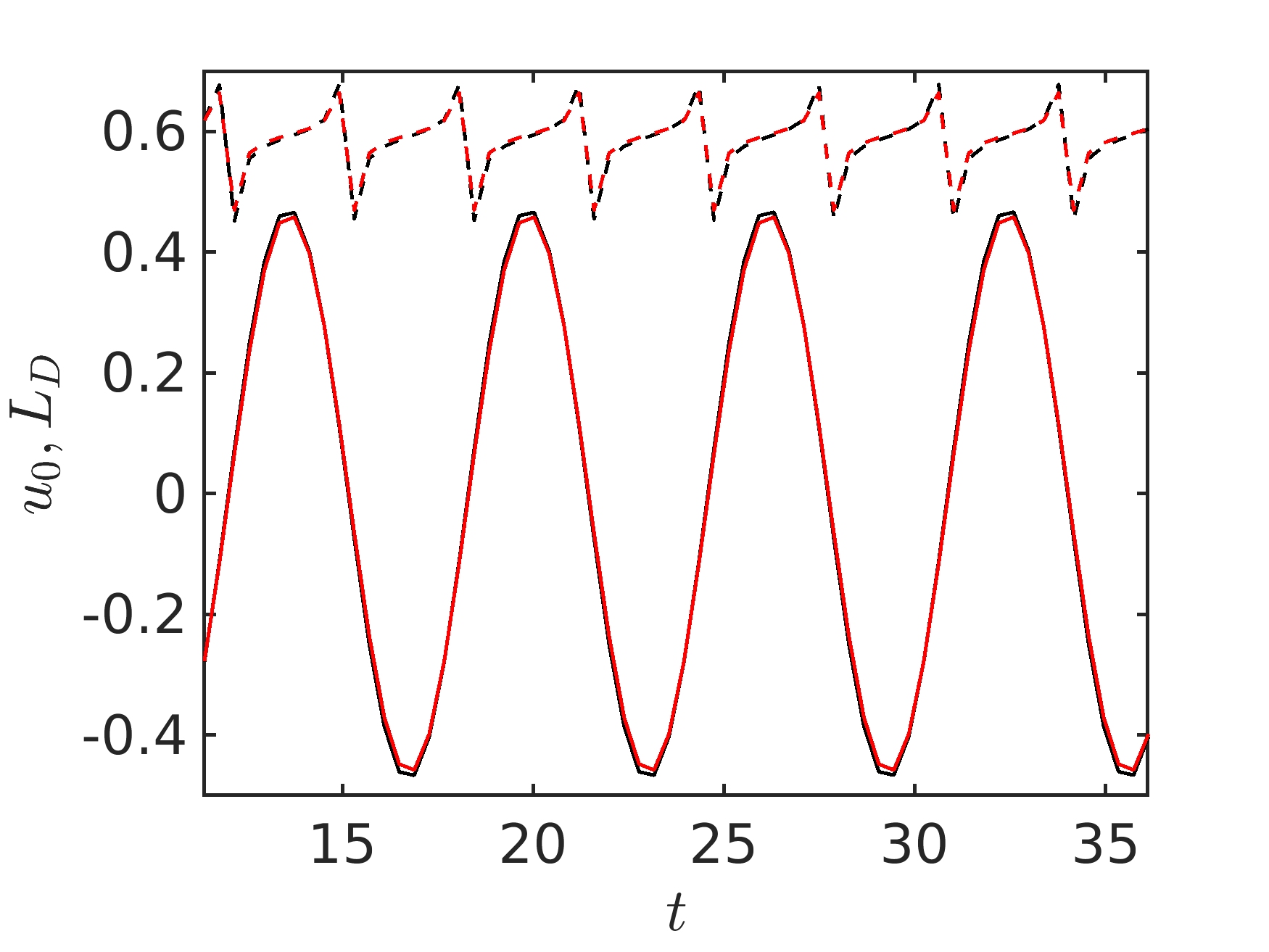}
    \raisebox{3cm}{\small$b)$}
    \includegraphics[width=0.3\textwidth]{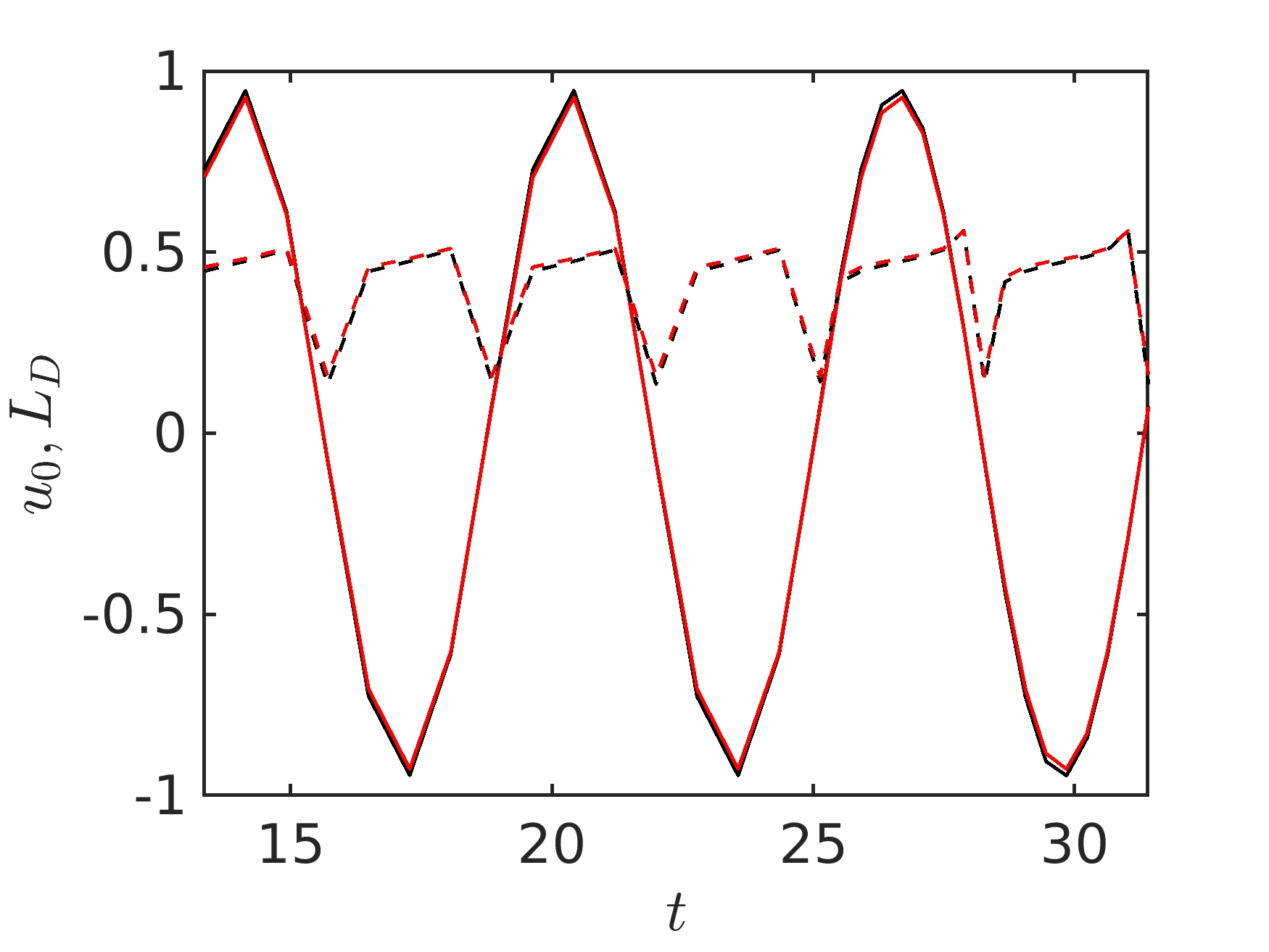}
    \raisebox{3cm}{\small$c)$}
    \includegraphics[width=0.3\textwidth]{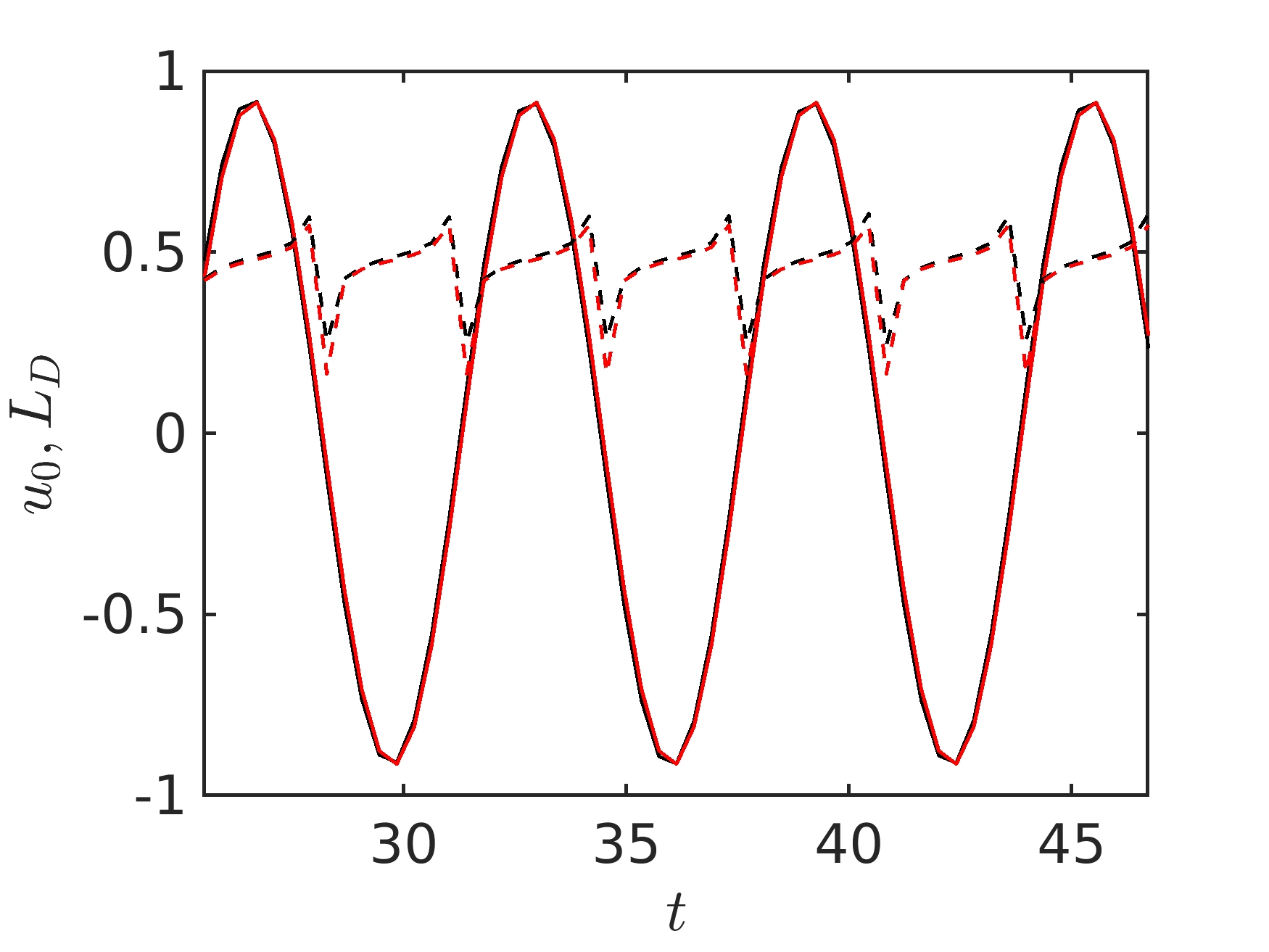}
    \caption{Comparison between model predictions (red) and DNS results (black) for (a) run~$6$ ($\mathrm{Re}_\delta=164$, $\delta=0.96$), (b) run~$3$ ($\mathrm{Re}_\delta=41$, $\delta=0.12$), and (c) run~$4$ ($\mathrm{Re}_\delta=164$, $\delta=0.12$). Solid and dashed lines represent the zeroth velocity moment $u_0$ and the drag lever arm $L_D$, respectively.}
    \label{fig:fg_comp}
\end{figure}

\begin{figure}
    \centering
    \raisebox{5cm}{\small$a)$}
    \includegraphics[width=0.465\textwidth]{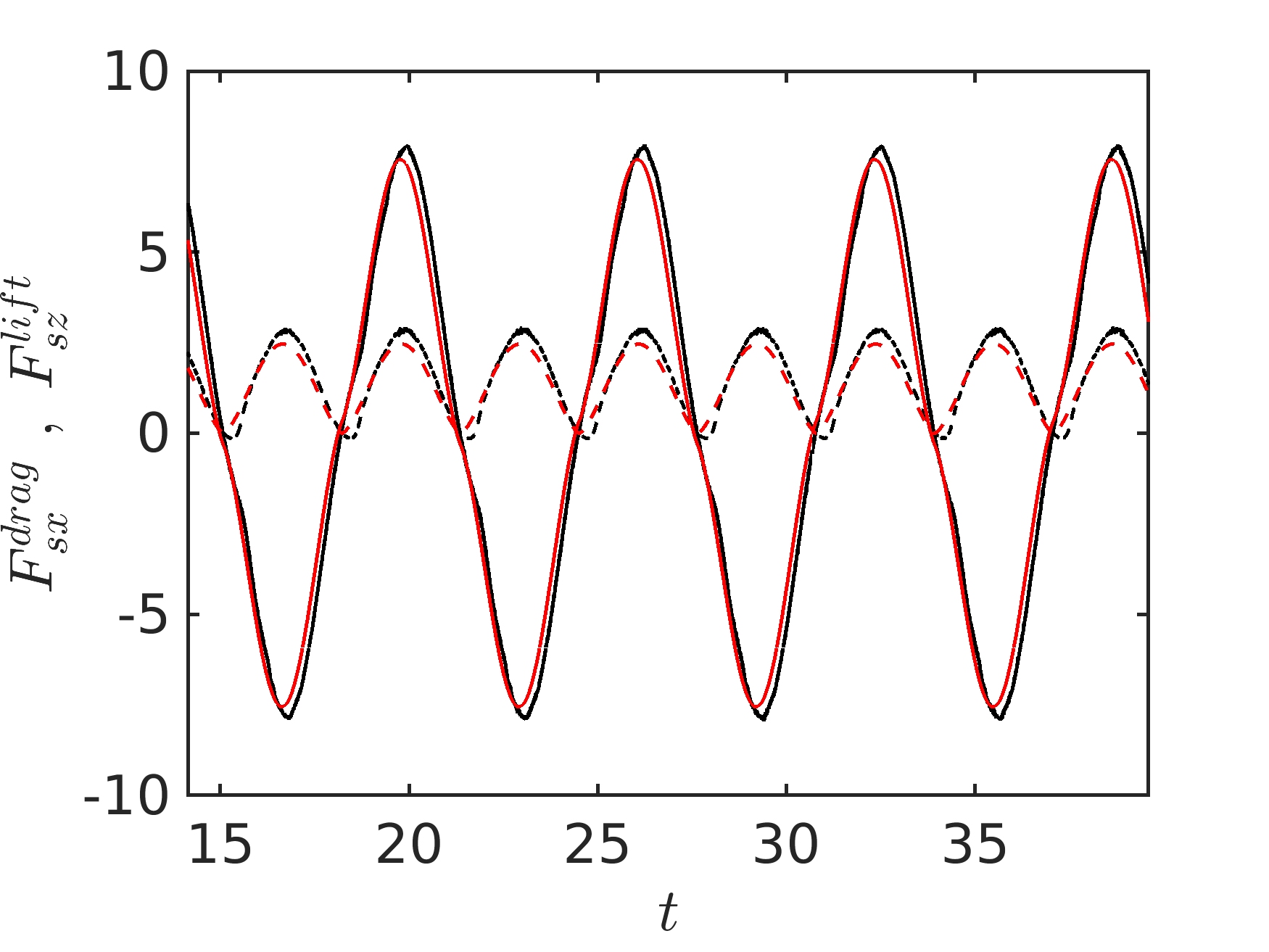}
    \raisebox{5cm}{\small$b)$}
    \includegraphics[width=0.465\textwidth]{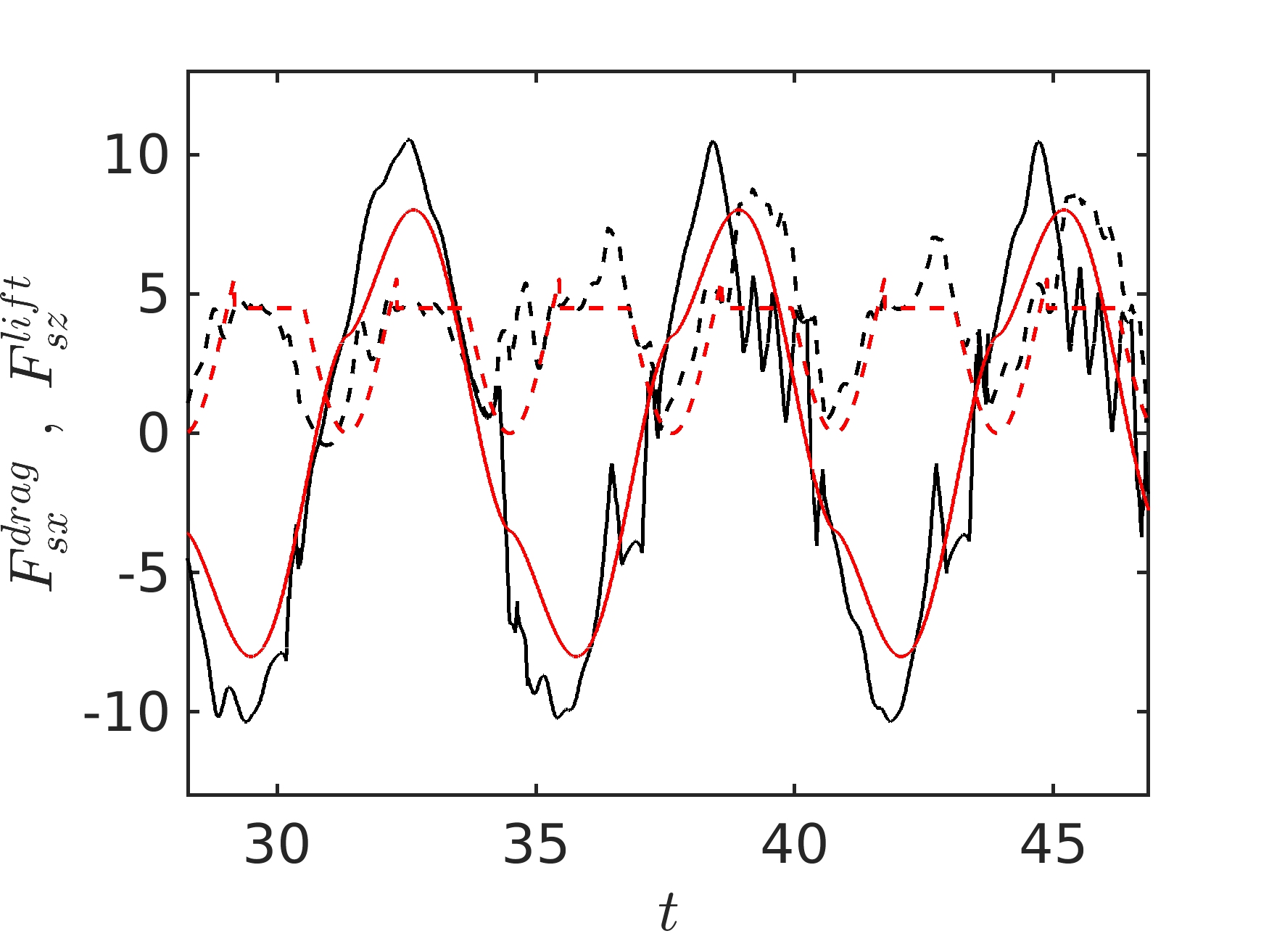}
    \caption{Comparison between model predictions (red) and DNS results (black) of the hydrodynamic forces acting on the particle for (a) run 6 and (b) run 4. Solid and dashed lines indicate drag and lift force components, respectively.
    In run~$6$, the Stokes profile is referenced from a virtual origin located at $z_0=z_b+0.15~D$.}
    \label{fig:drag_lift}
\end{figure}

The analytical predictions for the zeroth moment of the velocity $u_0$ and the drag lever arm $L_D$ show excellent agreement with DNS results from runs~$3$, $4$, and $6$ (figure~\ref{fig:fg_comp}). This confirms the accuracy of the expressions derived from the Stokes profile referenced to an effective zero level. 
Furthermore, figure~\ref{fig:drag_lift} shows that the same model accurately reproduces the time-resolved drag and lift forces for both small and large particle cases ($\delta=0.96$ and $\delta=0.12$, respectively). This close agreement between model and DNS results across a range of flow conditions supports the validity of the force expressions and associated torque calculations within the present parameter space.

In addition to viscous drag, the unsteady nature of the oscillatory flow gives rise to additional hydrodynamic forces, namely those due to the imposed external pressure gradients and the added mass effect. These also contribute to the horizontal torque on the particle. 
The torques associated with these forces are expressed by %
\begin{align}
    T_{\nabla p} &= F_{\nabla p} L_{\nabla p} = 
    \rho_f V_s U_0 \omega \cos(\omega t)\,L, \\
    T_\mathrm{AM} &\approx \frac{1}{2}\rho_fV_s\frac{\partial u_c}{\partial t} L,
\end{align}
respectively.

\begin{figure}
    \centering    
    \raisebox{4cm}{\small$a)$}
    \includegraphics[width=0.47\textwidth]{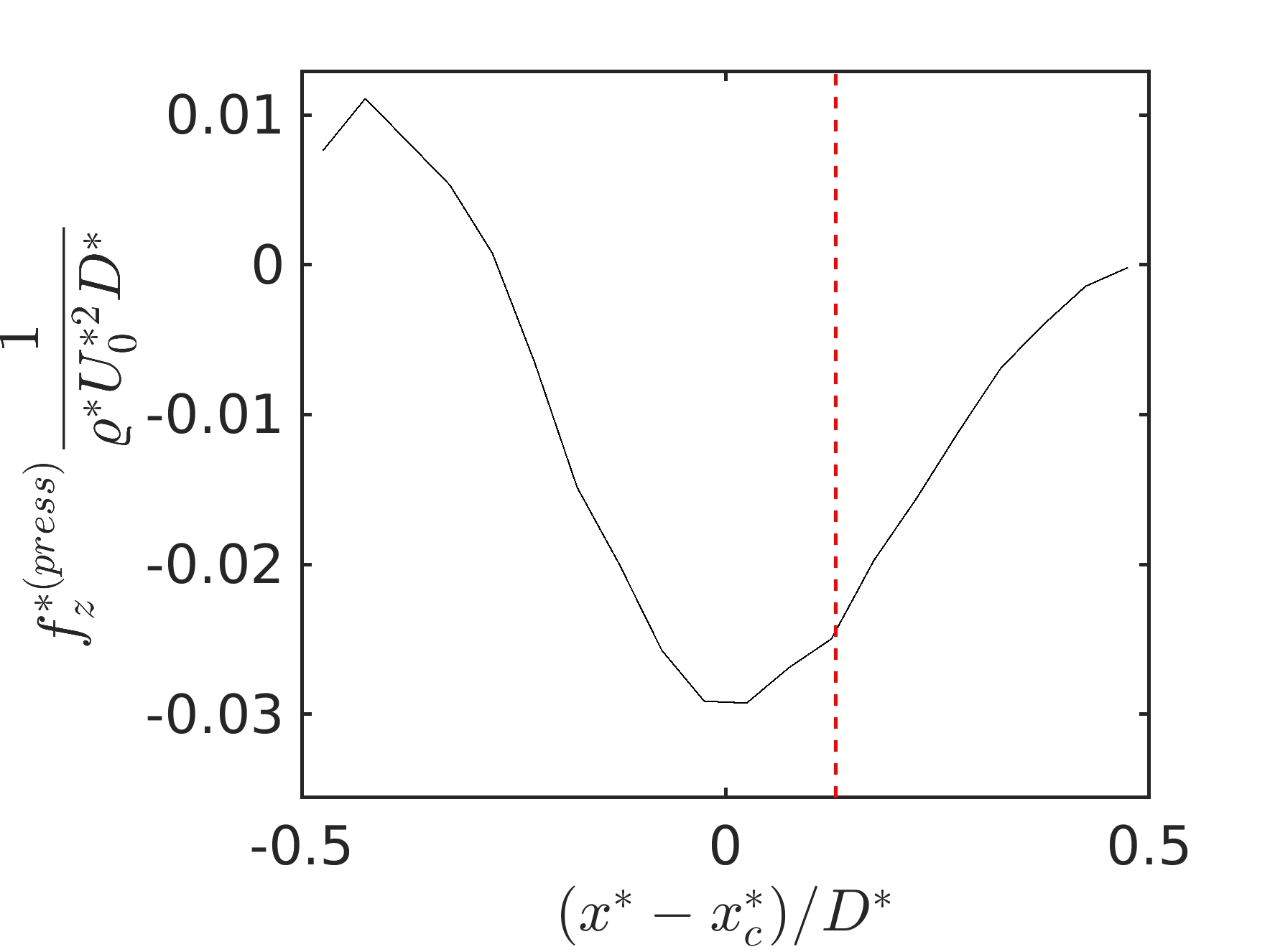}
    \raisebox{4cm}{\small$b)$}
    \includegraphics[width=0.42\textwidth]{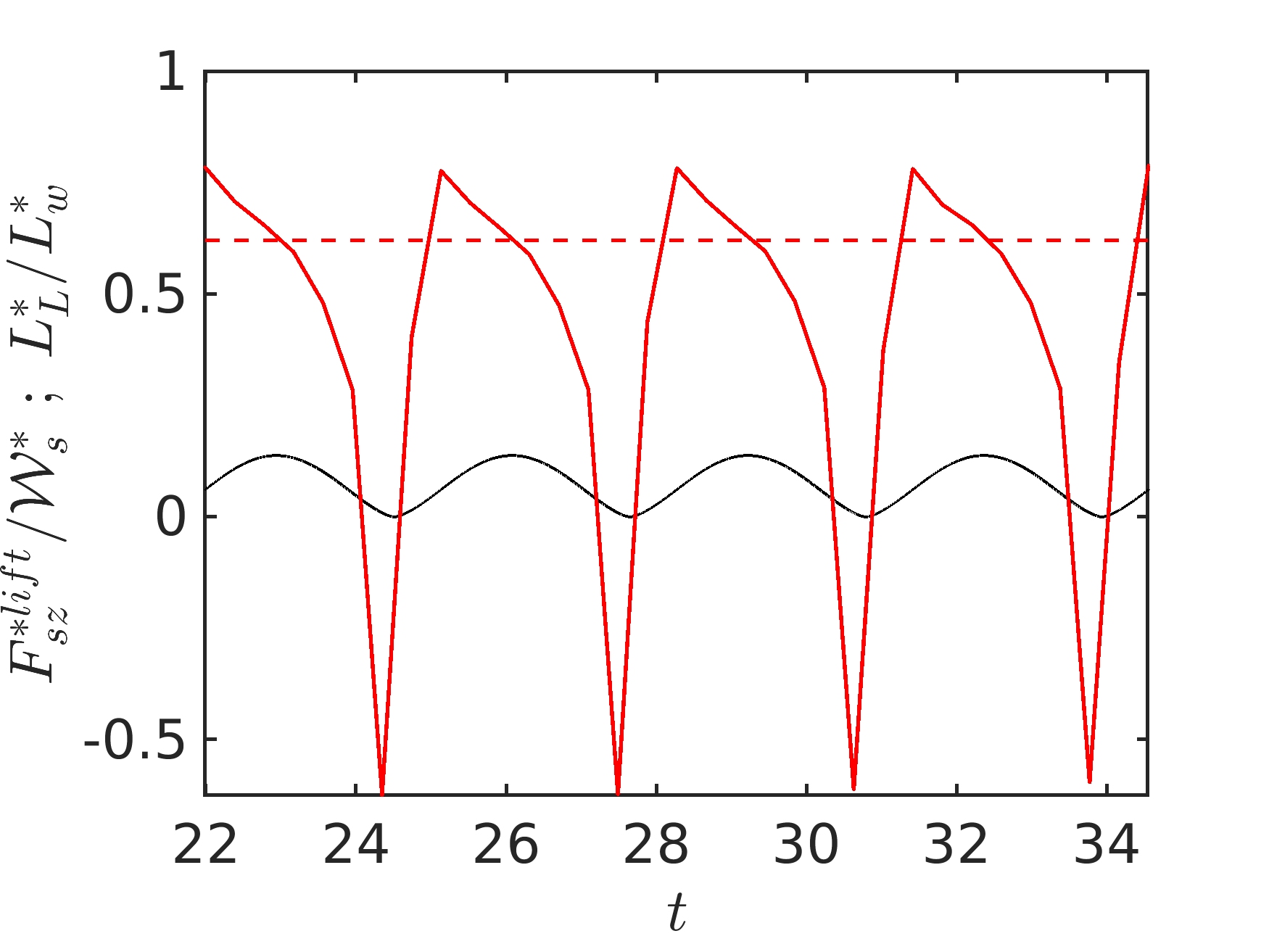}\\
    \raisebox{4cm}{\small$c)$}
    \includegraphics[width=0.47\textwidth]{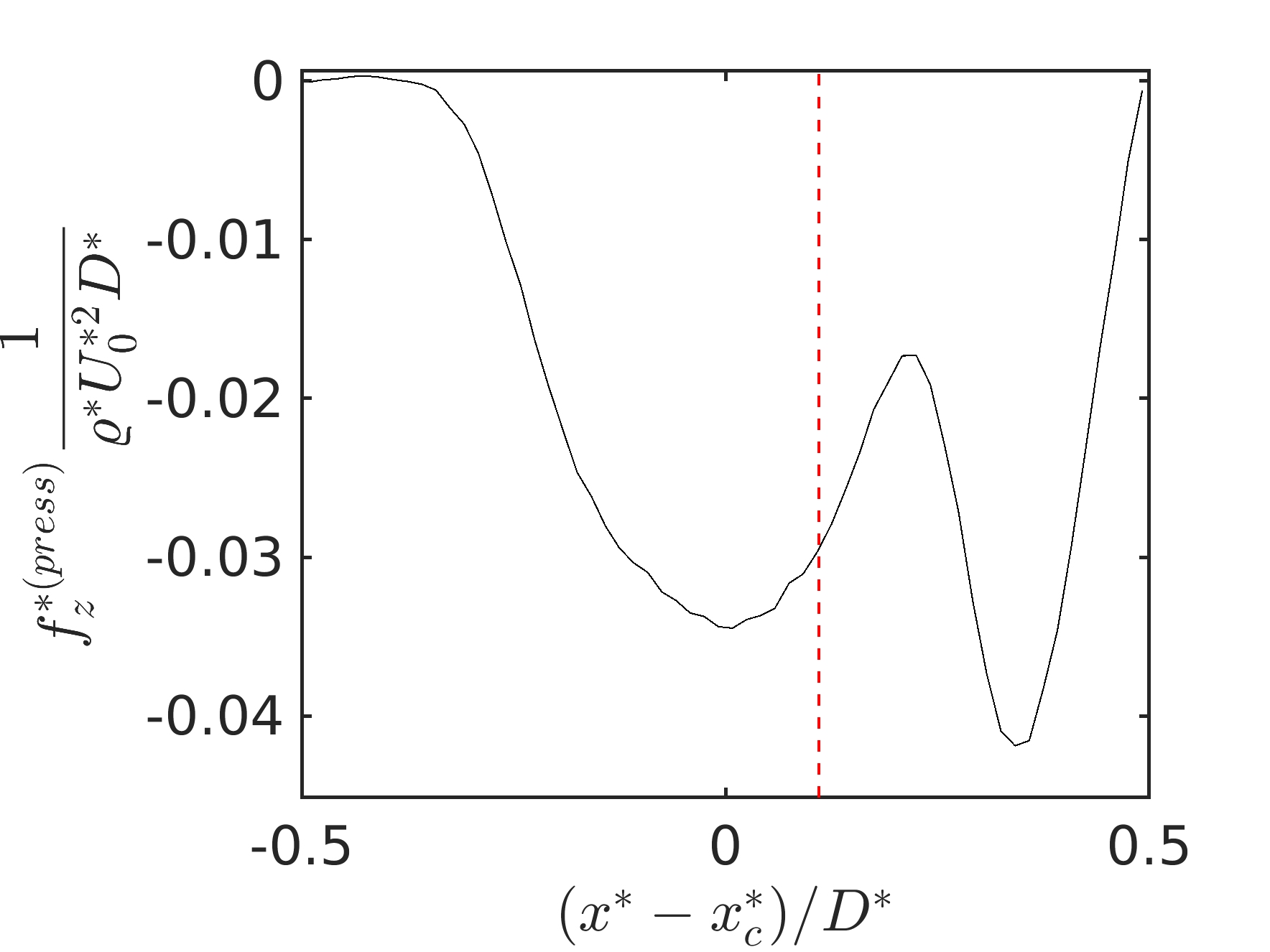}
    \raisebox{4cm}{\small$d)$}
    \includegraphics[width=0.42\textwidth]{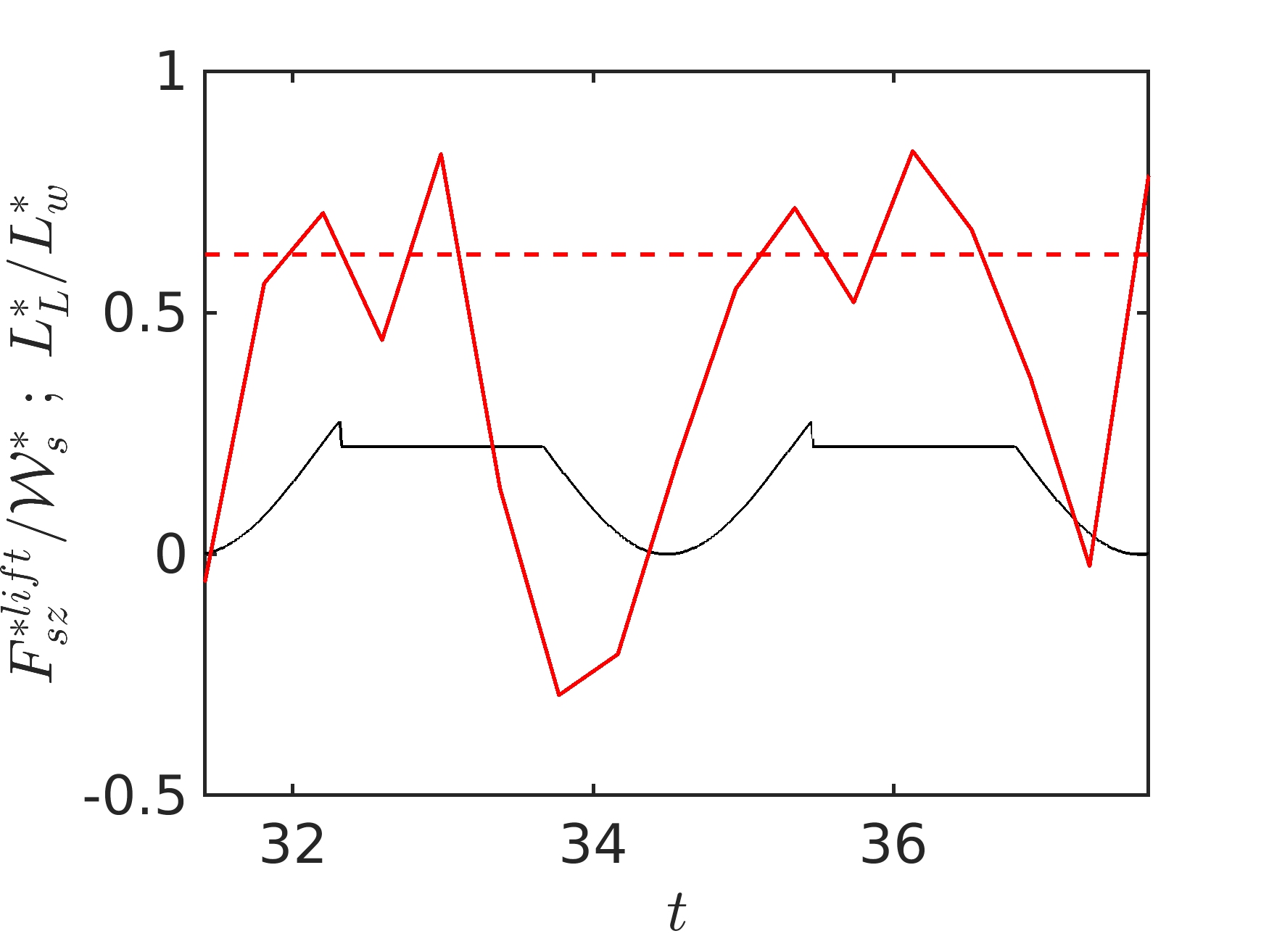}
    \caption{Panels (a) and (c) show the spatial distribution of the vertical force density $f_z$ (defined in \eqref{eq:fz}) along the normalised horizontal coordinate for runs 6 and 4, respectively, at the phase of maximum velocity far above the substrate. The vertical red dashed lines mark the location of the resultant lift force, as defined by \eqref{eq:Ll}.
    Panels (b) and (d) show the temporal evolution of the lift lever arm $L_L$ normalised by the weight lever arm $L_W$ (red), and the lift force $F_\mathrm{lift}$ normalised by the submerged weight $\mathcal{W_s}$ (black), for the same runs. The horizontal red dashed lines indicate the phase-averaged value of the normalised lever arm during the phases of maximum lift force.}
    \label{fig:lift_larm}
\end{figure}

The torques associated with forces in the vertical direction include the gravitational torque, depending on the submerged weight of the particle, and the lift force. The corresponding lever arm associated with the particle's weight is $L_\mathrm{w}=D/4$, while the lever arm of the lift force, $L_L$, was obtained directly from the DNS. %
In all cases, the lift-force lever arm was found to be smaller than $L_w$. During phases of the oscillation cycle when the lift force is significant, $L_L\simeq 0.62~L_w$ corresponding to a dimensionless value $L_L/D=0.155$ (see figure~\ref{fig:lift_larm}). 
This reduced value indicates that the lift force acts further from the particle centroid than the gravitational force, which is reflected in the net torque balance.
In fact, during these phases, the stress distribution over the sphere surface is asymmetric between the upstream and downstream sides. As a result, the net force is shifted downstream, as illustrated in figures~\ref{fig:lift_larm}a~and~c, which correspond to relatively small and large particles, respectively. %

Consequently, the torques due to the submerged weight and the lift force are given by
\begin{align}
    T_{By} &= - \frac{{\cal W}_sD}{2} 
    \:,\\
    T_{Sy}^{\mathrm{lift}} &= 3.22 C'' \rho_f\nu^2\sqrt{\mathrm{Re}_s^3} L_L 
    \label{eq:w_adim}
\end{align}
where ${\cal W}_s$ is the submerged weight of the particle.
%

Finally, the ratio $\Upsilon$ between destabilising torques (due to hydrodynamic drag, pressure gradient, added mass, and lift force) and the stabilising torque (due to the particle's submerged weight) is defined as
\begin{align}\label{eq:upsilon}
    \Upsilon 
    &\equiv 
    \dfrac{T_{Sy}^\mathrm{drag}+T^{\nabla p}_{Sy}+T_{Sy}^\mathrm{AM}+T_{Sy}^\mathrm{lift}}{T_{By}} \nonumber\\
    &= 
    4\Gamma \left[
    9C'\delta^2H
    g\left(\omega t,\frac{H}{\delta}\right)\,L_D
    + 
    h(\omega t,\zeta)\,L 
    + 
    \frac{9.66}{\pi\sqrt{2}}\,C^{\prime\prime}\dfrac{\delta^3}{\mathrm{Re}_\delta}\sqrt{\mathrm{Re}_s^3}L_L
    \right],
\end{align}
where the function $h(\omega t,\zeta)$ groups the contributions of the pressure gradient and added mass:
\begin{equation}
    h(\omega t,\zeta) = \frac{3}{2}\cos(\omega t)-\frac{1}{2}e^{-\zeta}\cos\left(\omega t-\zeta\right).
\end{equation}
The onset of particle emotion is predicted when $\Upsilon$ reaches or exceeds unity: 
\begin{equation}\label{eq:upsilon_geq1}
\Upsilon\left(\omega t;\,\mathrm{Re}_\delta,\,\delta,\,\Gamma,\,s,\,H,\,L\right) \geq 1.
\end{equation}

    

\subsection{Substrate effects at moderate Reynolds numbers}\label{sec:modRe}
The model described so far relies on the assumption that the ambient flow velocity is well approximated by the analytical solution for Stokes' oscillatory boundary layer, which is valid under laminar, smooth-wall flow conditions. This assumption holds when the viscous boundary layer thickness $\delta$ is sufficiently large compared to the substrate roughness. 

\begin{figure}
    \centering
    \includegraphics[width=0.47\textwidth]{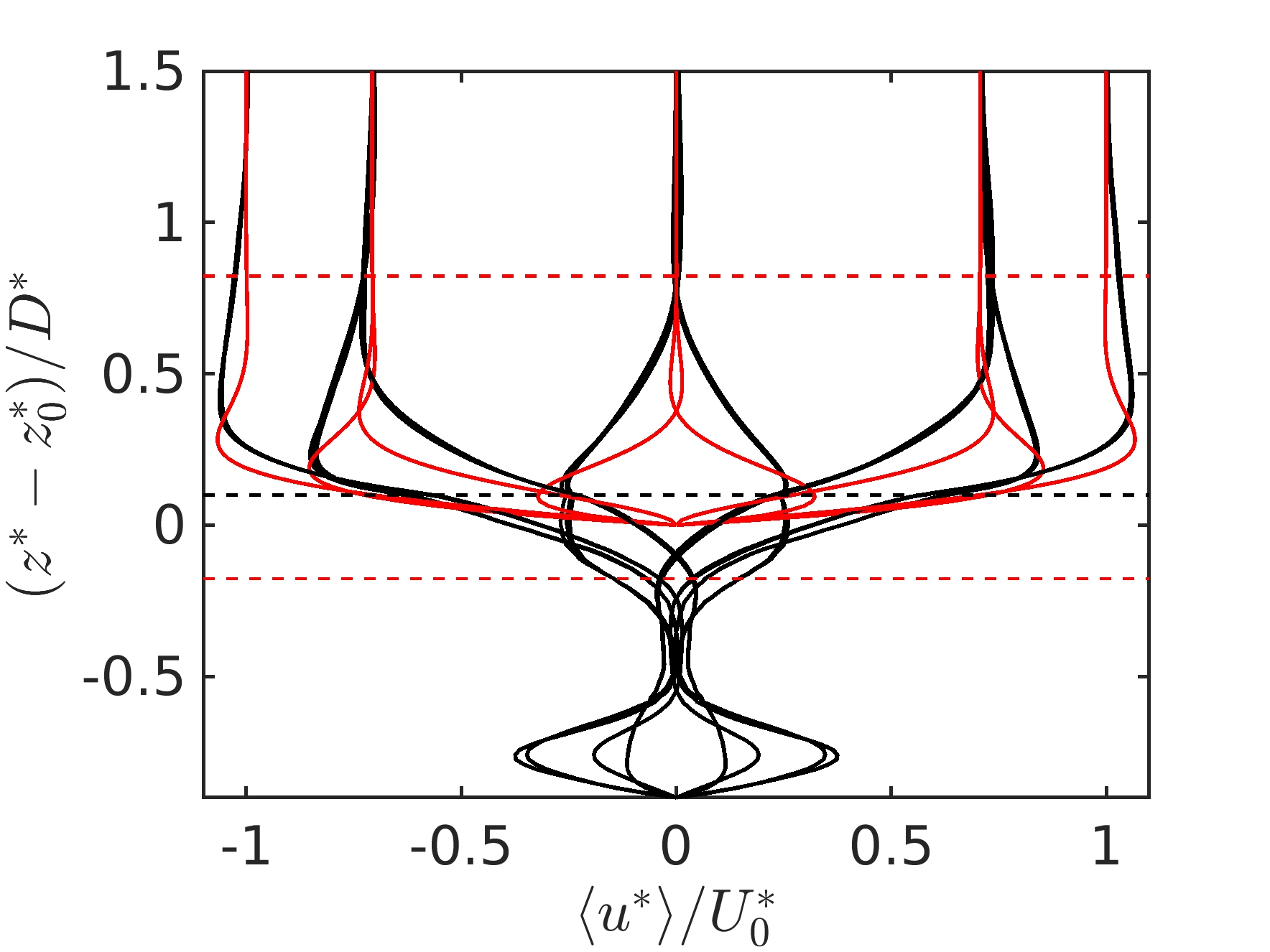}
    \includegraphics[width=0.47\textwidth]{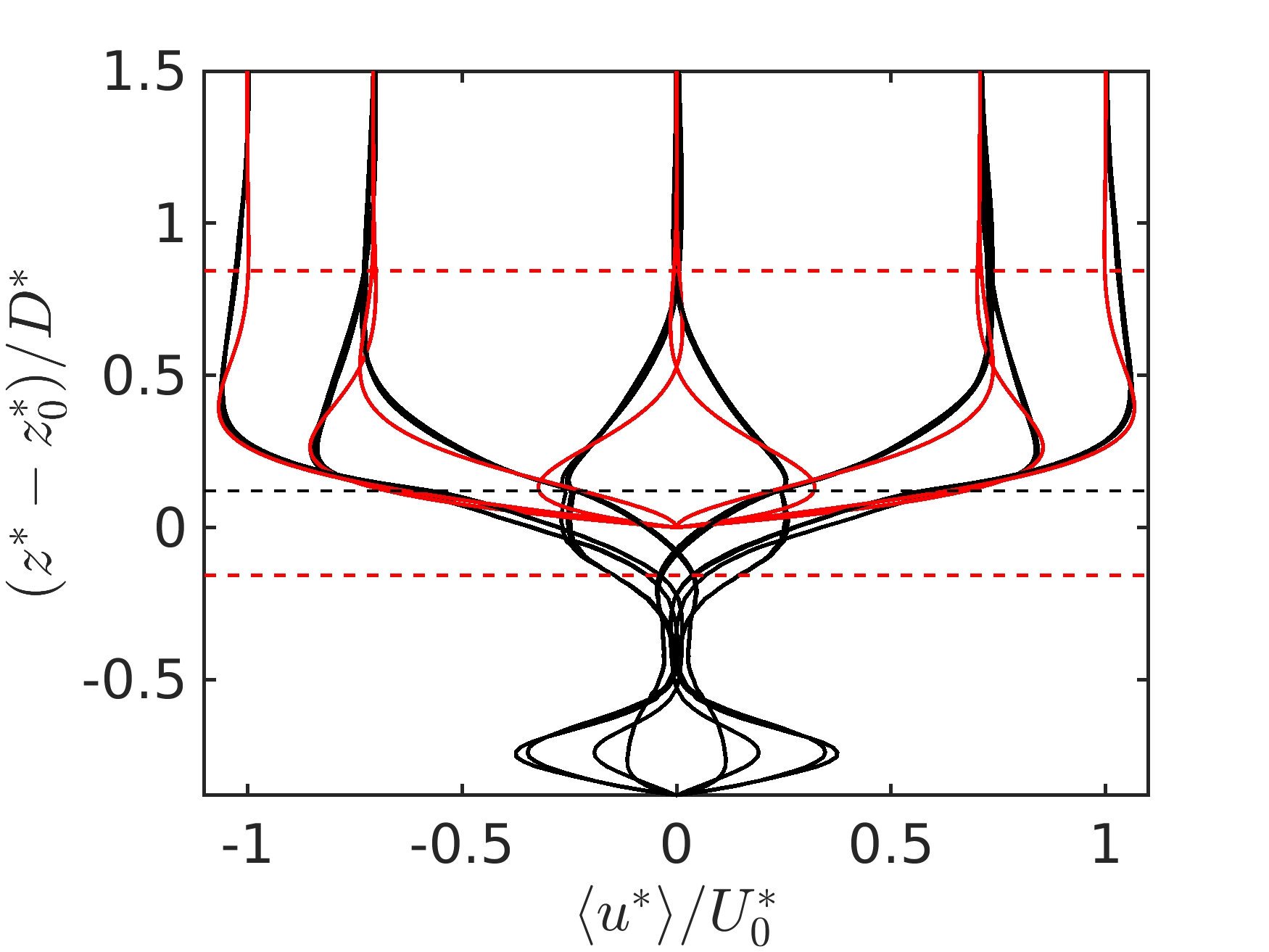}
    \caption{Velocity profiles from run~$4$ (black curves) compared with analytical Stokes profiles (red curves) shifted upward by $z_0$. The Stokes profiles are scaled using either (a) the viscous length scale $\delta$, or (b) the amplified length scale $\beta\delta$, where the amplification coefficient is set to $\beta=1.22$. %
    }
    \label{fig:dns_beta}
\end{figure}

However, for $\delta<1$ and moderate values of $\mathrm{Re}_\delta$, the flow can become increasingly influenced by the bottom roughness. In this regime, the effective length scale of the flow grows beyond $\delta$ and can approach the particle diameter $D$ as $\mathrm{Re}_\delta$ increases \citep{jensen1989turbulent}.
Consequently, the actual velocity profile deviates from the Stokes solution, as shown in figure~\ref{fig:dns_beta}a. %

Nonetheless, as long as the flow does not become turbulent, a reasonable approximation of the velocity profile can be recovered by introducing an effective boundary layer thickness, defined as $\delta_S=\beta\sqrt{2\nu/\omega}$, where $\beta>1$ is an empirical correction factor. %
For example, in run~$4$, characterised by $\mathrm{Re}_D>1500$, setting $\beta = 1.22$ yields a much improved agreement between the model predictions and DNS results (see figure~\ref{fig:dns_beta}b).
This empirical correction extends the applicability of the torque-based motion criterion discussed in \S~\ref{sec:torque} to moderate $\mathrm{Re}_D$ values, while retaining the same theoretical framework.

Finally, the present approach can be generalised to different substrate configurations. For instance, in a compact square arrangement, the lever arm of the hydrodynamic torque is given by $L=(\sqrt{2}/4)D$, while the lever arm of the submerged weight $L_\mathrm{w}$ varies between $D/4$ and $(\sqrt{2}/4)D$, depending on the flow orientation relative to the substrate geometry.
In a closely-packed hexagonal arrangement, these values become $L = (\sqrt{6}/6)D$ and $L_\mathrm{w} = [1,\,2]\,(\sqrt{3}/12)D$.
Further investigations will be required to incorporate the effects of substrate geometry and packing density (i.e., inter-particle spacing), which introduce additional degrees of freedom to the problem.

\section{Contact model}\label{sec:contact}
\begin{figure}
    \centering
    {\color{black!}\input{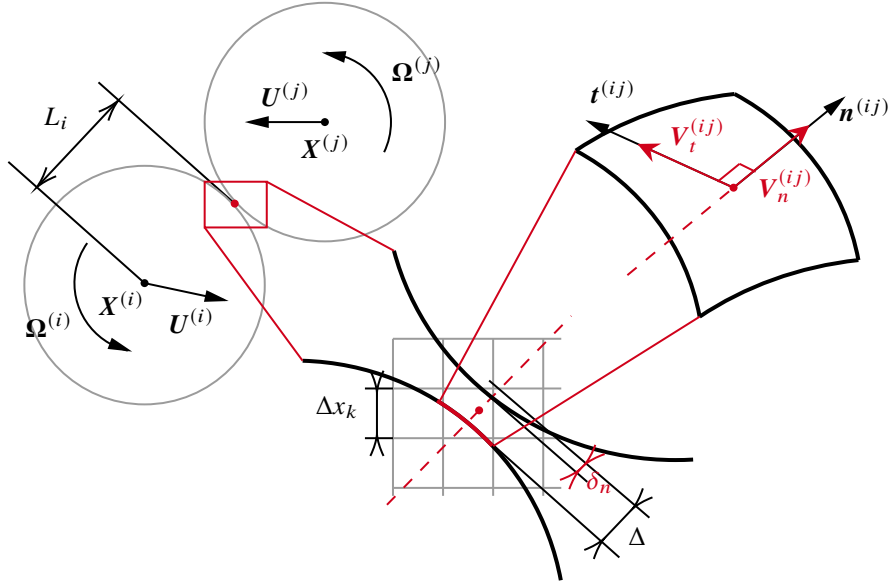}}
    \caption{Schematic of the inter-particle contact kinematics. Particles $i$ and $j$ are assumed to have equal radii.}
    \label{fig:cntmod}
\end{figure}

A soft-sphere collision model is employed to describe particle contact dynamics.
%
%
When two spherical particles, with diameter $D$ and centre positions at $\boldsymbol{X}^{(i)}$ and $\boldsymbol{X}^{(j)}$, approach each other, contact occurs when the distance between their centres, $\vert\boldsymbol{X}^{(i)}-\boldsymbol{X}^{(j)}\vert$, becomes smaller than $D+\Delta$. Here, $\Delta$ denotes a safety distance, which is taken to be equal to the grid spacing $\Delta x$ used in the IBM, as illustrated in figure~\ref{fig:cntmod}. %

The force developing at the contact point, $\boldsymbol{F}^{(ij)}$, is decomposed into surface normal and tangential components, $\boldsymbol{F}_n^{(ij)}$ and $\boldsymbol{F}_t^{(ij)}$, respectively. Each of these components is modelled as a spring-dashpot system, where the spring term accounts for the elastic contributions, while the dashpot term represents the dissipative, inelastic contributions. %
Thus, the normal and tangential components of the force that the $i$th particle exerts on the $j$th particle are given by the sum of their respective elastic and dissipative contributions:
\begin{align}
    \boldsymbol{F}_n^{(ij)} &= \boldsymbol{F}_n^\mathrm{el} + \boldsymbol{F}_n^\mathrm{dis}\label{eq:ctc_n}\:\:,\\
    \boldsymbol{F}_t^{(ij)} &= \boldsymbol{F}_t^\mathrm{el} + \boldsymbol{F}_t^\mathrm{dis}\label{eq:ctc_t}\:\:,
\end{align}
respectively. %
The dissipative contributions $\boldsymbol{F}_n^\mathrm{dis}$ and $\boldsymbol{F}_t^\mathrm{dis}$ are modelled as the product of the respective components of the relative velocity $\boldsymbol{V}^{(ij)}$ at the contact point, the effective mass (which is $\frac{1}{2}m$ for particles of equal mass $m$), and the respective damping coefficients, $\mu_{n}$ and $\mu_{t}=0.5\,\mu_{n}$. %
The relative velocity $\boldsymbol{V}^{(ij)}$ is equal to 
\begin{equation}
    \boldsymbol{V}^{(ij)} 
    = 
    \boldsymbol{V}^{(i)}-\boldsymbol{V}^{(j)} + 
    L^{(ij)}(\boldsymbol{\Omega}^{(i)} + \boldsymbol{\Omega}^{(j)})
    \times 
    \boldsymbol{n}^{(ij)}
    \:\:,
\end{equation}
where $\boldsymbol{V}^{(i)}$ and $\boldsymbol{V}^{(j)}$ are the linear velocities of the particle centres, $L^{(ij)}$ is the distance from the contact point to the centre of the $i$th particle, $\boldsymbol{\Omega}^{(i)}$ and $\boldsymbol{\Omega}^{(j)}$ are the angular velocities of the particles, and $\boldsymbol{n}^{(ij)}$ is the surface-normal unit vector pointing towards the $j$th particle. %
The damping coefficients depend on the normal restitution coefficient, $e_n$, the effective mass, and the normal stiffness coefficient $k_{n}$ \citep{silbert2001}. %
In particular, the normal damping coefficient is assumed to be twice the value of the tangential damping coefficient. %
The parameters for the contact model are summarised in table~\ref{tab:ctcparams}. %

The normal and tangential elastic contributions to the contact force are proportional to the respective normal and tangential displacements, $\boldsymbol{\delta}_n^{(ij)}$ and $\boldsymbol{\delta}_t^{(ij)}$, at the contact point, and are given by %
\begin{align}
    \boldsymbol{F}_n^\mathrm{el} &= -\boldsymbol{\delta}_n^{(ij)} k_n\label{eq:elctc_n}\\
    \boldsymbol{F}_t^\mathrm{el} &= -\boldsymbol{\delta}_t^{(ij)} k_t\label{eq:elctc_t}
    \:\:,
\end{align}
where $\boldsymbol{\delta}_n^{(ij)}=\delta_n\,\boldsymbol{n}^{ij}$, and the normal deformation is $\delta_n=d+\Delta-2\,L^{(ij)}$ (see figure~\ref{fig:cntmod}). %
The tangential stiffness coefficient $k_t$ is assumed to be $k_t=(2/5)k_n$, following the approach by \citet{silbert2001}. %
The tangential displacement is computed by accounting for the possibility that elastic energy accumulated during the contact can be released when the corresponding tangential elastic force exceeds the Coulomb friction threshold. %
The particles do not slide, as long as the condition
\begin{equation}
    \vert\boldsymbol{F}_t^{(ij)}\vert\leq \mu_c^{(ij)}\,\vert\boldsymbol{F}_n^{(ij)}\vert
    \label{eq:slip}
\end{equation}
is satisfied, where $\mu_c^{(ij)}$ denotes the Coulomb coefficient of dynamic friction, and $\boldsymbol{F}_t^{(ij)}$ is the static friction force determined by \eqref{eq:ctc_t}, following a widely adopted approach in granular flow models \citep[e.g.][]{syamlal1993mfix}. %

To compute the elastic contribution to the static friction force, $\boldsymbol{F}_t^\mathrm{el}$, the tangential displacement $\boldsymbol{\delta}^{(ij)}_t$ is updated at each time step $\Delta t$ according to the following expression: %
\begin{equation}
    \boldsymbol{\delta}^{(ij)}_t(t+\Delta t) = \hat{\boldsymbol{\delta}}^{(ij)}_t(t) + \delta_t\,\boldsymbol{t}^{(ij)}
    \:\:,
    \label{eq:delt}
\end{equation}
where $\delta_t=V_{t}^{(ij)}\Delta t$ and $\boldsymbol{t}^{(ij)}=\boldsymbol{V}_{t}^{(ij)}/\vert\boldsymbol{V}_{t}^{(ij)}\vert$ denote the increment of the tangential relative displacement and the displacement unit vector at time $t$, respectively. %
The initial value of $\delta_t$ is given by the smaller of two values: $V_{t}^{(ij)}\Delta t$ or $\delta_nV_{t}^{(ij)}/V_{n}^{(ij)}$, where $V_{n}^{(ij)}=\boldsymbol{V}^{(ij)}\cdot\boldsymbol{n}^{(ij)}$ represents the normal component of the relative velocity. %
In \eqref{eq:delt}, the accumulated tangential displacement $\boldsymbol{\delta}^{(ij)}_t(t)$ is adjusted for rotation of the contact plane during the time interval $\Delta t$ by the expression %
\begin{align}
    \hat{\boldsymbol{\delta}}^{(ij)}_t(t) 
    &= 
    (\boldsymbol{\delta}^{(ij)}_t(t)\cdot\boldsymbol{\zeta})\boldsymbol{\zeta}
    + 
    (\boldsymbol{\delta}^{(ij)}_t(t)\cdot\boldsymbol{t}_\mathrm{old})\boldsymbol{t}_\mathrm{new}
    \:\:,
\end{align}
where the unit vectors $\boldsymbol{\zeta}$, $\boldsymbol{t}_\mathrm{old}$, and $\boldsymbol{t}_\mathrm{new}$ are defined by
\begin{align}
    \boldsymbol{\zeta} &= \dfrac{\boldsymbol{n}^{(ij)}(t)\times\boldsymbol{n}^{(ij)}(t+\Delta t)}{\vert \boldsymbol{n}^{(ij)}(t)\times\boldsymbol{n}^{(ij)}(t+\Delta t)\vert} \:,\:
    \boldsymbol{t}_\mathrm{old} &= \dfrac{\boldsymbol{\zeta}\times\boldsymbol{n}^{(ij)}(t)}{\vert \boldsymbol{\zeta}\times\boldsymbol{n}^{(ij)}(t)\vert} \:,\:
    \boldsymbol{t}_\mathrm{new} &= \dfrac{\boldsymbol{\zeta}\times\boldsymbol{n}^{(ij)}(t+\Delta t)}{\vert \boldsymbol{\zeta}\times\boldsymbol{n}^{(ij)}(t+\Delta t)\vert}\:\:.
\end{align}
When the condition~\eqref{eq:slip} no longer holds, the two particles slide relative to each other at the contact point, and the static friction force, expressed by \eqref{eq:ctc_t}, is replaced by the dynamic friction force
\begin{equation}
  \boldsymbol{F}_t^{(ij)} =\left\lbrace
    \begin{array}{ll}
         -\mu_c^{(ij)}\, \vert\boldsymbol{F}_{n}^{(ij)}\vert\,\boldsymbol{t}^{(ij)} & \text{if}\:\boldsymbol{t}^{(ij)} \neq 0 \\
         -\mu_c^{(ij)}\, \vert\boldsymbol{F}_{n}^{(ij)}\vert\,\frac{\boldsymbol{\delta}_t}{\vert\boldsymbol{\delta}_t\vert} & \text{if}\:\boldsymbol{t}^{(ij)} = 0,\:\boldsymbol{\delta}_t\neq 0\\
         0 & \text{otherwise}.
    \end{array}
    \right.
\end{equation}
Finally, the total contact force and torque acting on the $i$th particle are obtained by summing the contributions from all surrounding particles in contact, as given by %
\begin{align}
    \boldsymbol{F}_c^{(i)} &= \displaystyle\sum_{j\neq i} \boldsymbol{F}^{(ij)}\:\:,\\
    \boldsymbol{T}_c^{(i)} &= \displaystyle\sum_{j\neq i} \left(L^{(ij)}\boldsymbol{n}^{(ij)}\times\boldsymbol{F}^{(ij)}\right)
    \:\:,
\end{align}
respectively. %
To solve the contact dynamics effectively, the fixed time-substep $\Delta t_{ctc}$ used to solve \eqref{eq:theory_particle_trans_dimless} and \eqref{eq:theory_particle_rot_dimless} is chosen to be ${\mathcal O}(10^2)$ smaller than the fixed time-step $\Delta t$ adopted to advance the flow field. %
\begin{table}
    \centering
    \begin{tabular}{cccccc}
        $\delta$ & $\mu_c$ & $\tilde k_n$ & $e_n$ & $\Delta/\Delta x$ & $D/\Delta x$ \\
        \hline
        $0.1$ & $0.4$ & $1\cdot10^2$ & $0.3$ & $1.0$ & $63$ \\
        $1.0$ & $0.4$ & $2\cdot10^3$ & $0.3$ & $1.0$ & $16$ \\
    \end{tabular}
    \caption{Contact model parameters. The parameters denote the Coulomb coefficient of dynamic friction $\mu_c$, the dimensionless normal stiffness coefficient $\tilde k_n=6\,k_n^*\Delta x/(\pi\,\varrho_sgD^{3})$, the dry restitution coefficient $e_n$, the dimensionless penalty force range $\Delta/\Delta x$, and number of grid points per particle diameter $D/\Delta x$.}
    \label{tab:ctcparams}
\end{table}
\bibliographystyle{apalike}
\bibliography{main}

\end{document}